\documentclass[pra,twocolumn,superscriptaddress]{revtex4-1}
\usepackage{makeidx}
\usepackage[utf8]{inputenc}
\usepackage{graphicx}
\usepackage{amsmath}
\usepackage{mathrsfs}
\usepackage{graphicx}
\usepackage{subfigure}
\usepackage{dcolumn}
\usepackage{bm}
\usepackage{bbm}
\usepackage{color}
\usepackage[dvipsnames]{xcolor}
\usepackage{esint}
\usepackage{lineno}
\usepackage[breaklinks,
            colorlinks,
            urlcolor=blue,
            linkcolor=blue,
            anchorcolor=blue,
            citecolor=blue]{hyperref}

\begin{document}

\title{Quantum transistor realized with a single $\Lambda$-level atom coupled to the microtoroidal cavity }

\author{Davit Aghamalyan}
\affiliation{ Centre for Quantum Technologies, National University of
Singapore, 3 Science Drive 2, Singapore 117543}
\author{Jia-Bin You}
\affiliation{Institute of High Performance Computing, Agency for Science, Technology and Research (A*STAR), 1 Fusionopolis Way, \#16-16 Connexis, Singapore 138632}
\author{Hong-Son Chu}
\affiliation{Institute of High Performance Computing, Agency for Science, Technology and Research (A*STAR), 1 Fusionopolis Way, \#16-16 Connexis, Singapore 138632}
\author{Ching Eng Png}
\affiliation{Institute of High Performance Computing, Agency for Science, Technology and Research (A*STAR), 1 Fusionopolis Way, \#16-16 Connexis, Singapore 138632}
\author{Leonid Krivitsky}
\affiliation{Institute of Materials Research and Engineering, Agency for Science, Technology and Research (A*STAR) 138634, Singapore}
\author{Leong Chuan Kwek}
\affiliation{MajuLab, CNRS-UNS-NUS-NTU International Joint Research Unit, UMI 3654, Singapore}
\affiliation{ Centre for Quantum Technologies, National University of
Singapore, 3 Science Drive 2, Singapore 117543}
\affiliation{National Institute of
Education and Institute of Advanced Studies, Nanyang Technological
University, 1 Nanyang Walk, Singapore 637616}

\begin{abstract}

We propose a realization of the quantum transistor for coherent light fields for the fibre-coupled microdisk cavities. We demonstrate by combining numerical and analytical methods that both in strong coupling and bad cavity limits it is possible to change system's behaviour from being fully transparent to being fully reflective by varying the amplitude of the external control field. We remark that tuning the amplitude of the control field is significantly easier in the experimental setting than tuning cavity-atom coupling strength which was suggested in ref.\cite{PA14} for two-level atoms and works only in the strong coupling limit. We also demonstrate the possibility of controlling the statistics of the input coherent field with the control field which opens the venue for obtaining quantum states of light.


\end{abstract}

\maketitle


\section{Introduction}
Quantum networks\cite{Kimble2008} provide a prominent template for the design and realization of scalable quantum information processing systems. Quantum network consists of nodes, which are formed with a physical system such as atoms. Nodes are then linked together through the quantum channel, and this usually is done with the help of photons referred in this context as ``flying qubits". The interaction between light and matter establishes the transfer and the manipulation of information between the ``flying qubits" and the nodes. Quantum networks may eventually play an important role in the future implementation of quantum computation, communication, and metrology\cite{Nielsen2000,Zoller2005,Gisin2007,Giovanetti2004,Giovanetti2010}.

Trapped atoms in Fabry-Perot cavities have been one of the most fruitful systems for testing fundamentals of quantum optics in cavity QED setup\cite{Haroche1989,WV2006,SZ97,C91}. Single atoms in Fabry-Perot cavities have been demonstrated to be good candidates for a quantum network\cite{Raimond2001,BPK12,Cirac1997,RR2015,WV2006}, however, it turns out that these cavities fail to realize large scale networking. To overcome this issue, several types of microchip-based systems(microdisk, micropillar, micro bottle, and photonic crystal cavities)\cite{Vahala2004} have been engineered and successfully utilized for implementing cavity-QED type of experiments\cite{A2006,A2009,S2007,OJV2013,kippenberg2005,S2014,Dayan2008} by coupling them with trapped cold atoms, quantum dots. Numerical and theoretical methods have also been developed to understand the optical properties of these systems \cite{KDL1999,Spillane2005,SP07,S2006}.

Microtoroidal and microdisk cavities hold a promise to realize scalable quantum networks and are fascinating platforms for realizing quantum optical experiments since the electrical field, with its small mode volume, reaches high values inside the cavity resulting in a large light-matter coupling. Experimentally, the strong coupling regime has been successfully reported for such systems\cite{A2006,A2009,S2007,OJV2013}. Due to their small losses, these systems have high quality factors (Q) and, in one experiment,   $Q$ as large as $4\times 10^{8}$ have been realized \cite{kippenberg2005}. Moreover, by coupling tapered fibre with ring resonator the efficiency of coupling light in and out of the microtoroidal resonator can achieve up to 0.997  as demonstrated experimentally in Ref. \cite{A2006}.

Photonic quantum devices \cite{Brien2009} are necessary components for implementing functional quantum network, and they play an important role in storing the quantum states of light, as well in controlling the propagation of light. Switching the direction of light propagation is one of the most important operations that need to be performed in the quantum network. To achieve this task a quantum light transistor\cite{V2012,B2009,YY2009,W2013} ] is needed and is implemented by changing an external parameter, which results in ``on" or ``off" state of the transistor, much like a gate valve in a water pipe. If this device is implemented solely through optical means, then this kind of switch is called ``all optical switch" \cite{DI2005,B2009,YY2009,W2013}. Quantum transistors act as ``gate valves" for  quantum states of light\cite{Ste07,SZ97}.

Both theoretical proposals \cite{Kyriienko2016,Hong2008} along with actual experimental implementations \cite{V2012,W2013,B2009} for realizing single-photon transistor have been put forward.

In this paper, we focus on the realization of a quantum transistor for an incoming coherent field. An interesting result that quantum communication between two atomic ensembles can be achieved by means of only coherent laser fields has been theoretically proposed\cite{Duan2000} and later an entanglement between two atomic ensembles has been experimentally demonstrated  in  Ref.\cite{Polzik}. These findings demonstrate that quantum network can be formed with only coherent laser fields which overcomes the difficulty of creating quantum states of light for realizing quantum communication.

In the Ref. \cite{PA14}, Parkins and Aoki suggested an interesting scheme for the coherent light quantum switch by utilizing clockwise and anti-clockwise cavity mods of the whispering gallery modes(WGM) of the ring resonator. They showed that, under certain parameter regime (strong cavity-fibre coupling along with strong cavity-atom coupling), it is possible to achieve a coherent light switch by tuning cavity-atom interaction strength $g$ going from weak coupling to strong coupling regimes.

Here, we highlight, that controlling the interaction strength $g$ in the actual experiments could be very challenging because one needs to modify the distance between the atoms and ring cavity to modify the evanescent coupling. Moreover,  in order to have a functional switch, it is desirable to have an easily tunable external parameter. To address this issue, we propose to replace the two-level atom with a three-level atom in a $\Lambda$-level configuration and the ``gate valve" is implemented by tuning the amplitude of the control field. In several theoretical articles~\cite{LY2012,Hong2008,YL2011} the interaction of three-level atom has been theoretically investigated, however in all these papers the typical EIT condition of zero two-photon detuning has been assumed. Contrary, for our protocol it is crucial to have non zero two-photon detuning, otherwise because of the coherent population trapping mechanism system behaves as transparent for any value of control field, because of the optical pumping into the dark state\cite{Ste07,SZ97}. We demonstrate that by choosing suitably two-photon detuning, an all-optical switch can be efficiently implemented in our system.

In this manuscript, we argue that a quantum switch, controlled by varying the amplitude of the external field, is easier to implement experimentally compared to the previous proposals. Moreover, our protocol for a transistor works even for the bad-cavity limit, which overcomes experimental effort to bring the system in the strong coupling regime. However, it is important to point out that contrary to the strong coupling regime where reflected light does not change its statistics, in the bad cavity limit it becomes strongly quantum after being reflected.

The manuscript is outlined as follows. In the section \ref{Model}, we provide a theoretical description of the system and set the stage for the numerical simulations of the master equation that governs the systems dynamics.  In the section \ref{quantum switch}, we  demonstrate, both numerically and analytically, that our system functions as a quantum transistor for an incoming coherent field (even within the bad cavity limit). In  section  \ref{photonstatistics}, we study the statistics of the transmitted and reflected fields, in strong coupling and bad cavity limits. Section \ref{conclusion} is devoted to the conclusions. In  Appendix \ref{appendix}  analytical results for the bad-cavity limit are derived using adiabatic elimination of the cavity modes.

\section{The system and the master equation formalism}
\label{Model}
\begin{figure}
\includegraphics[width=8cm]{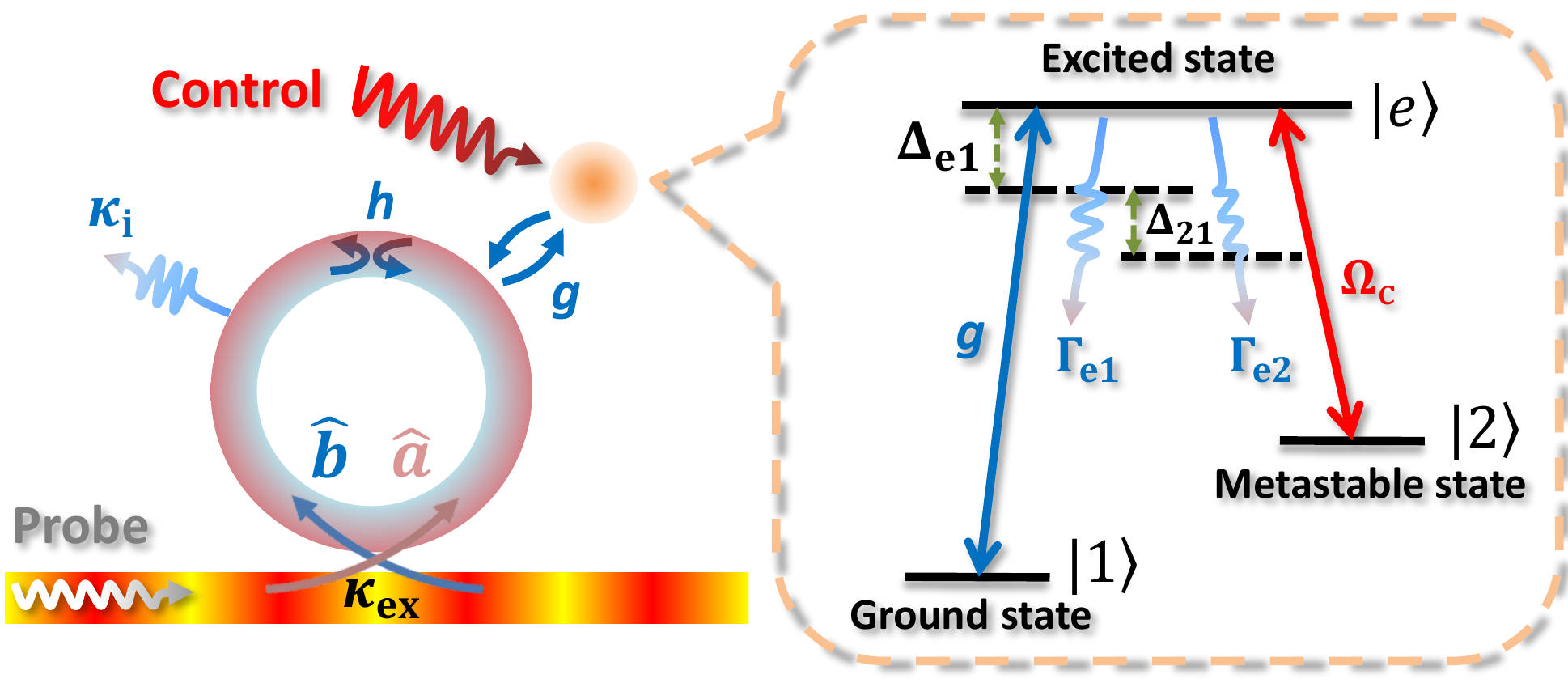}
\caption{ Scheme of a three-level atom coupled to a ring cavity and a tapered-fiber. Input fields $a_{in,ex},b_{in,ex}$ propagate through the fiber which is coupled with a rate $k_{ex}$ with a microtoroid cavity which has a resonant frequency of $ \omega_{r} $. Coherent probe field of frequency $\omega_{p}$ drives the mode $a$ with strength $\Omega_{p}$. Two counter-propagating WGM modes $a$ and $b$ are assumed to be coupled with a strength $h$ due to the scattering from imperfections. Both modes can leak out of the cavity with a rate $k_{i}$, and the outgoing fields resulting from the fiber are given by the $a_{out,ex},b_{out,ex}$ and related to the input and intra-cavity fields through the conventional input-output relations. Degenerate cavity modes $a$ and $b$  are  coupled with the three-level atom and drive the transition $1-e$. Control field with amplitude $\Omega_{c}$ and the frequency $\omega_c$ drives the transition $2-e$. Atomic populations of the excited state $ e$  can decay  through two decay channels either to the state $ 1  $ or to the state $ 2 $.}
\label{scheme}
\end{figure}
A schematic representation of the system along with main parameter definitions is given in  Fig.~\ref{scheme}. It is important to point out that once the anti-clockwise WGM mode $a$ is created, there are two different mechanisms that can give rise to the clockwise mode $b$. The first mechanism is the evanescent coupling with a strength $g$ with the two-level atom since atom can re-emit the photon in both directions: clockwise and anti-clockwise. The second mechanism is a result of the inhomogeneity in the dielectric media and is described by the parameter $h$(for more details, see Ref~\cite{SP07}). In this paper we will focus mainly on the case when $h=0$.


In a rotating frame $U(t)=e^{i\omega_{p}t(a^{\dag}a+b^{\dag}b-\sigma_{11})-i\omega_{c}t\sigma_{22}}$, the Hamiltonian for the system takes form
\begin{equation}
\label{atom-ring}
\begin{split}
H&=\Delta_{r}(a^{\dag}a+b^{\dag}b)+h(a^{\dag}b+b^{\dag}a)+\Delta_{e1}\sigma_{ee}+\Delta_{21}\sigma_{22}\\
&+(g^{*}a^{\dag}\sigma_{1e}+ga\sigma_{e1})+(gb^{\dag}\sigma_{1e}+g^{*}b\sigma_{e1})\\
&+\Omega_{p}(a+a^{\dag})+\Omega_{c}(\sigma_{2e}+\sigma_{e2}),\\
\end{split}
\end{equation}
where $\Delta_{r}=\omega_{r}-\omega_{p}$, $\Delta_{e1}=\omega_{e1}-\omega_{p}$ and $\Delta_{21}=\omega_{21}+\omega_{c}-\omega_{p}$. After introducing the dissipative channels the system is described by the Lindblad master equation (here we assume zero temperature thermal reservoir):
\begin{equation}
\begin{split}
\dot{\rho}&=-i[H,\rho]+\kappa\mathcal{D}[a]\rho+\kappa\mathcal{D}[b]\rho\\
&+\frac{\Gamma_{e1}}{2}\mathcal{D}[\sigma_{1e}]\rho+\frac{\Gamma_{e2}}{2}\mathcal{D}[\sigma_{2e}]\rho,\\
\end{split}
\label{mast.eq}
\end{equation}
where $\mathcal{D}[\hat{o}]\rho=2\hat{o}\rho\hat{o}^{\dag}-\hat{o}^{\dag}\hat{o}\rho-\rho\hat{o}^{\dag}\hat{o}$ is the Lindblad superoperator and $\kappa=\kappa_{\text{ex}}+\kappa_{\text{i}}$, $\Gamma_{e1}$, $\Gamma_{e2}$  are the decay rates of cavity and atom respectively.

\paragraph{Input-output formulation of the system.} The input and output fields which are schematically shown on Fig.\ref{scheme} and are related through the input-output relations (see the Chapter 7 of Ref.~\cite{WM94}),  given in the Heisenberg picture by the following expressions:
\begin{eqnarray}
a_{out,ex}(\tau)&=&-a_{in,ex}(\tau)+\sqrt{2\kappa_{ex}}a(\tau),  \\
b_{out,ex}(\tau)&=&-b_{in,ex}(\tau)+\sqrt{2\kappa_{ex}}b(\tau),
\end{eqnarray}
where input and output fields have delta function commutation relations in time. The field $\Omega_{p}$ in the Hamiltonian corresponds to coherent field incoming from the left and input field incoming  from the right is assumed to be in the vacuum state, which is given by the average values of the input operators:
\begin{equation}
\langle a_{in} \rangle=-\frac{i\Omega_{p}}{\sqrt{2\kappa_{ex}}}, \quad \langle b_{in} \rangle=0.
\end{equation}
The transmission and the reflection coefficients, normalized to the input photon flux number, are given by
\begin{equation}
\label{transmission}
T=\frac{\langle a^{\dagger}_{out,ex} a_{out,ex} \rangle }{|\Omega_{p}|^{2}/2\kappa_{ex}},\quad R=\frac{\langle a^{\dagger}_{out,ex} a_{out,ex}\rangle}{|\Omega_{p}|^{2}/2\kappa_{ex}}.
\end{equation}

\paragraph{Normal mode decomposition.} In order to achieve a better understanding of the system, it is instructive to rewrite the Hamiltonian in terms of the normal modes of cavity $A$ and $B$, defined as
\begin{equation}
A=\frac{a+b}{\sqrt{2}}, \quad B=\frac{a-b}{\sqrt{2}}
\end{equation}
After expressing $a$ and $b$ through the normal modes, the Hamiltonian of the system reads as
\begin{align}
\label{splitting}
\nonumber
 H_{N.M.} &= \Delta_{e1}\sigma_{ee}+\Delta_{21}\sigma_{22}+\Omega_{c}(\sigma_{2e}+\sigma_{e2})+(\Delta_{r}+h) A^{\dagger}A \\ \nonumber
 &+\frac{1}{\sqrt{2}}(\Omega^{*}_{p}A+\Omega_{p}A^{\dagger})+(g_{A}^{*}A^{\dagger}\sigma_{1e}+g_{A}  \sigma_{1e}A)\\  \nonumber
 &+(\Delta_{r}-h) B^{\dagger} B+\frac{1}{\sqrt{2}}(\Omega^{*}_{p}B+\Omega_{p}B^{\dagger})  \\
 &+(g_{B}^{*}B^{\dagger}\sigma_{1e}+g_{B}  \sigma_{1e}B)
 \end{align}
where $g_{A}$ and $g_{B}$, are given by:
\begin{align}
 g_{A}&=\sqrt{2}g_{0}f(r)\cos{(kx)} \\
 g_{B}&=\sqrt{2}g_{0}f(r)\sin{(kx)}
 \label{dec}
\end{align}
Eqs.~(\ref{dec}) show that by properly choosing the location of the atom along the ring cavity, it is possible to decouple one of the cavity modes. In the rest of the manuscript  assume that $\sin{(kx)}=0$, so that the mode $B$ is decoupled from the atom. It is easy to notice from the expression for $ H_{N.M.} $ that there are no terms in the Hamiltonian that couple mode $B$ with the atom or with other normal mode $A$ (that terms are given by the two last lines in the Eq.~\ref{splitting} and  we denote that part of Hamiltonian as $H_{B}$). This in turn implies that systems $\Sigma+A$(here $\Sigma$ represents the subspace of a two-level atom ) and $B$ are non-interacting and  the full system Hamiltonian and the density matrix are given by
\begin{align}
H_{N.M. }&=H_{\Sigma+A}+H_{B},\\
\label{dec1}
\rho &=\rho_{\Sigma+A} \otimes \rho_{B}.
\end{align}
Next, we proceed to write the master equation of the system in the normal mode basis
\begin{align}
\label{masteq2}
\dot{\rho}&=-i [H_{N.M.},\rho]+\kappa {\cal D}[A]\rho+\kappa {\cal D}[B]\rho \\ \nonumber
                &+\frac{\Gamma_{e1}}{2}\mathcal{D}[\sigma_{1e}]\rho+\frac{\Gamma_{e2}}{2}\mathcal{D}[\sigma_{2e}]\rho
\end{align}
After substituting Eq.~(\ref{dec1}) into Eq.~(\ref{masteq2}), and tracing out separately  the subsystems $\Sigma+A$ and $B$, equations for the respective subsystems take the following form:
\begin{align}
 \label{masteq3}
 \dot{\rho_{\Sigma+A}}&=-i [H_{\Sigma+A},\rho_{\Sigma+A}]+\kappa {\cal D}[A]\rho_{\Sigma+A} \\ \nonumber
 \label{masteq4}
 &+\frac{\Gamma_{e1}}{2}\mathcal{D}[\sigma_{1e}]\rho_{\Sigma+A}+\frac{\Gamma_{e2}}{2}\mathcal{D}[\sigma_{2e}]\rho_{\Sigma+A}, \\
\dot{\rho_{B}}&=-i [H_{B},\rho_{B}]+\kappa{\cal D}[B]\rho_{B},
 \end{align}
\paragraph{Remark.} It is important to notice that Eqs.~(\ref{masteq3}) and ~(\ref{masteq4}) present a significant numerical advantage compared to the Eq.~(\ref{mast.eq}), since in the first case full system density  matrix is obtained as(as a tensor product) by solving two separate equations for density matrices of dimension $~O(n)$ contrary to the second case where  one equation for the full system density matrix of the dimension  $~O(n^{2})$ has to be solved , here $n$ shows truncation number of the Fock state for the cavity modes.

\section{Quantum transistor}
\label{quantum switch}
\begin{figure}
\label{intenisties}
\begin{tabular}{ccccc}
\includegraphics[width=4cm]{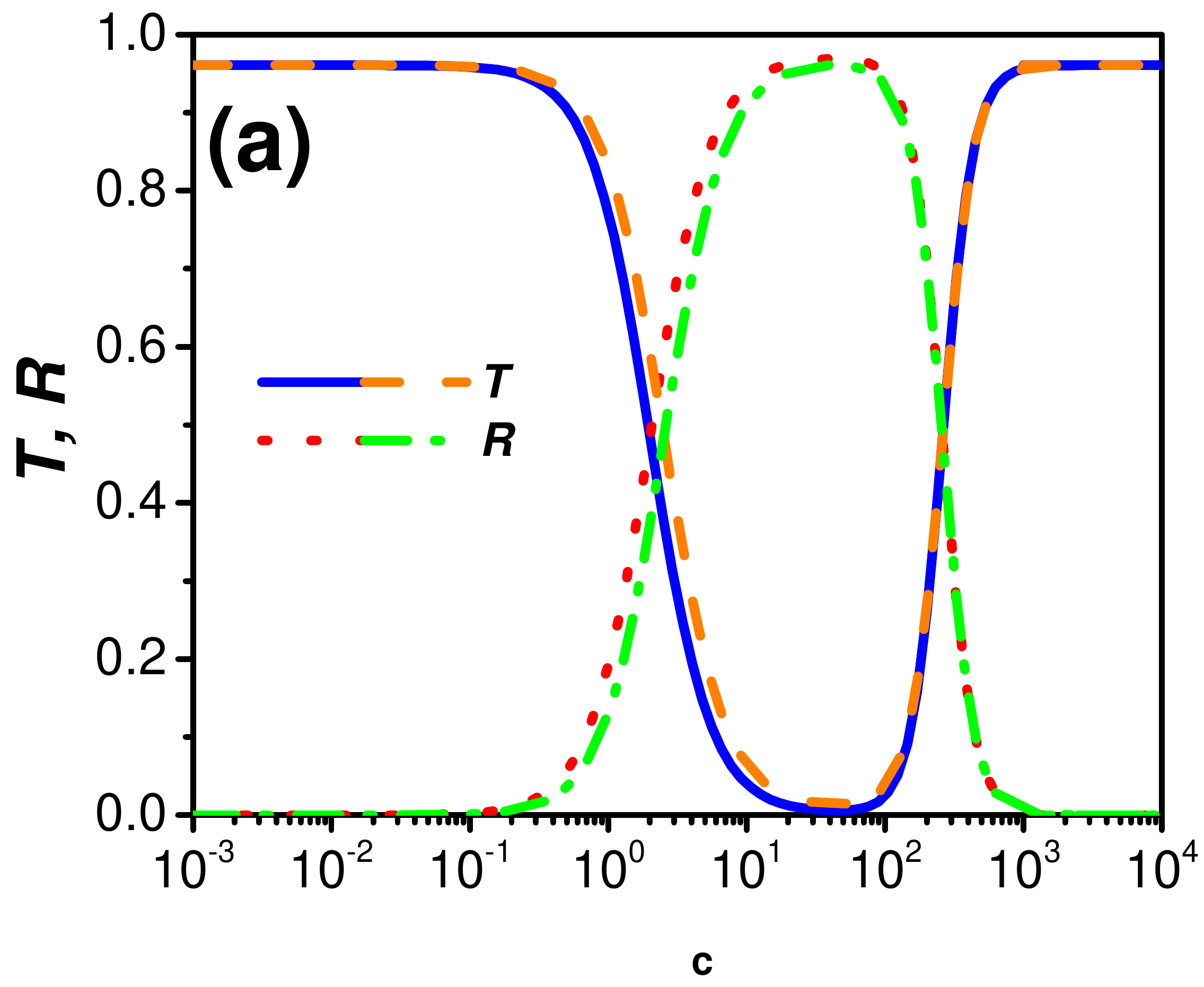} &
\includegraphics[width=4.1cm]{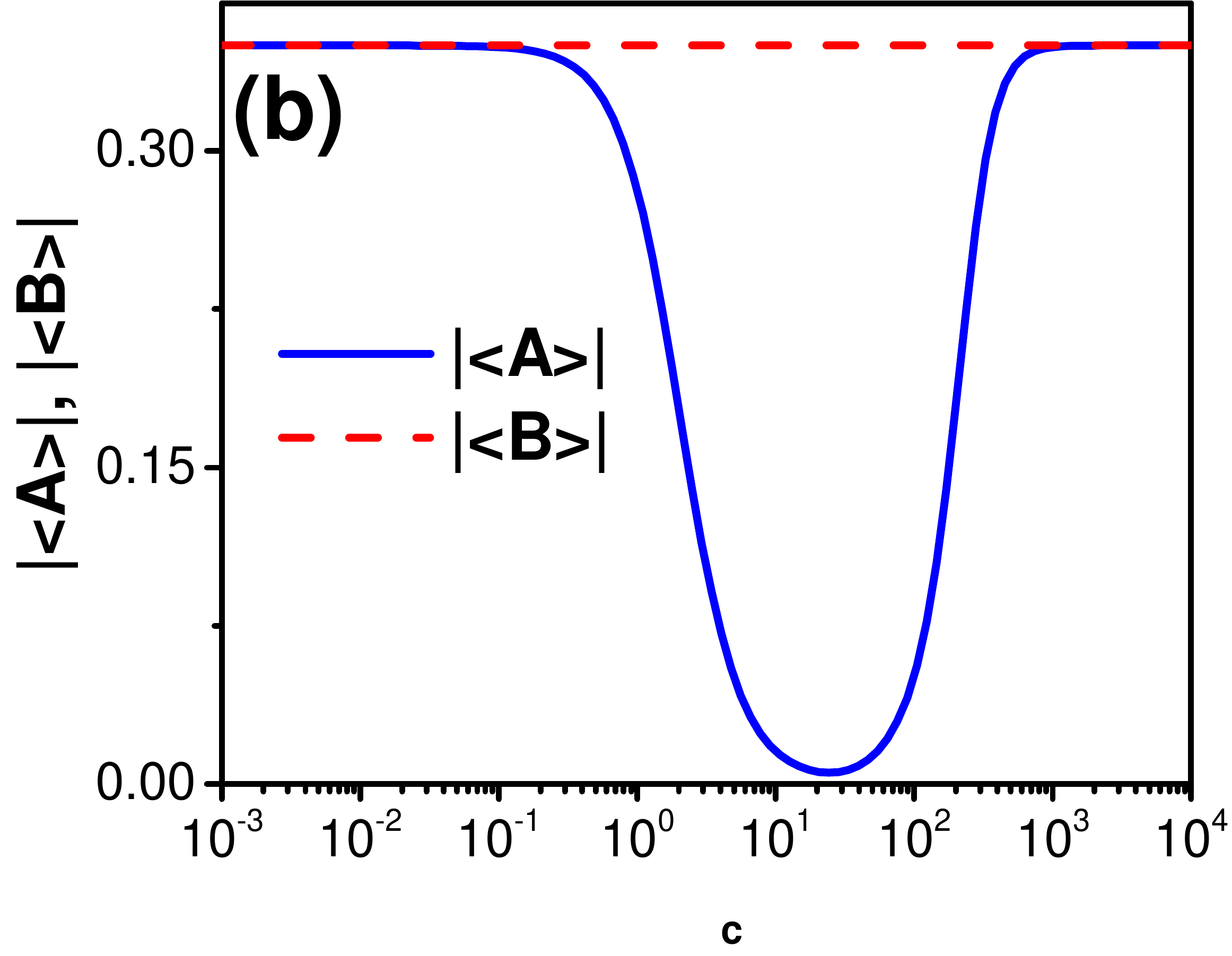} \\
\includegraphics[width=4cm]{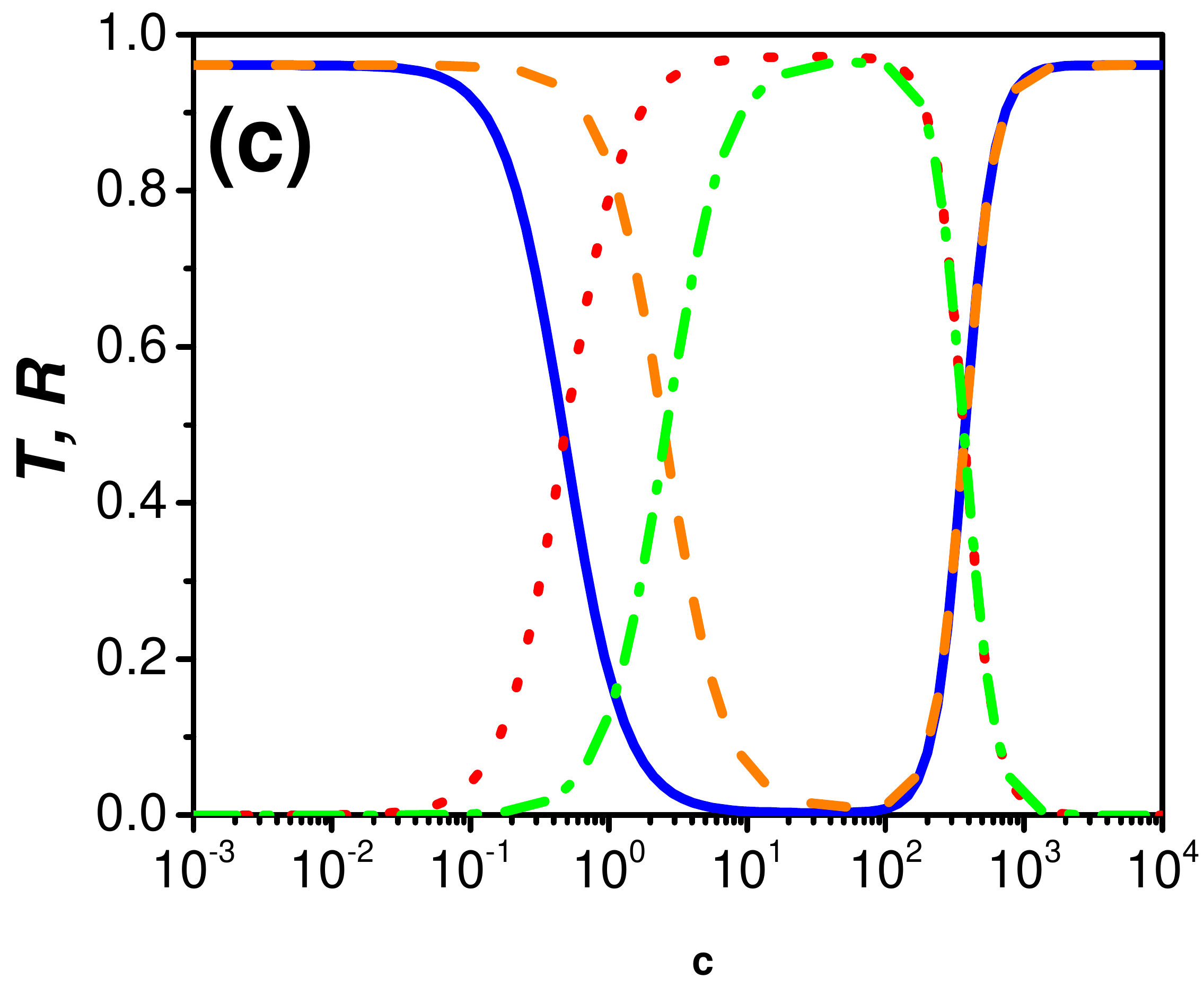} &
\includegraphics[width=4.1cm]{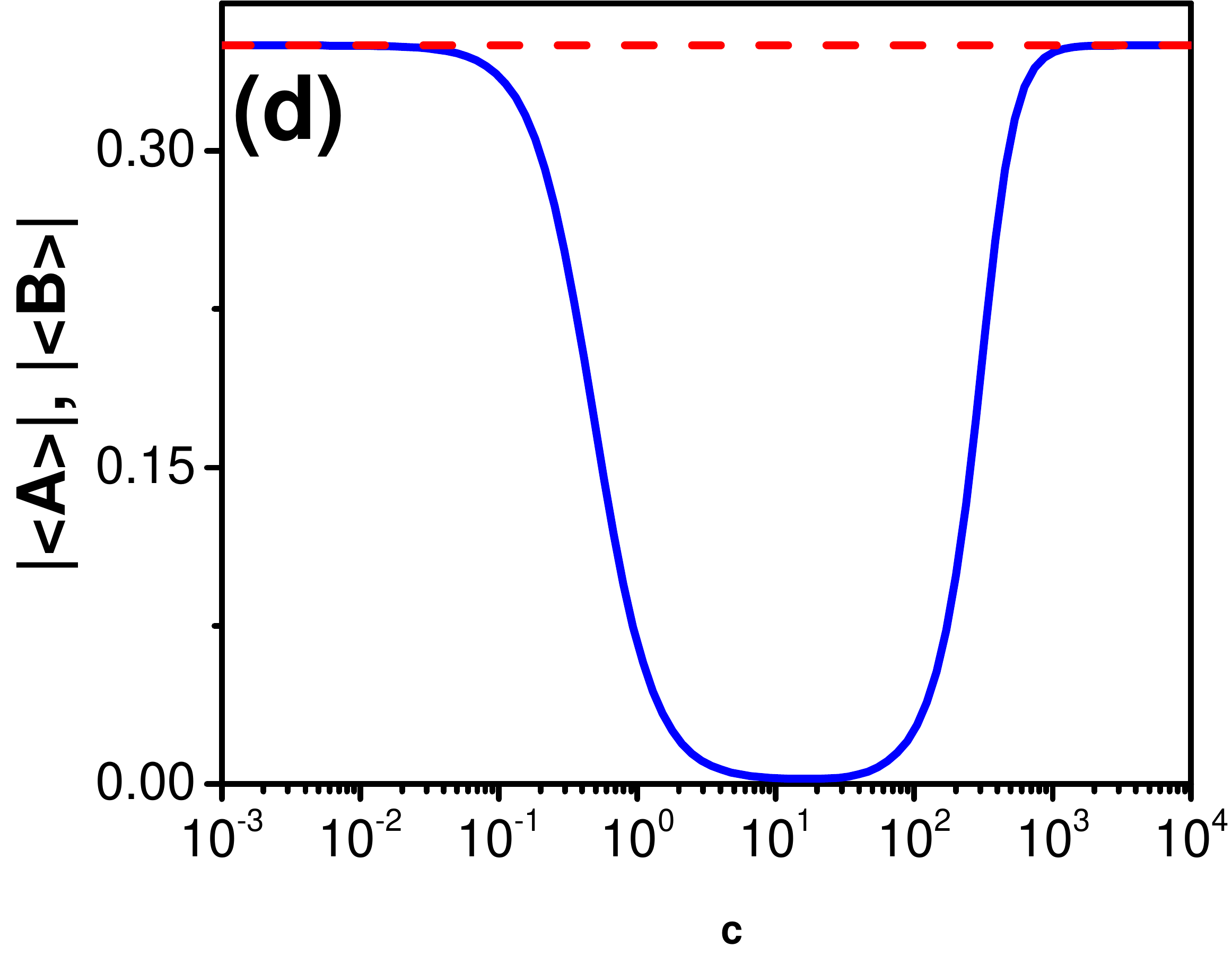} \\
\includegraphics[width=4cm]{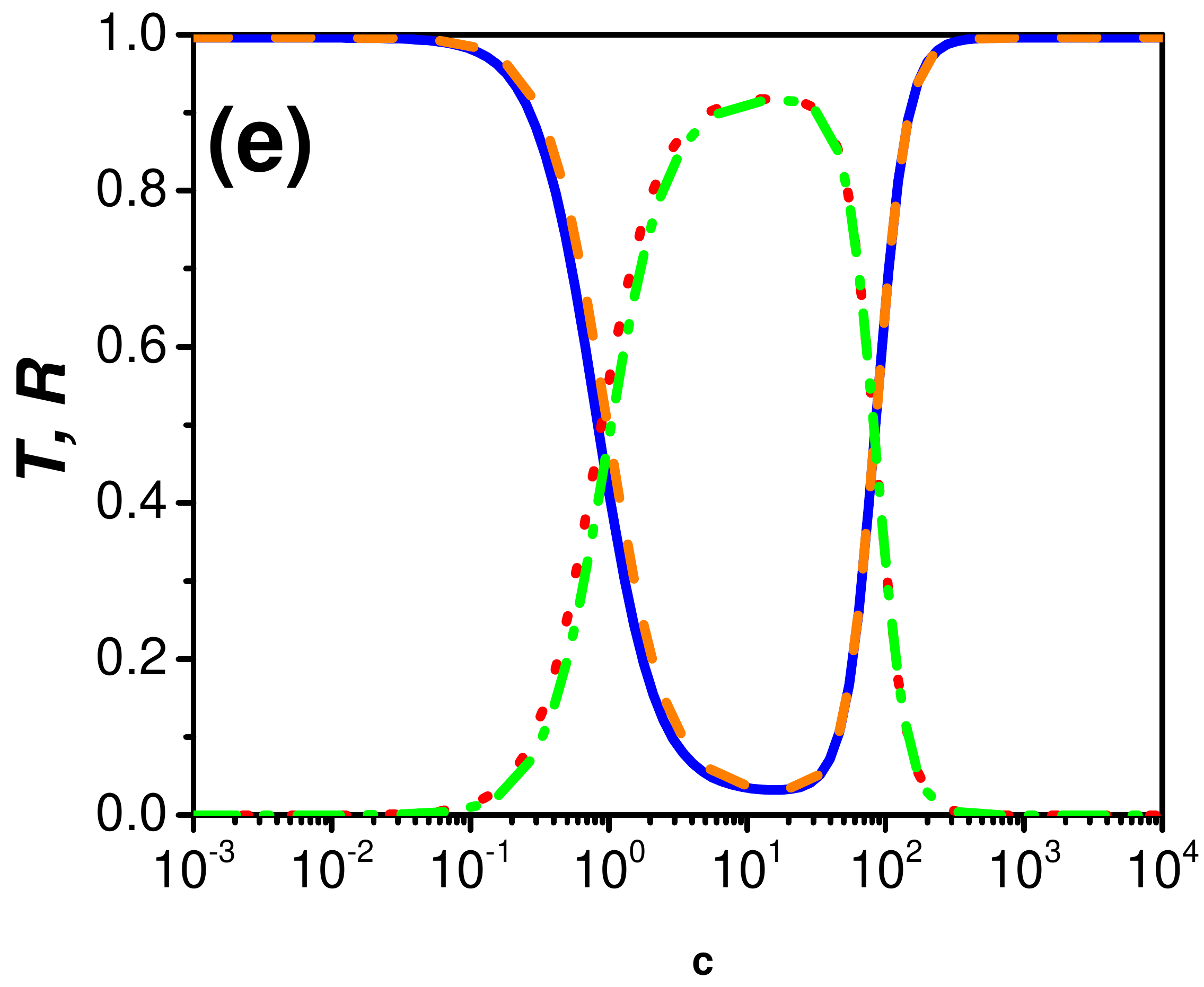} &
\includegraphics[width=4.1cm]{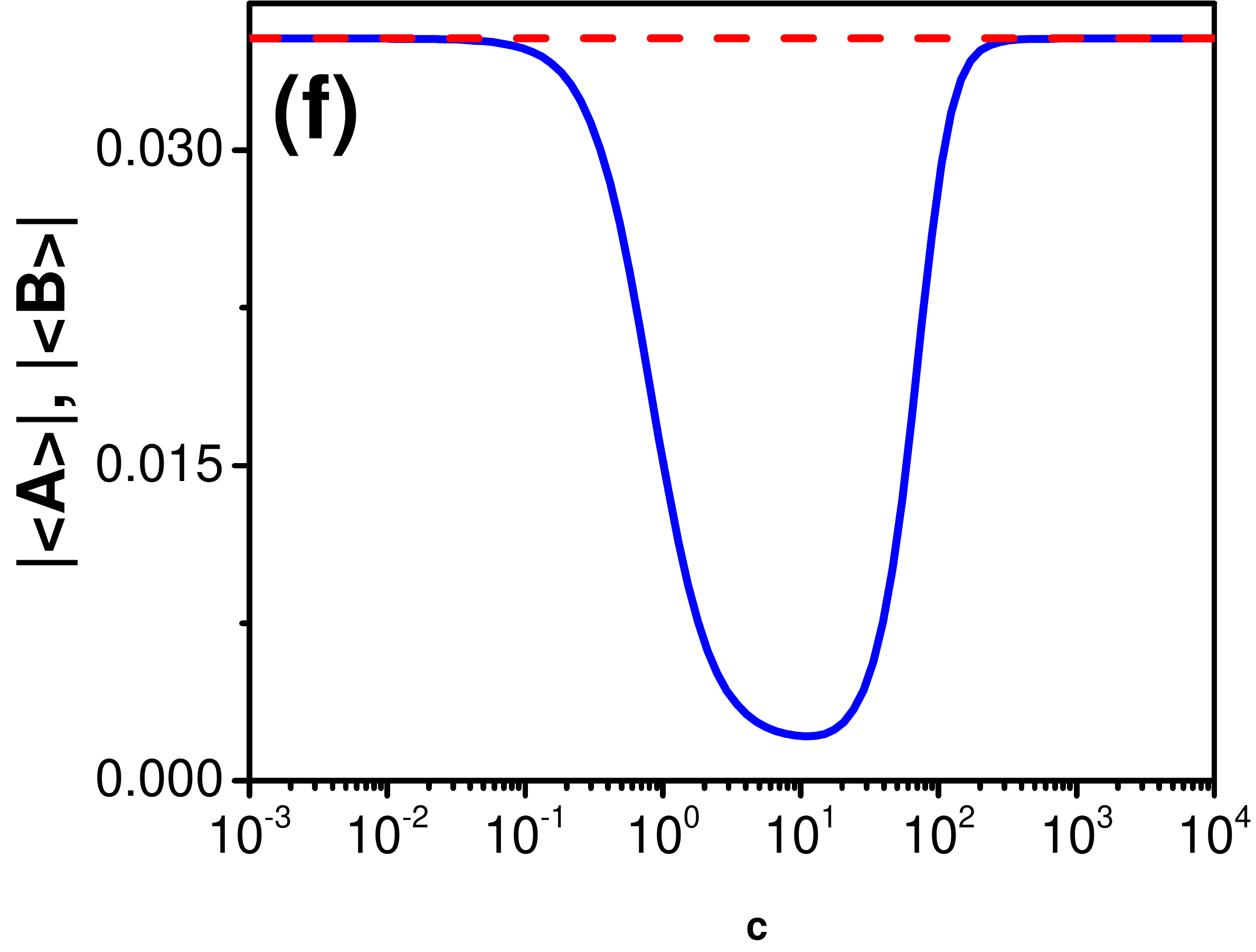} \\
\includegraphics[width=4cm]{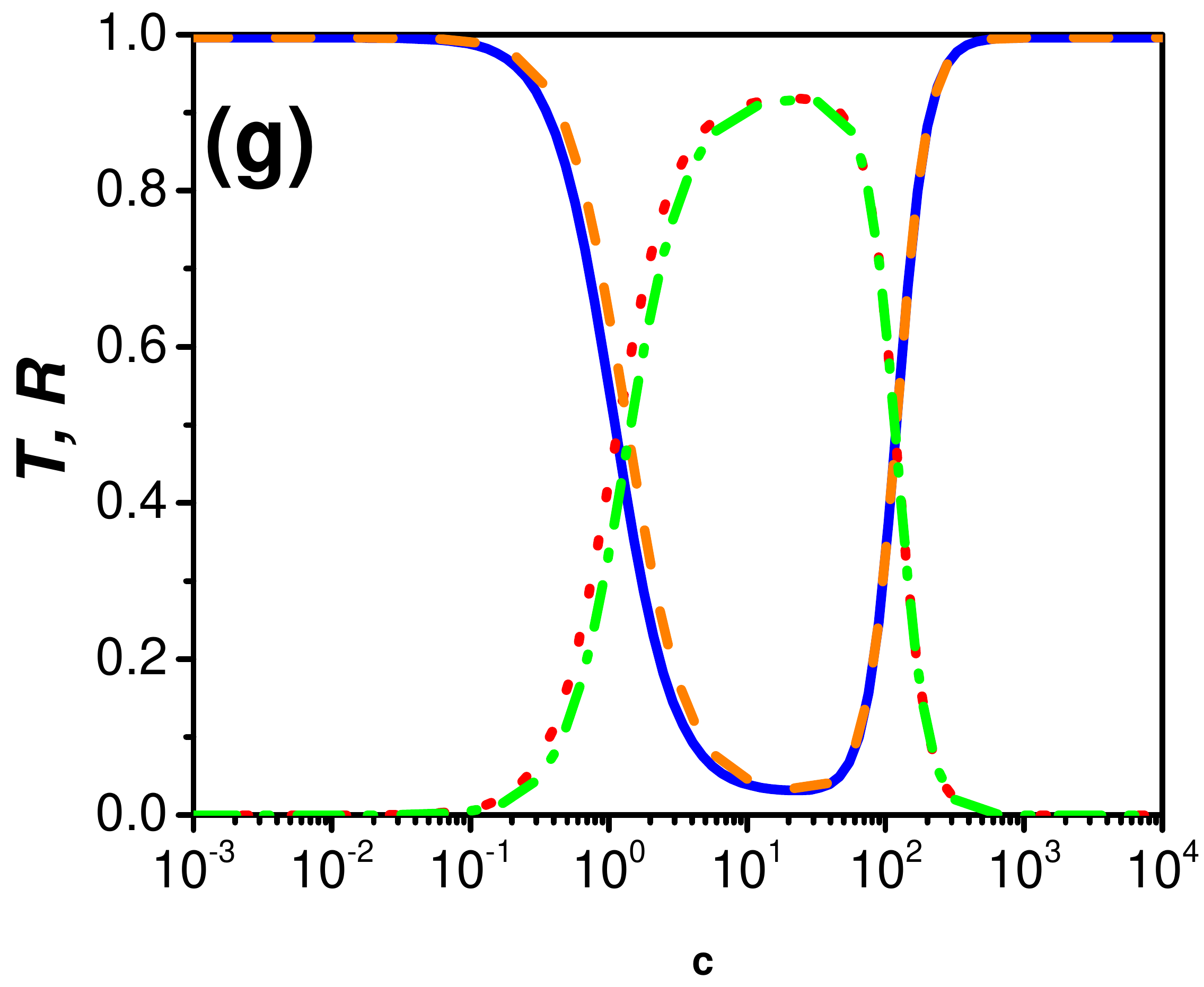} &
\includegraphics[width=4.1cm]{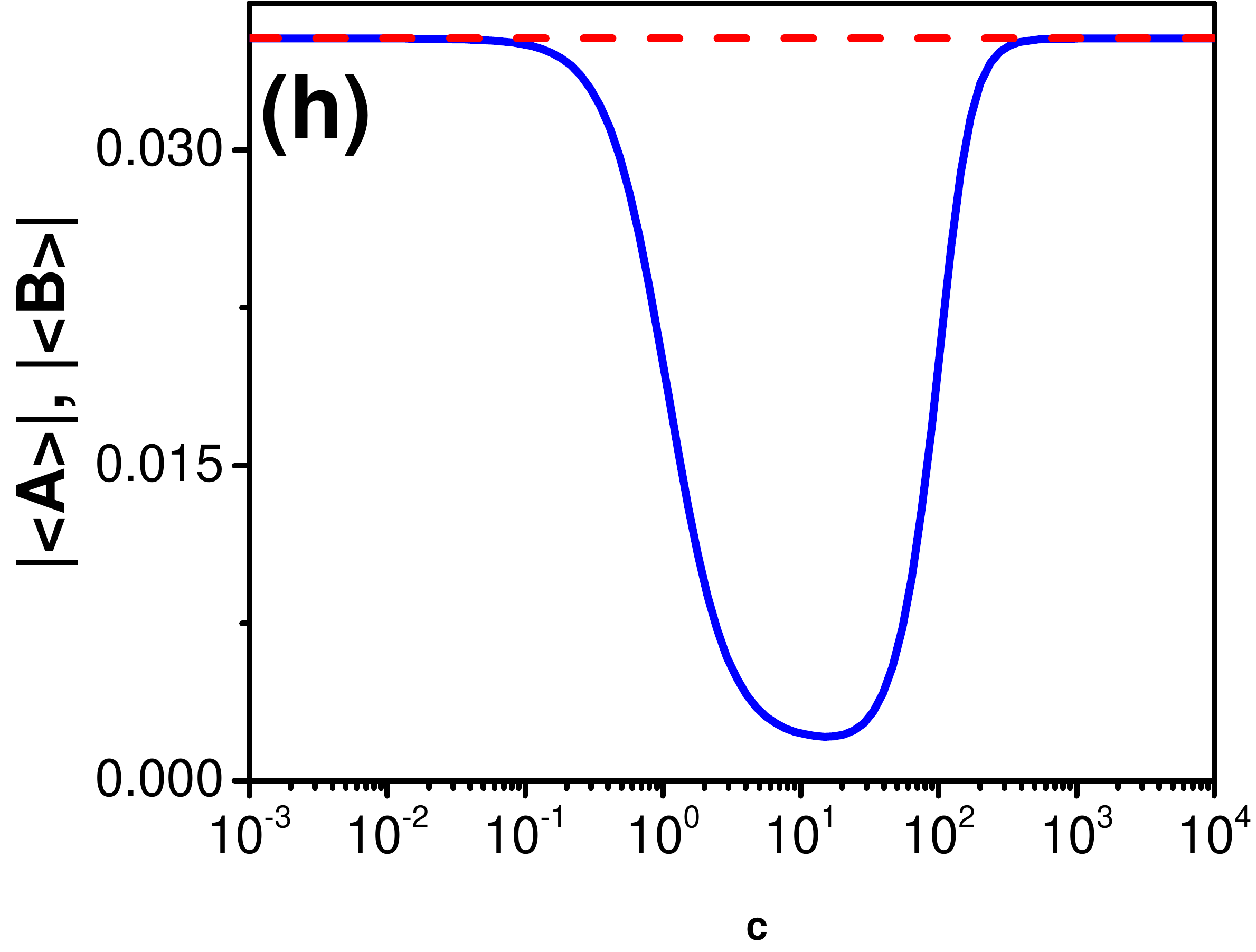} \\
\end{tabular}
\caption{(colour online). Normalized power transmission $T$, reflection $R$ and the populations of normal modes $A$ and $B$ as a function of control field strength $\Omega_{c}$. The blue and red solid lines are transmission and reflection functions resulting from the master equation simulations, respectively. The orange and green dashed lines are transmission and reflection functions resulting from adiabatic elimination, respectively. The parameters for strong coupling cases are $\{\Delta_{r},\Delta_{e1},h,g,\Omega_{p},\kappa_{\text{ex}},\kappa_{\text{i}},\Gamma_{e1},\Gamma_{e2}\}/2\pi=\{ 0,0,0,100,10,20,0.2,5.2,5.2\}$MHz and (a,b) $\Delta_{21}/2\pi=70$MHz; (c,d) $\Delta_{21}/2\pi=140$MHz. The parameters for bad cavity cases are $\{\Delta_{r},\Delta_{e1},h,g,\Omega_{p},\kappa_{\text{ex}},\kappa_{\text{i}},\Gamma_{e1},\Gamma_{e2}\}/2\pi=\{ 0,0,0,100,10,200,0.2,5.2,5.2\}$MHz and (e,f) $\Delta_{21}/2\pi=70$MHz; (g,h) $\Delta_{21}/2\pi=140$MHz.}
\label{switch_demo}
\end{figure}

The main result of this manuscript is shown in Fig.~\ref{switch_demo} and is obtained by numerical simulations of the master  Eqs. ~\ref{masteq3} and ~\ref{masteq4} which takes into account all dissipative channels. Here we use the superspace method which is outlined in a great detail in Ref.~\cite{Nav15}. Moreover, analytical results for the bad cavity limit($ g < \kappa_{ex}$), which are outlined in  detail in  section Appendix~\ref{appendix} material,  are also plotted in Fig.2 for comparison with numerics. In the Fig.~\ref{switch_demo} transmission and reflection are plotted as a function of the amplitude of the control field, for the set of parameters given in the caption. For the system parameters we use the realistic experimental values taken from the Ref.\cite{A2006}, where $\text{SiO}_{2}$ microtoroidal resonator was coupled with a cloud of cold cesium atoms. For some range of $\Omega_{c}$,  $T \approx 0$ and $R \approx 1$ (we remark that $T+R < 1$ due to the losses in the system), which means that the system works as a quantum transistor.
To gain better understanding about the behavior of transmitted and reflected intensities, in the right column of Fig.~\ref{switch_demo} we amplitudes  of the modes $A$ and $B$ as a function of $\Omega_{c}$. The mode $B$ is decoupled from an atom so its population stays constant, however, mode $A$ is strongly coupled to the atom, which in turn is coupled to the external control field, and  for some range of the control field amplitude the mode $A$ is going out of resonance, and this range coincides with the range where transmission goes to zero which is apparent by comparing the first and the second columns in  Fig.~\ref{switch_demo}. This behaviour can be easily explained by expressing the output field $a_{out,ex}=-a_{in,ex}+\sqrt{ \kappa}(A+B)$, through the normal modes, and taking into account that in the switch region $\langle A \rangle \approx 0$. Since the normal mode $B$ is decoupled from the atom, its average value can be obtained by solving steady state equations for the empty cavity($g=0$). As it is shown in the Appendix \ref{appendix}, it follows from the Eq.~(\ref{empty_cavity}), that in the case $\Delta_{r}=0$ and $h=0$, $\sqrt{\kappa}\langle B \rangle=\langle a_{in,ex}\rangle $. Under mean-field approximation $T \approx \langle a^{\dagger}_{out,ex} \rangle \langle a_{out,ex} \rangle \approx 0$, because of the destructive interference between the input field and the mode $B$. Here, to estimate the intensity we applied a mean-field approximation which is not assumed later in the manuscript. The condition $\langle A \rangle \approx 0$ implies that mode $a$ and $b$ have the same amplitude with opposite sign, this means that atom in a way is acting like a pump which is redistributing photon fluxes between these two modes and ones this two modes get equally populated system is acting as a "mirror". From Fig.~\ref{switch_demo} it is seen that for small value and large values of control fields atom is "effectively" getting decoupled from the cavity. For small values of $\Omega_c$, this happens simply because atom is getting optically pumped to the level $2$, since $|\omega_{1e}-\omega_{2e}|>>\gamma_{1e},\gamma_{2e} $. For the large values of $\Omega_c$, the atom-field dressed energy level gets detuned on the large amount $\approx \Omega_{c}>>\gamma_{1e},\gamma_{2e}$ and cavity mode gets out of resonance with the dressed light atom energy state. This statements are substantiated by analytical results for the bad cavity limit which are presented in Appendix \ref{appendix}. As it is demonstrated there for both limits of very small and very large $\Omega_c$, $\rho_{1e}\approx 0$, which means that absorption is vanishing and light is propagating through the cavity without "feeling" the atom.

An interesting feature of our system, that switch functionality regime can be made wider by changing the two-photon detuning, is apparent, for example, by comparing Fig.~\ref{switch_demo}(a), where $\Delta_{12}=70MHz$ with Fig.~\ref{switch_demo}(c)(strong-coupling limit), where $\Delta_{12}=140MHz$.  Moreover, from that figures we see that analytical curves,  given by the dashed lines, agree well with numerical simulations of master equations,  given by blue and red curves. This agreement quite remarkably  holds partially even in the strong coupling limit, which is apparent from Figs.~\ref{switch_demo}(a) and (c). As we can see from the second column of Fig.~\ref{switch_demo}, average value of the mode $A$ is one order of magnitude bigger in the bad-cavity limit, which results in having better switch in the strong coupling limit where transmission turns out to be smaller on one order of magnitude compared to the bad-cavity case. We comment, that the fact that mode $\langle A \rangle$ is bigger in strong coupling limit manifests itself in having different statistics for the reflected field in this limit compared to the strong coupling-limit, which we discuss in more detail in Section~\ref{photonstatistics}.

\begin{figure}
\begin{tabular}{ccccc}
\includegraphics[width=4cm]{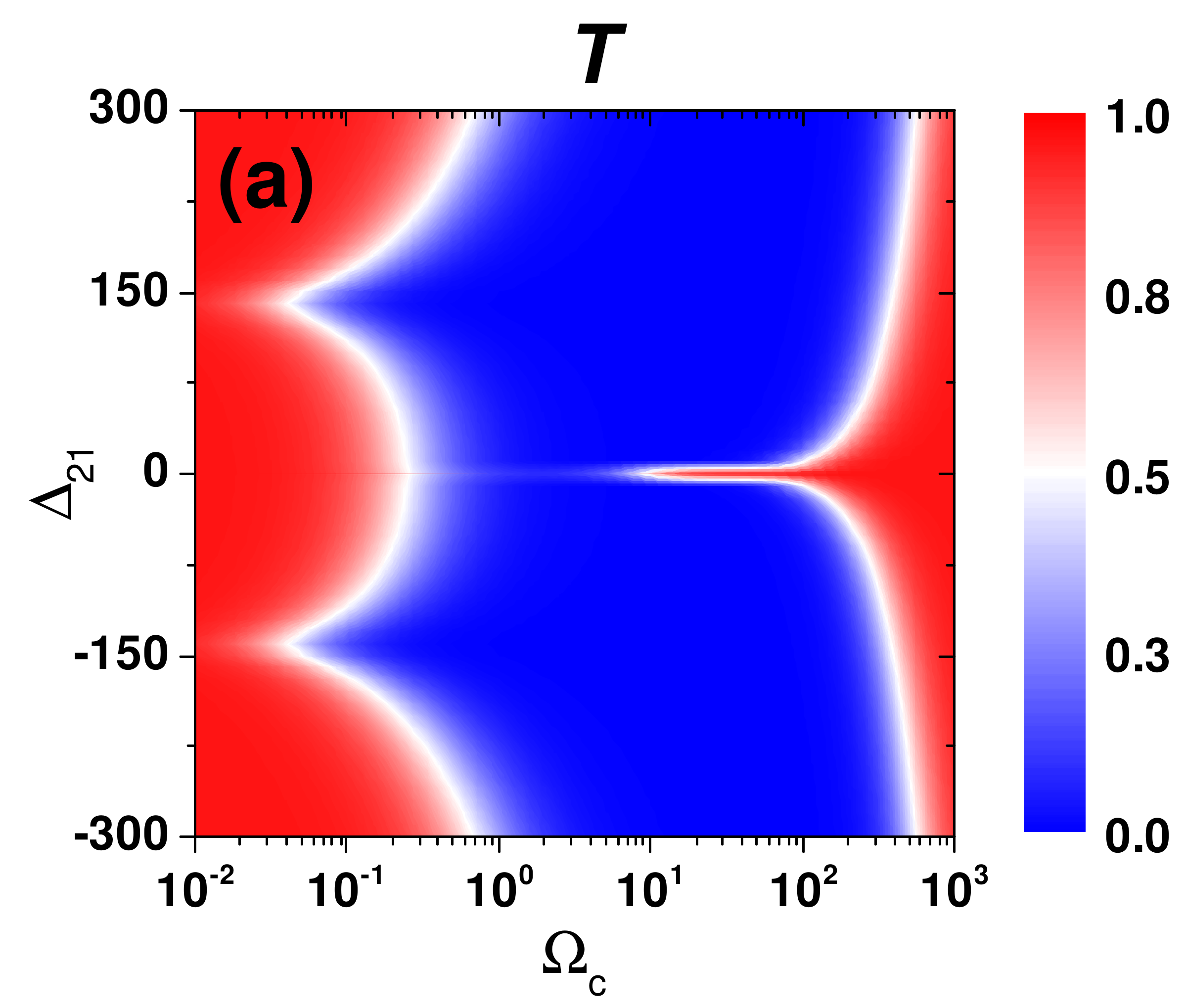} &
\includegraphics[width=4cm]{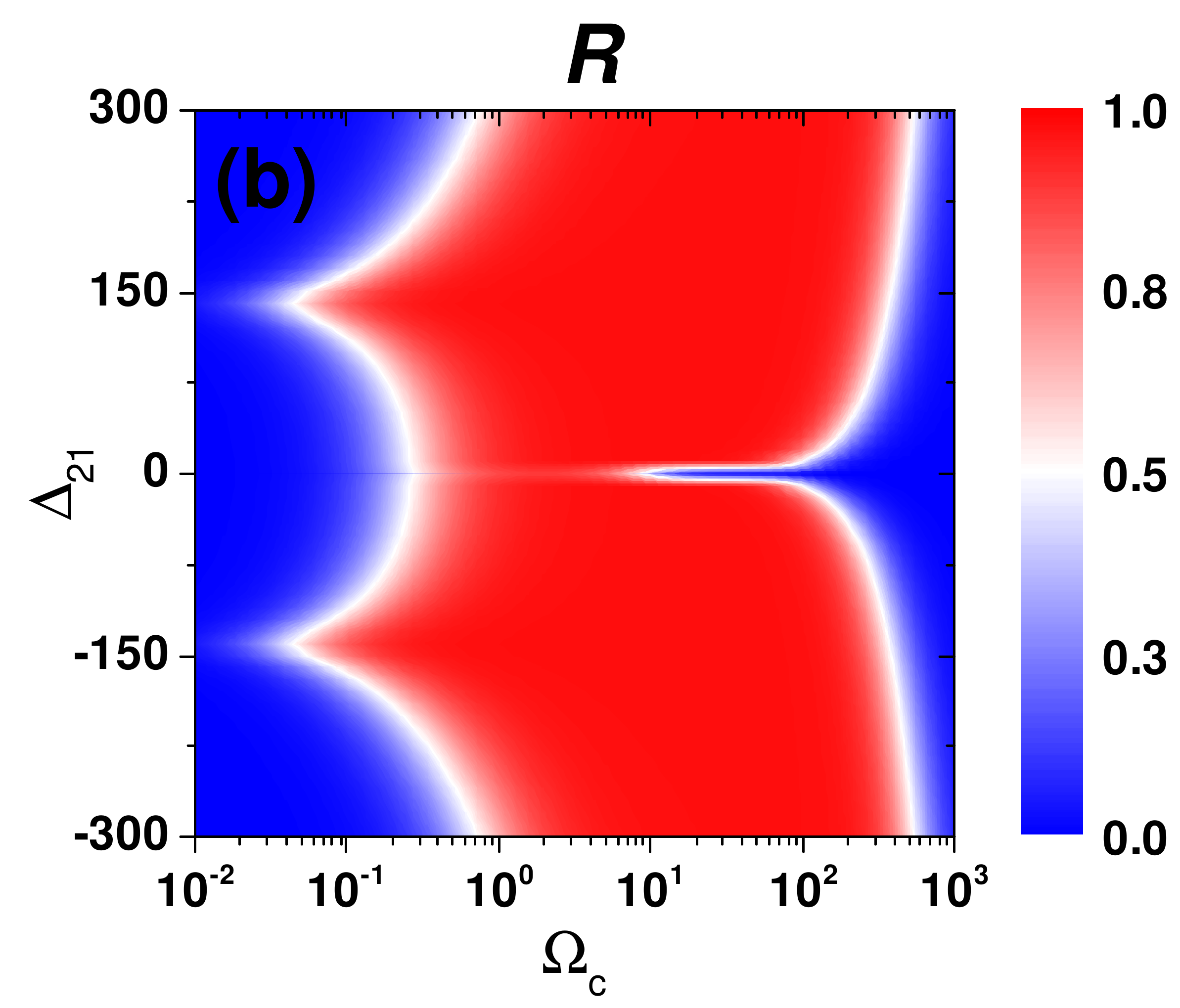} \\
\includegraphics[width=4cm]{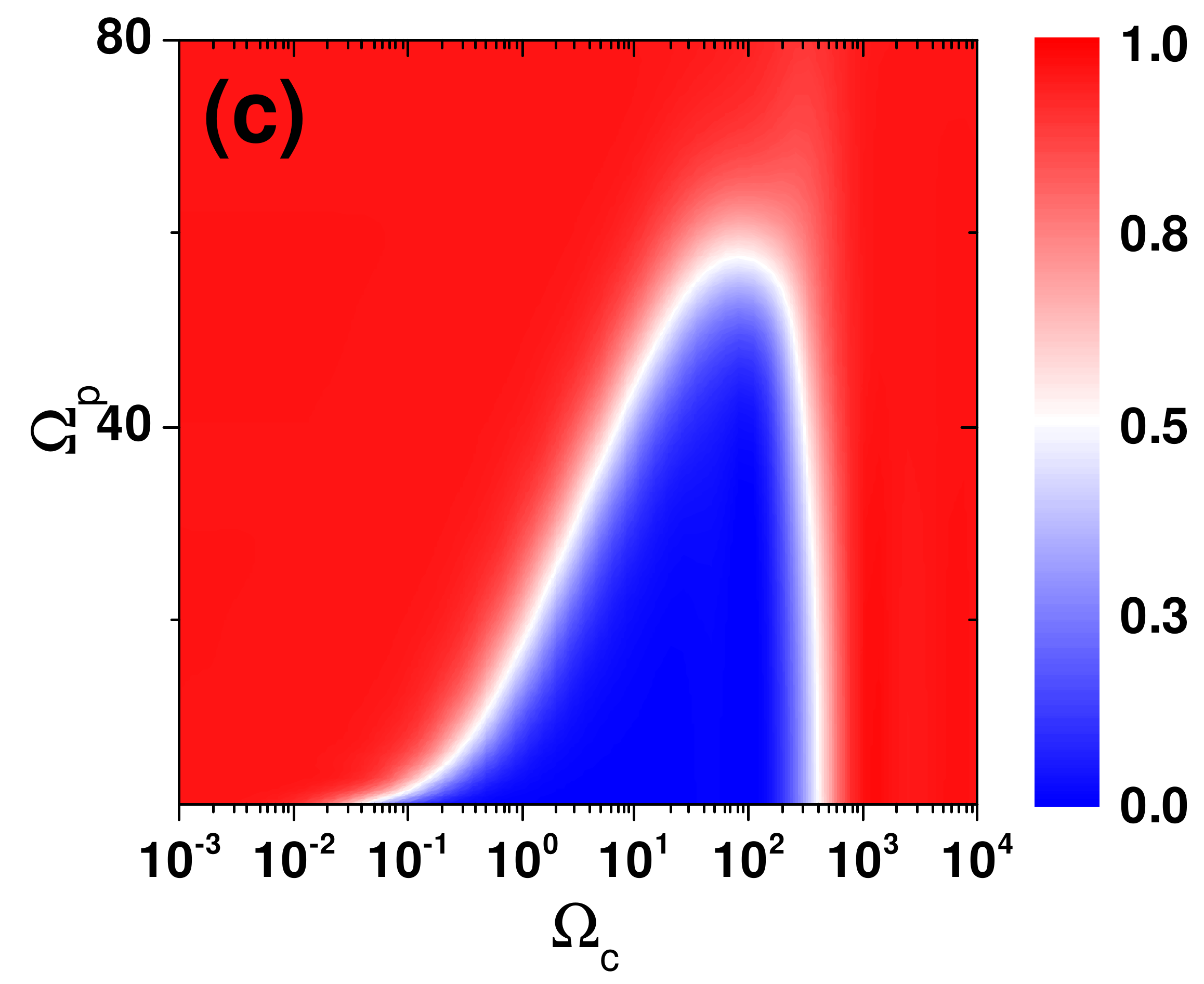} &
\includegraphics[width=4cm]{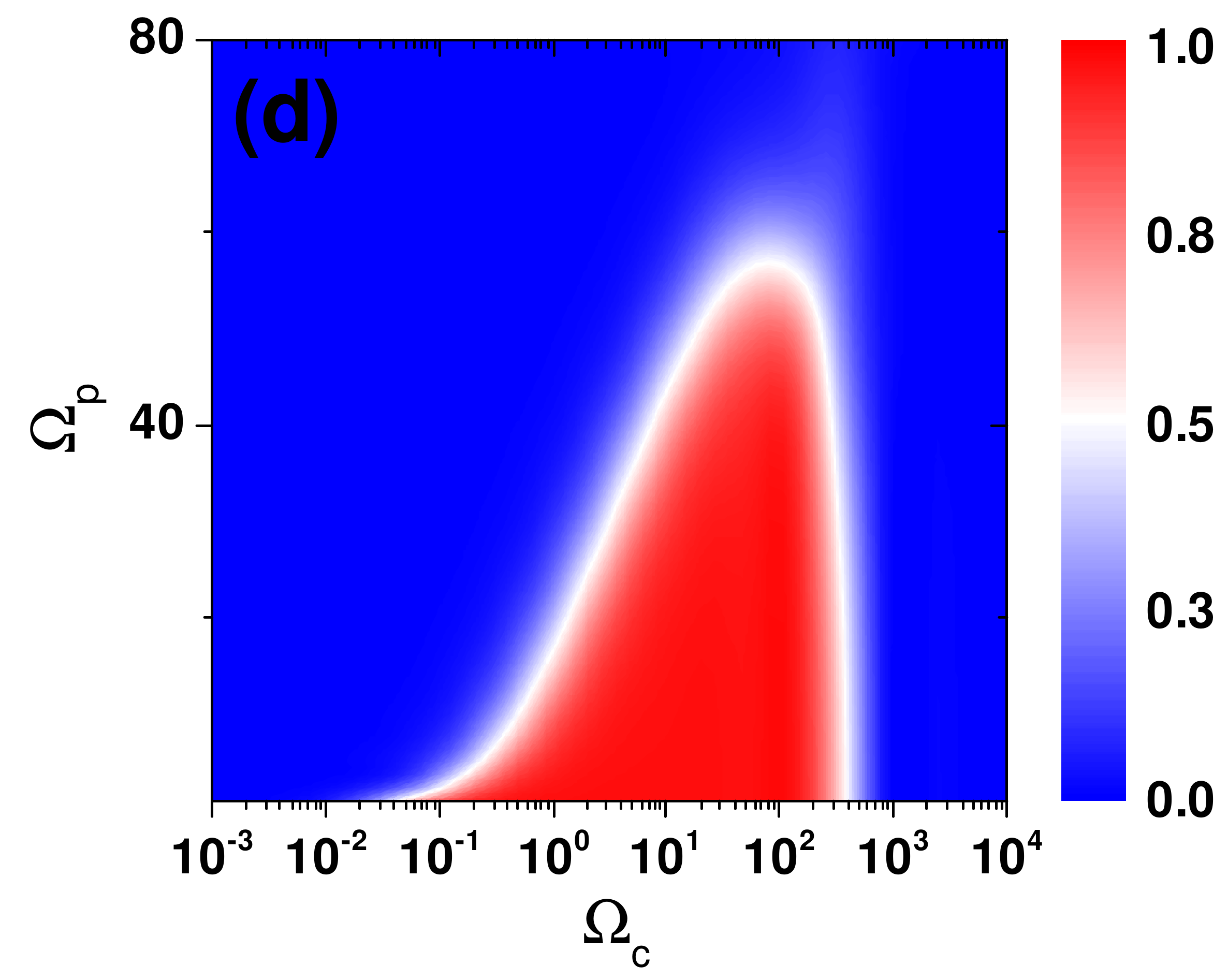} \\
\includegraphics[width=4cm]{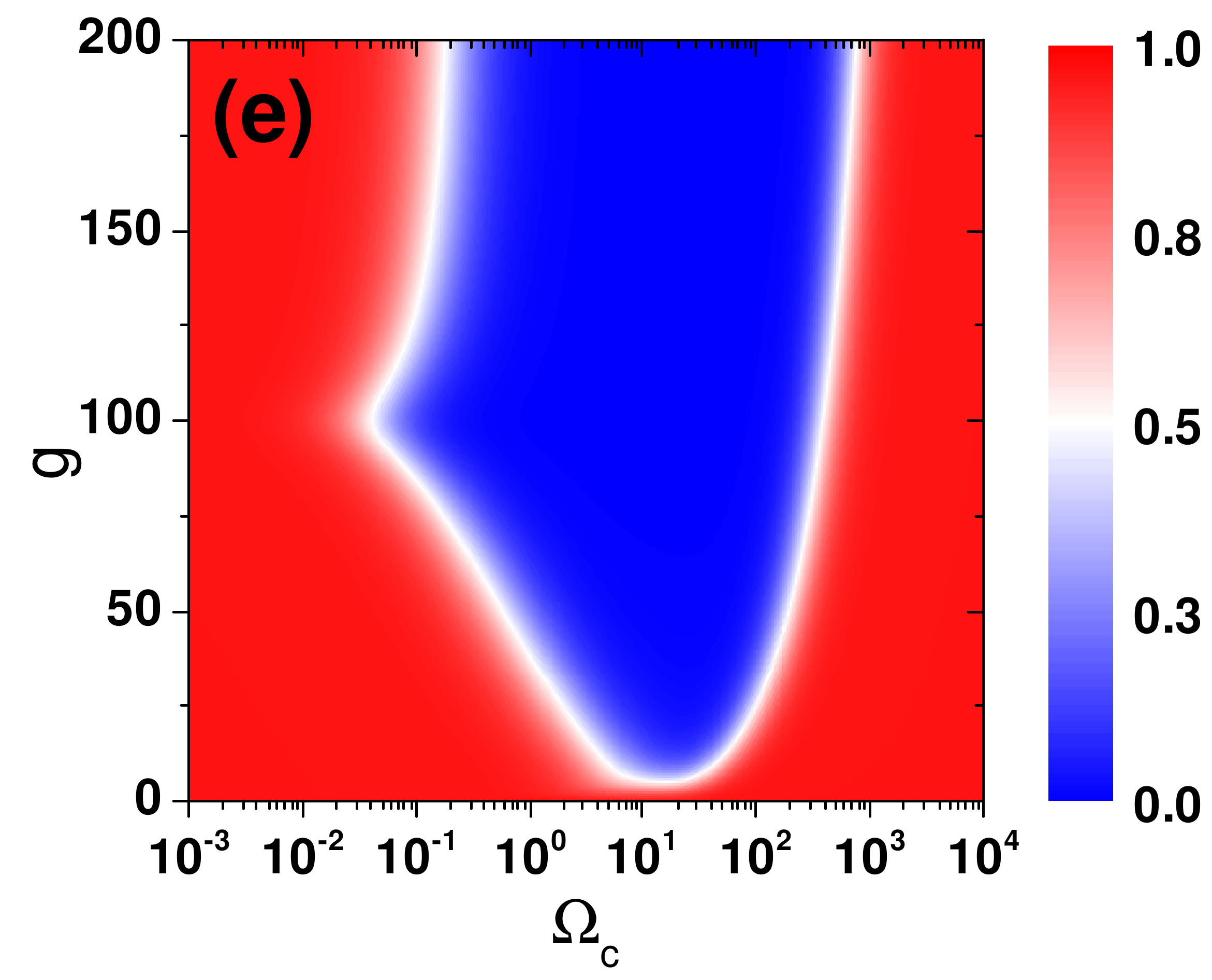} &
\includegraphics[width=4cm]{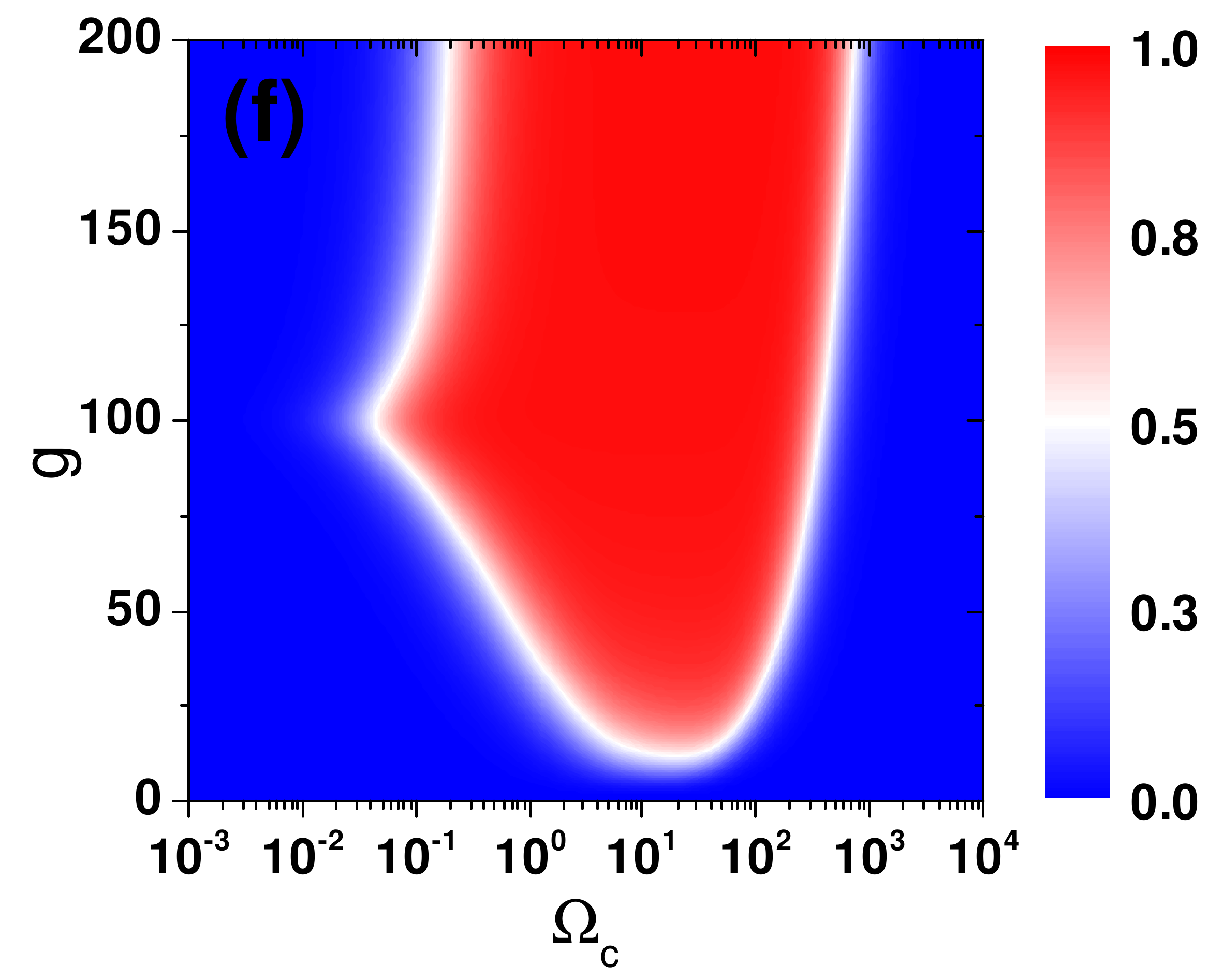} \\
\includegraphics[width=4cm]{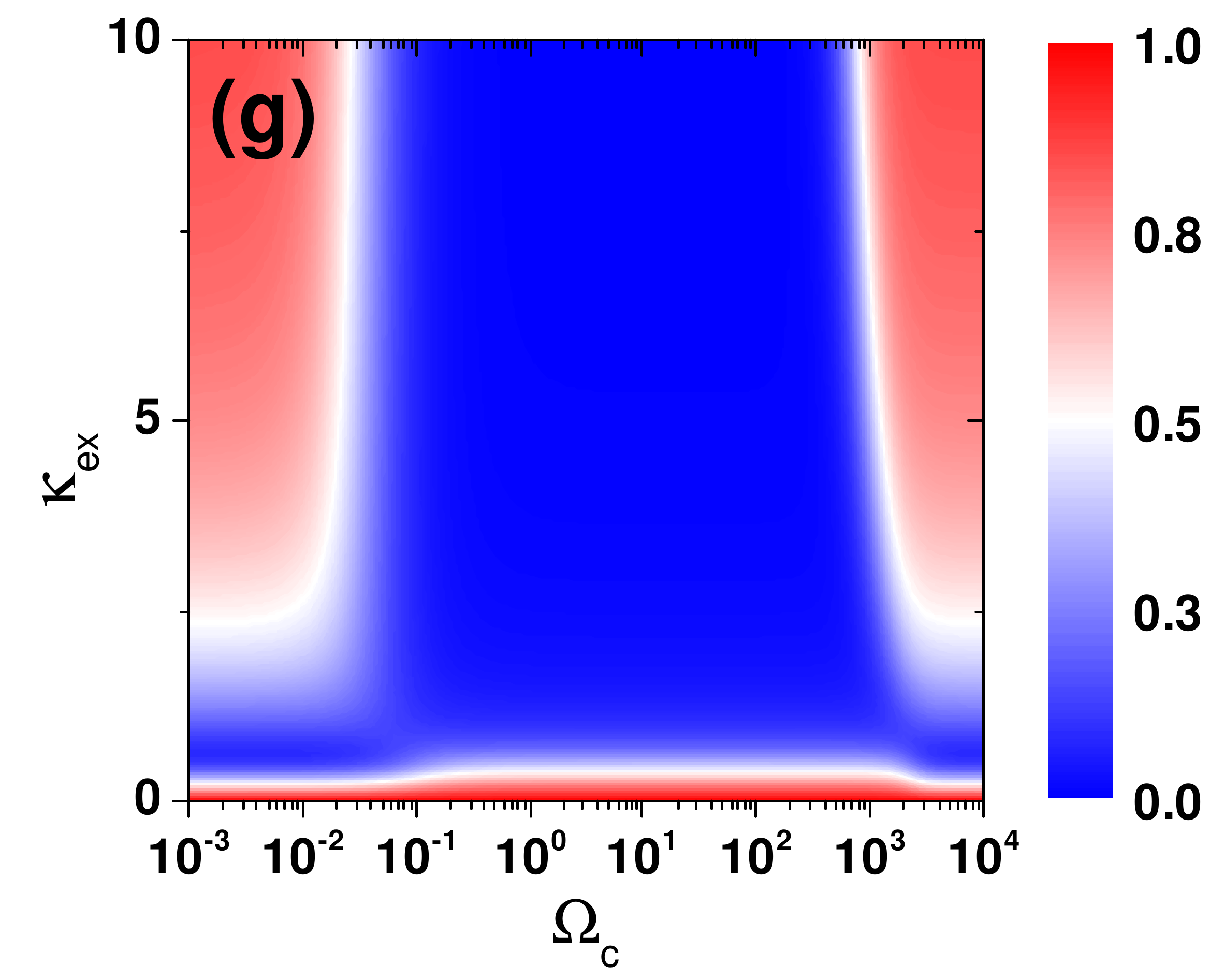} &
\includegraphics[width=4cm]{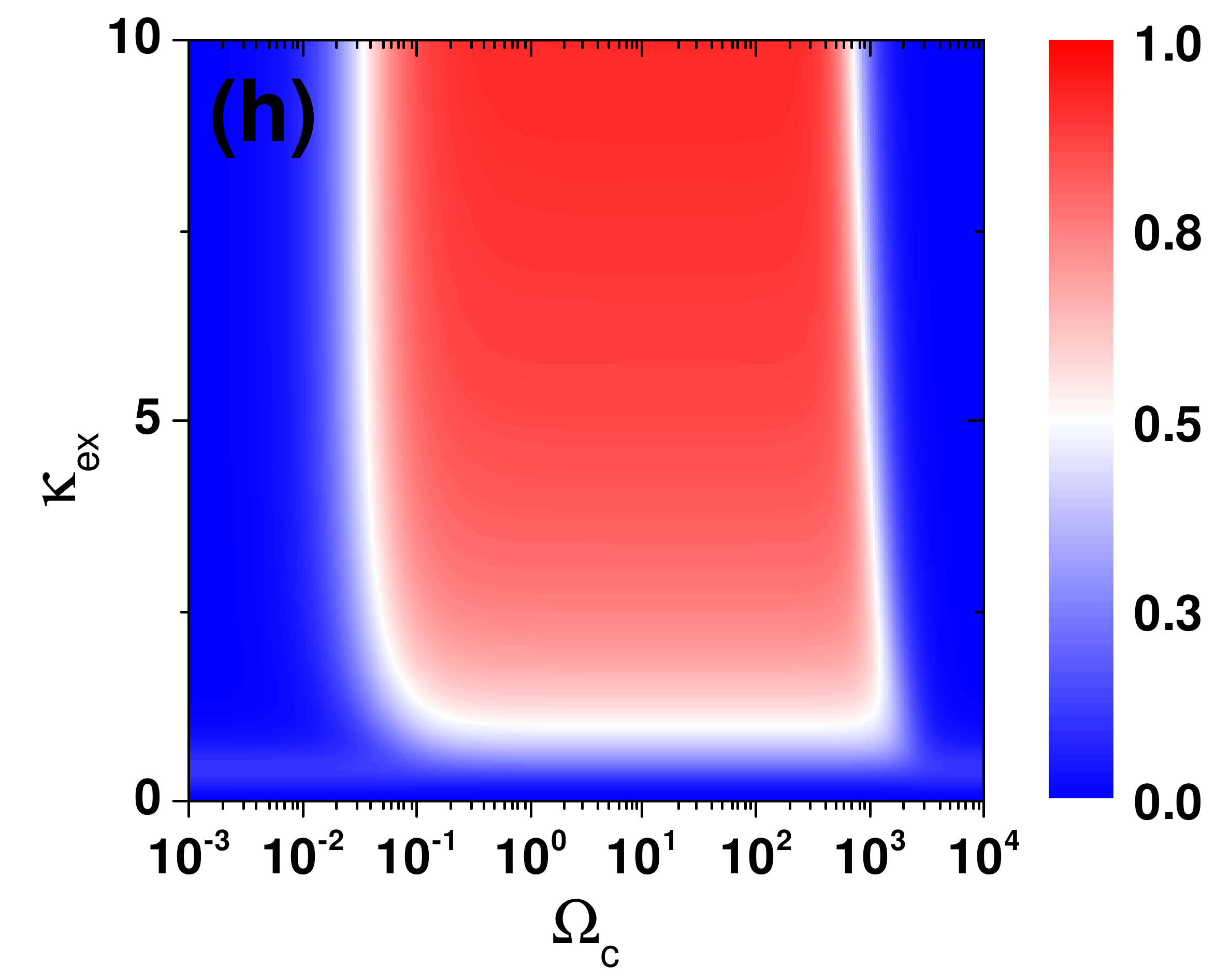} \\
\end{tabular}
\caption{(colour online). Contour plots for transmission and reflection profiles in strong coupling regime. The parameters are the same as Fig. \ref{switch_demo}(c) except $\Omega_{p}=1$.}
\label{switch_flux_strong}
\end{figure}

\begin{figure}
\begin{tabular}{ccccc}
\includegraphics[width=4cm]{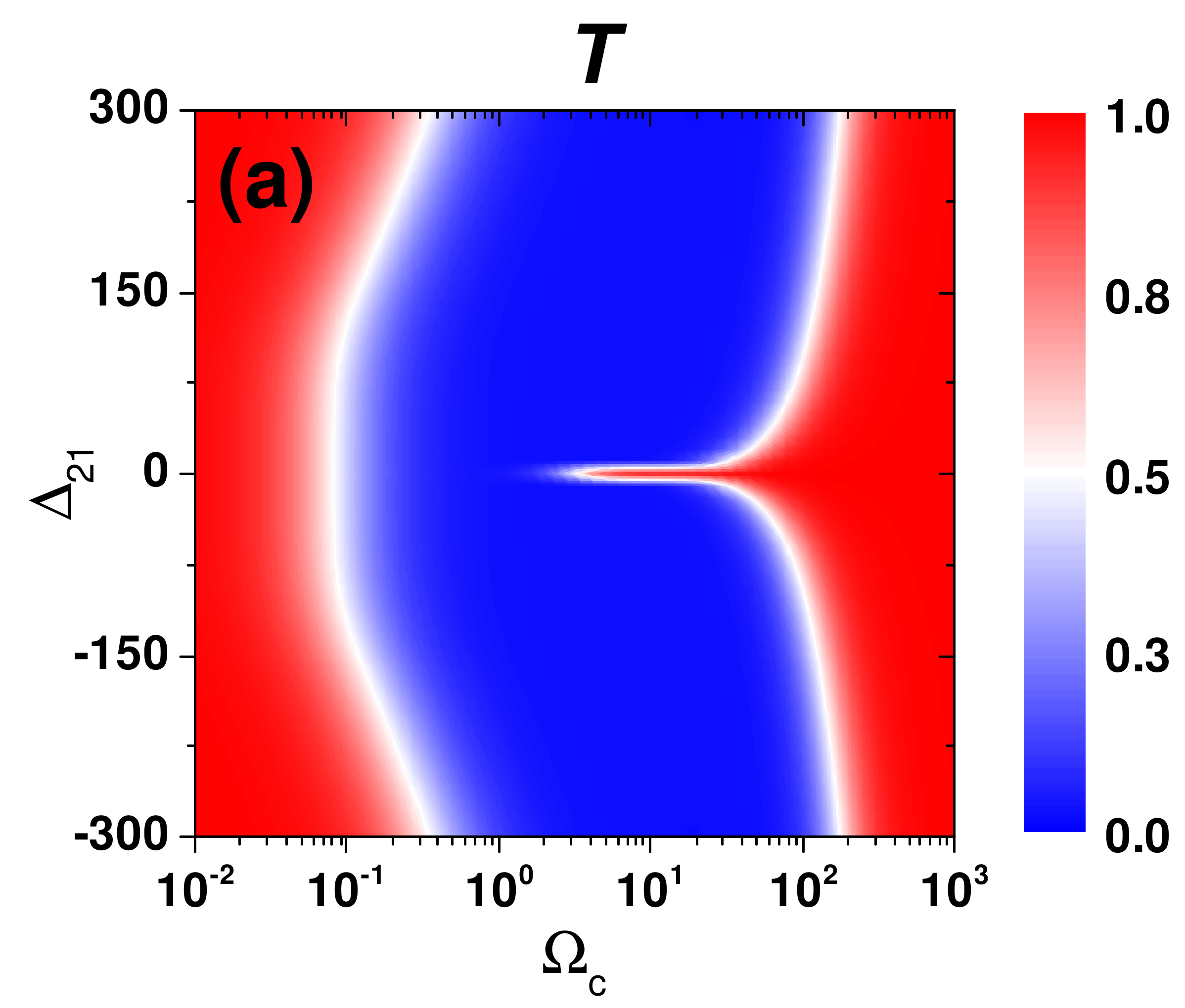} &
\includegraphics[width=4cm]{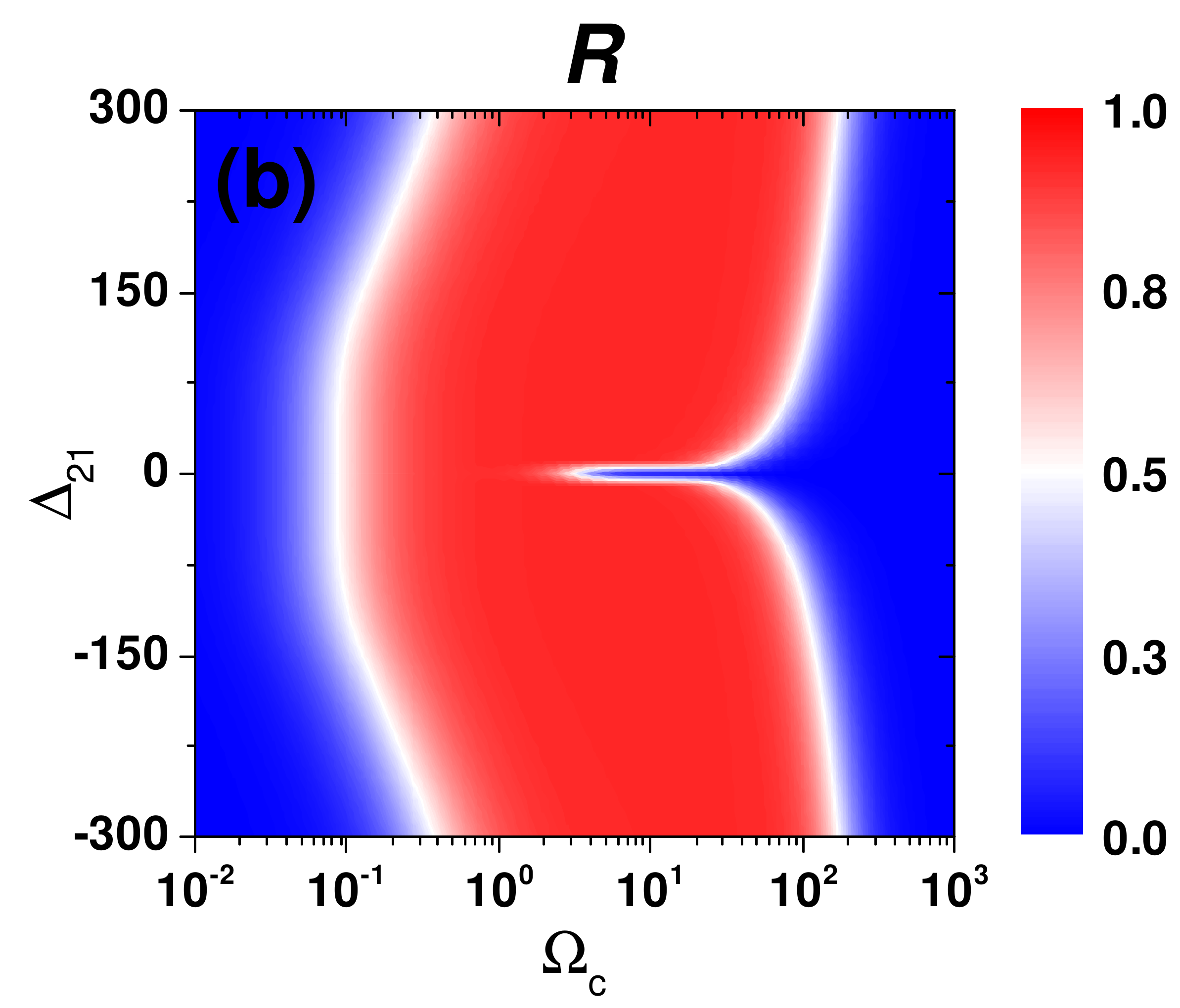} \\
\includegraphics[width=4cm]{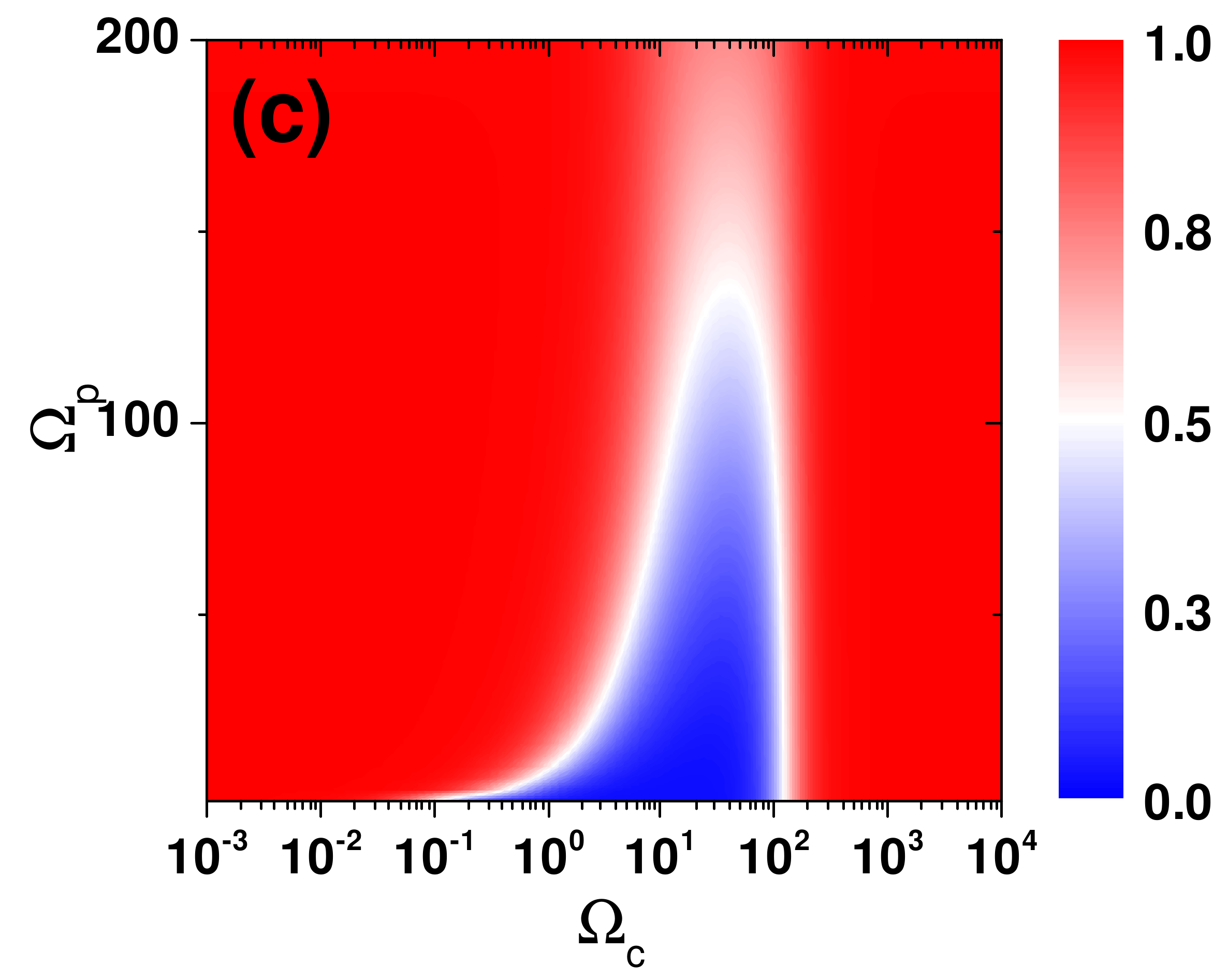} &
\includegraphics[width=4cm]{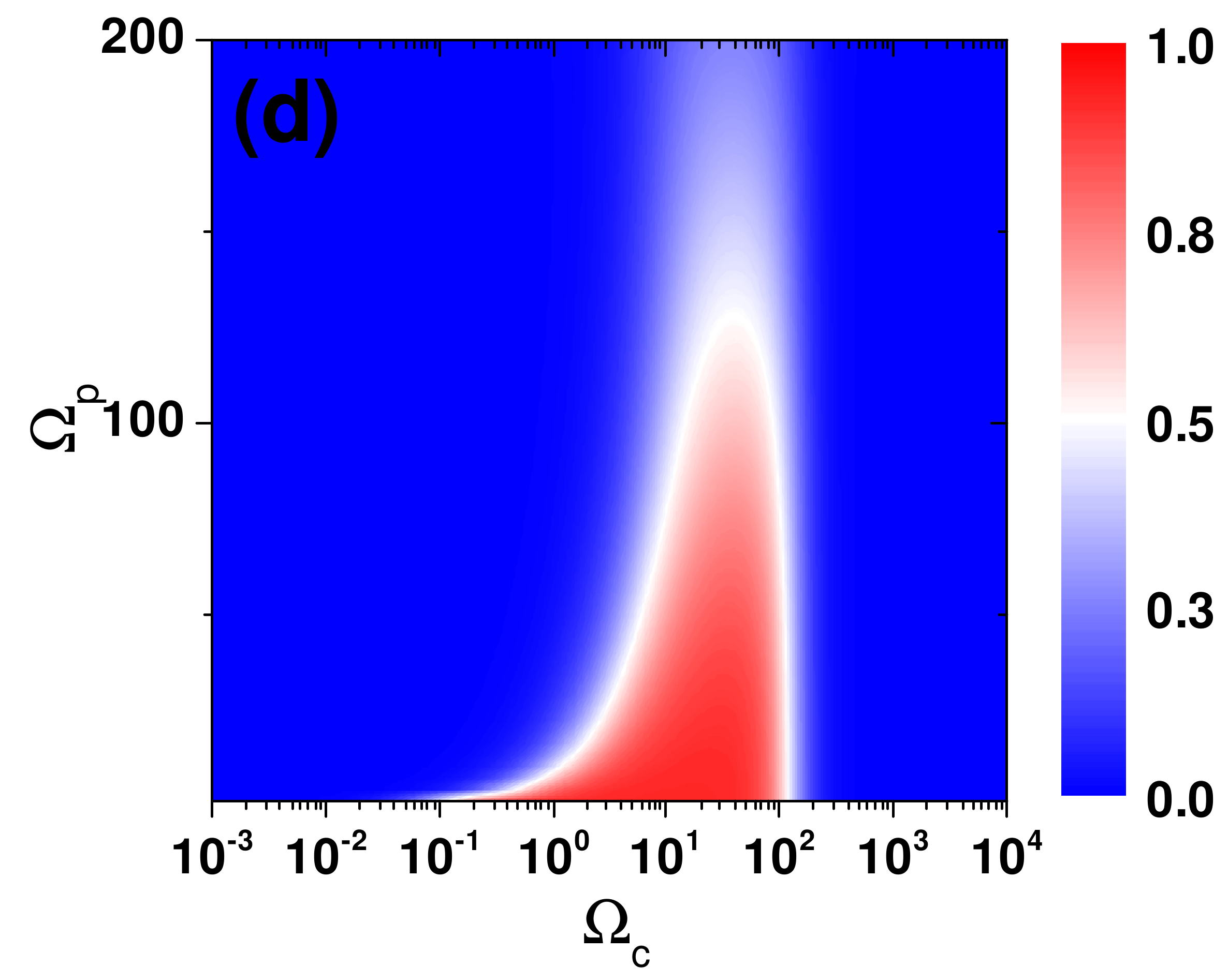} \\
\includegraphics[width=4cm]{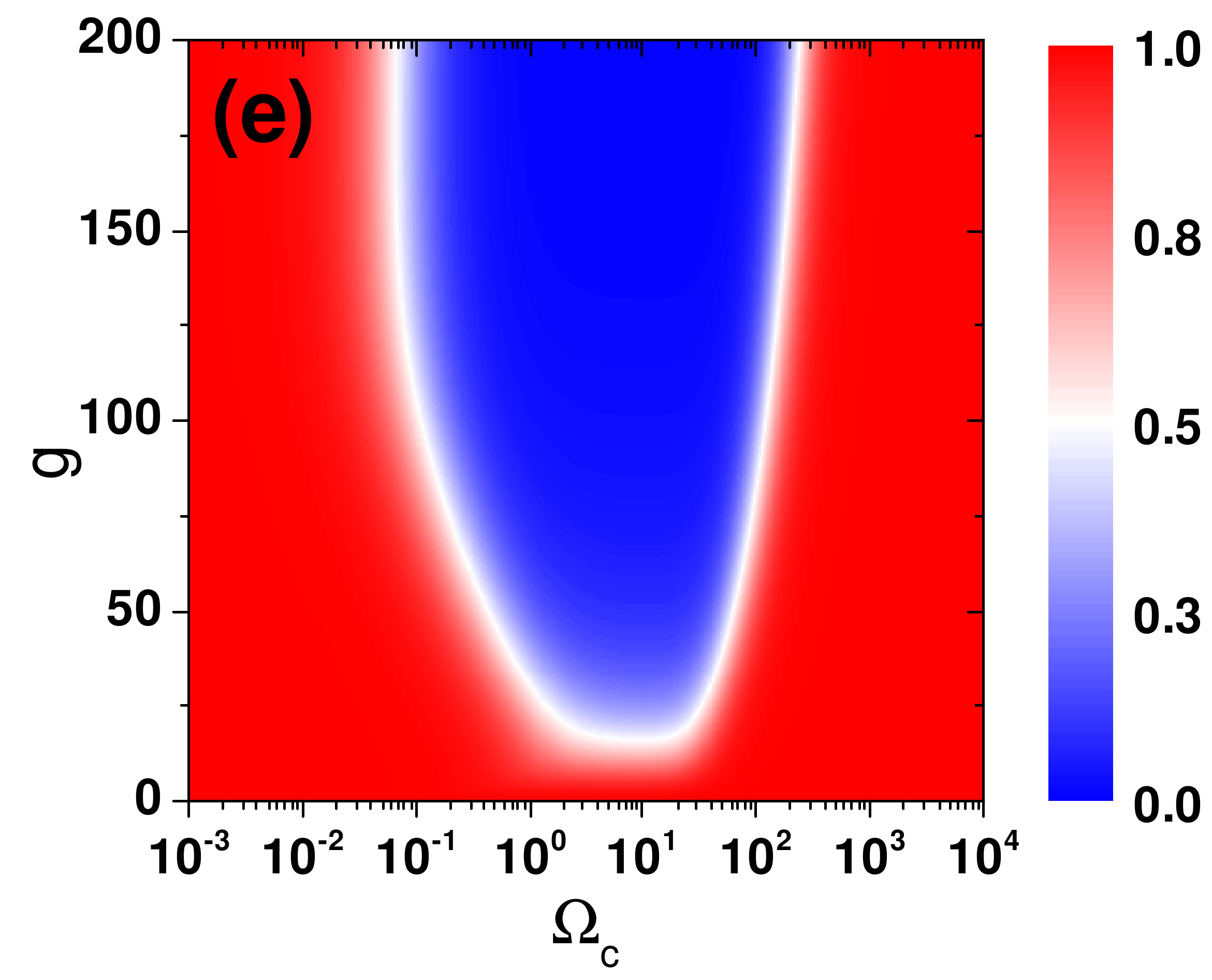} &
\includegraphics[width=4cm]{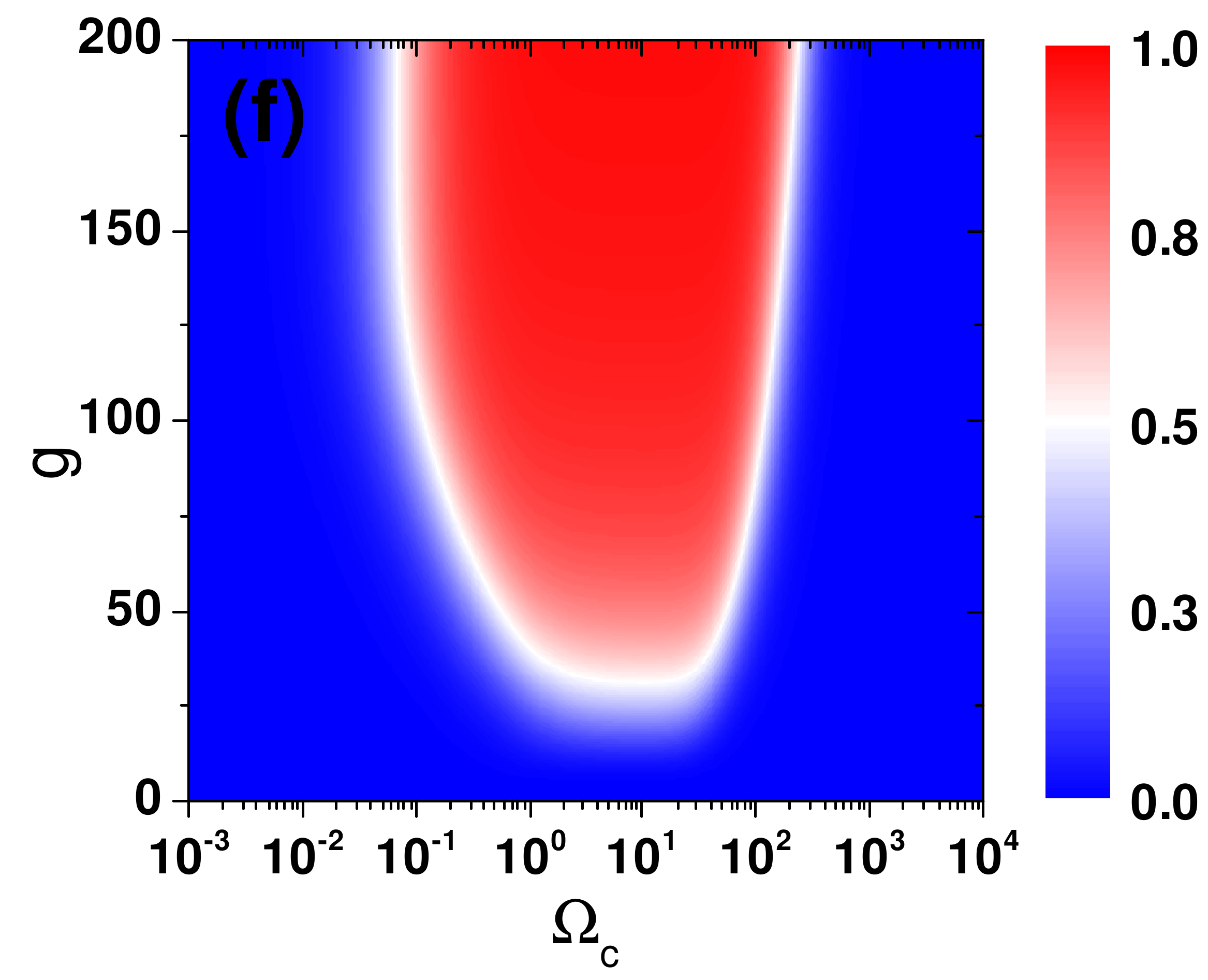} \\
\includegraphics[width=4cm]{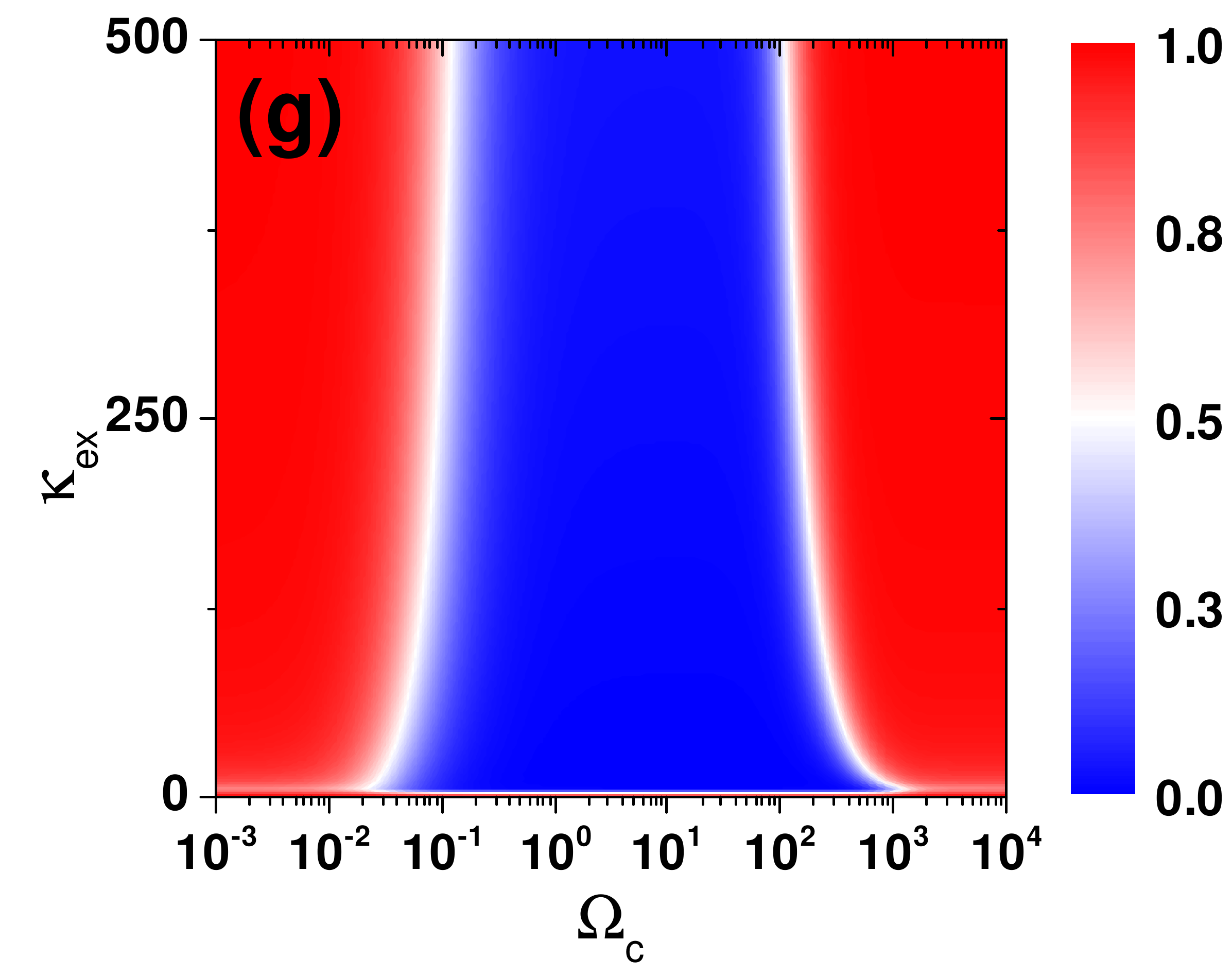} &
\includegraphics[width=4cm]{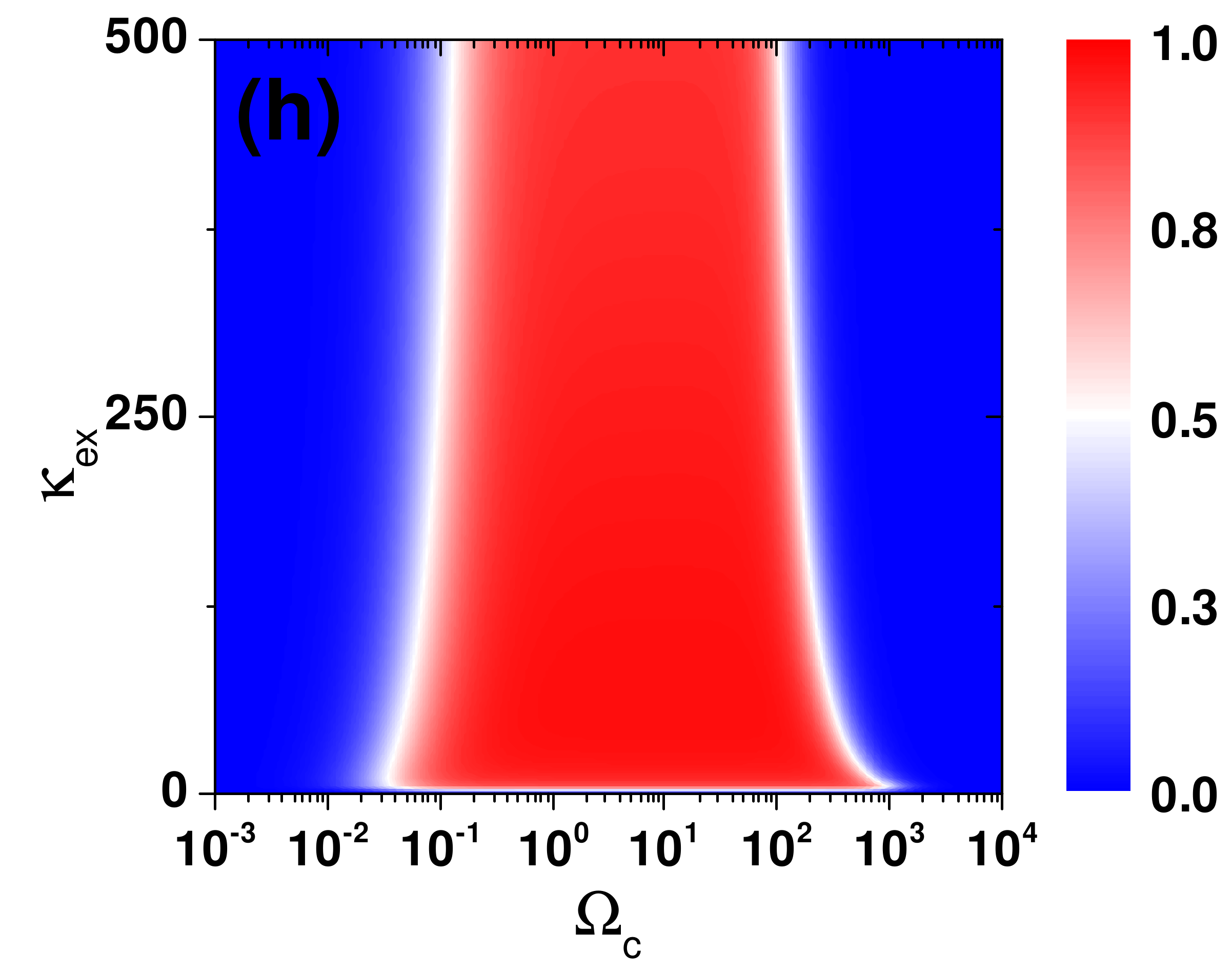} \\
\end{tabular}
\caption{(colour online). Contour plots for transmission and reflection profiles in bad-cavity regime. The parameters are the same as Fig. \ref{switch_demo}(g) except $\Omega_{p}=1$.}\label{switch_flux_bad}
\end{figure}

Bigger is the range of $\Omega_c$ over which the system works as a quantum transistor,  better is the quantum switch. To understand why is this the case it is constructive to consider the opposite limit when this range is extremely narrow, then experimental imperfections and noise can easily push the system out of the  regime of functionality. Bearing this in mind, we make a series of contour plots for exploring the parameter regimes where the "range of functionality" is  broad. In these contour plots, one axis represents the external control field and other axis denotes the physical parameter of interest. Fig.3 shows the series of contour plots where the left and right columns  show transmission and reflection intensities. Range of functionality is given by the length of horizontal line (for a given value of parameter along the $y$-axis) which has a dark/blue  colour corresponding to $T \approx 0$.
In Fig. \ref{switch_flux_strong}(a,b) we plot $T$ and $R$ in the $\Delta_{21}$-$\Omega_{p}$ plane, in the strong-coupling limit. For the small value of $\Omega_{c}$ we see double-peak structure which is a signature of vacuum Rabi splitting, because for the small value of control field excited state is splitting in the Jaynes-Cummings doublet, and peaks are located at $\approx \sqrt{2}g$. The factor $\sqrt{2}$ is a result of having standing waves in the microtoroidal cavity. From this figures, we conclude, that for the switch with a wide rang of functionality the optimal value of two photon detuning should be chosen equal $\Delta_{21}\approx \sqrt{2}g$. From Fig. \ref{switch_flux_bad}(a,b) we see that, in principle, many values of two photon detuning realize a good switch because in this case there is no vacuum Rabi splitting, and consequently no Rabi oscillations occur, as photons leave the cavity before being reabsorbed by the atom. However, for being consistent we also choose $\Delta_{21}\approx \sqrt{2}g$.

The transmission/reflection in $\Omega_{p}$-$\Omega_{c}$ plane is shown in Fig. \ref{switch_flux_strong}(c,d), in the strong-coupling limit. With increasing amplitude of the input drive field, the range over which switch functions (i.e. the dark/red region ) narrows until it completely disappears.  This behaviour occurs as a result of an atom being saturated on the $1-e$ transition. So the weaker is the amplitude of the input field, the better switch can function. We remark, that onset of saturation occurs for smaller value of $\Omega_{p}$, in the bad cavity limit (See Fig \ref{switch_flux_bad}(c,d), because in this limit atom gets saturated with relatively small number of photons. Moreover, in the bad-cavity limit the range of functionality for a given value of $\Omega_{p}$, is narrower compared to the strong-coupling limit.
The transmission/reflection in $g$-$\Omega_{p}$ plane is shown in Fig. \ref{switch_flux_strong}(e,f). For very small values of $g$, photons will pass through the cavity without interacting with the atom which means $T \approx 1$ and system does not perform as a  switch. We remark, that since here $h=0$,  the only way for producing anti-clockwise photons is through interaction with the atom. With increasing $g$ the range of functionality gets larger and has an optimal value. Thus, transmission eventually goes to zero with increasing $g$, once destructive interference occurs between input field and intracavity field amplitudes. Here, we see that for a fixed two photon detuning $\Delta_{21}$, there is an optimal value of $g$ given by $g \approx \Delta_{21}/\sqrt{2}$ in agreement with existence vacuum Rabi splitting as was discussed above.
Same kind of behaviour is observed in the bad-cavity limit(See~\ref{switch_flux_bad}(e,f), except in this regime there is no optimal value of $g$ for a given two-photon detuning, because of the absence of vacuum Rabi splitting.

Transmitted and reflected intensities in $k_{ex}$-$\Omega_{p}$ plane are shown in Fig. \ref{switch_flux_strong}(e,f). Here we see that system performs as a switch both in strong-coupling ($k_{ex}<100$) and bad-cavity  ($k_{ex}>100$) limits, showing wider range of functionality in the strong coupling limit in agreement with our findings for the saturation behaviour. In Fig. \ref{switch_flux_bad}(e,f) we show a zoom of Fig. \ref{switch_flux_strong}(e,f), as it reveals an interesting feature. System is performing as a switch only in the fiber-overcoupling regime which is given by the condition $k_{ex}>>k_{i}$, this condition ensures that most of the light is transferred in the cavity and then collected out of the cavity, which is obviously a necessary condition for strong light-matter interaction.


\section{photon statistics}
\label{photonstatistics}

\begin{figure}
\begin{tabular}{ccccc}
\includegraphics[width=4cm]{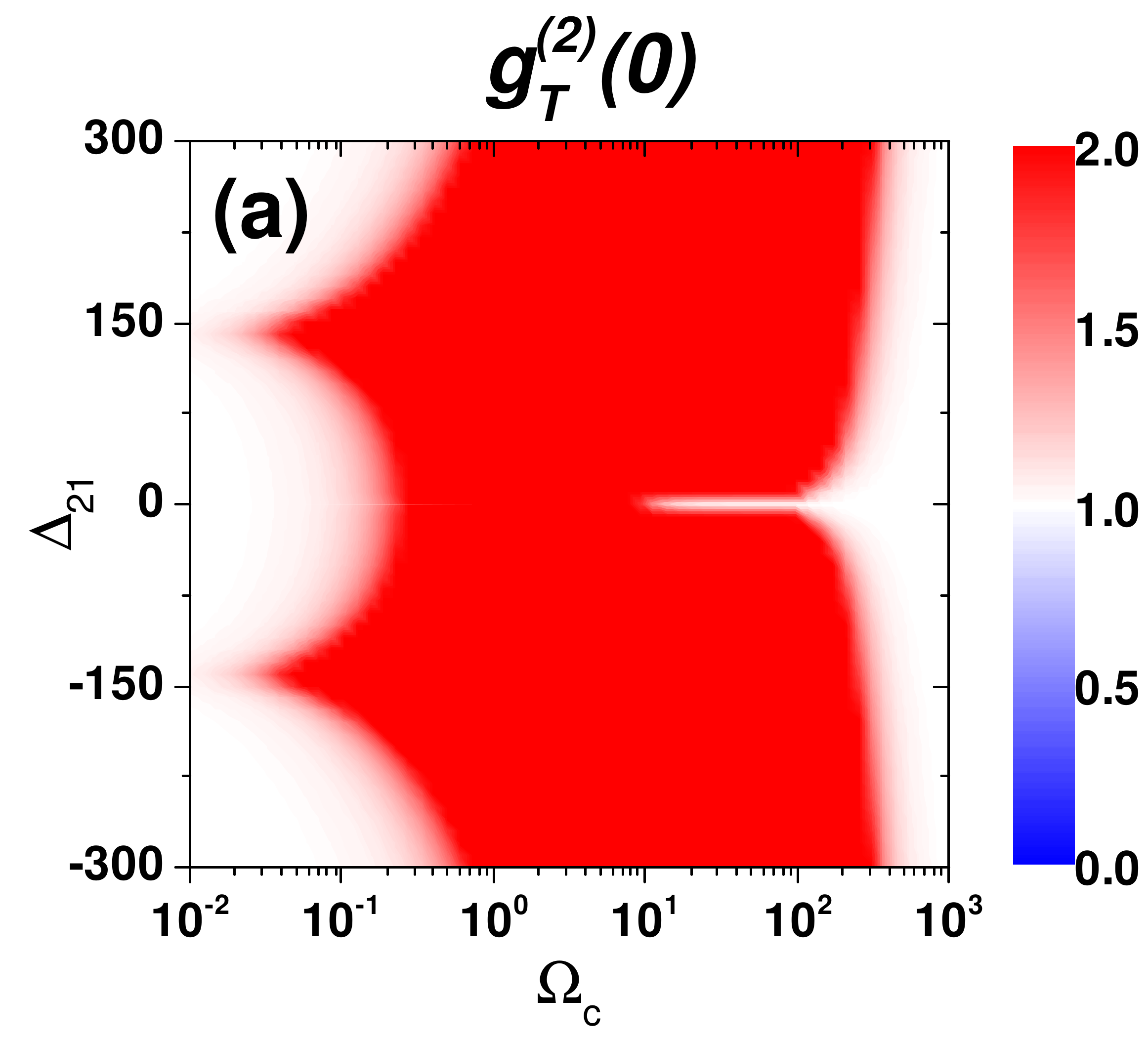} &
\includegraphics[width=4cm]{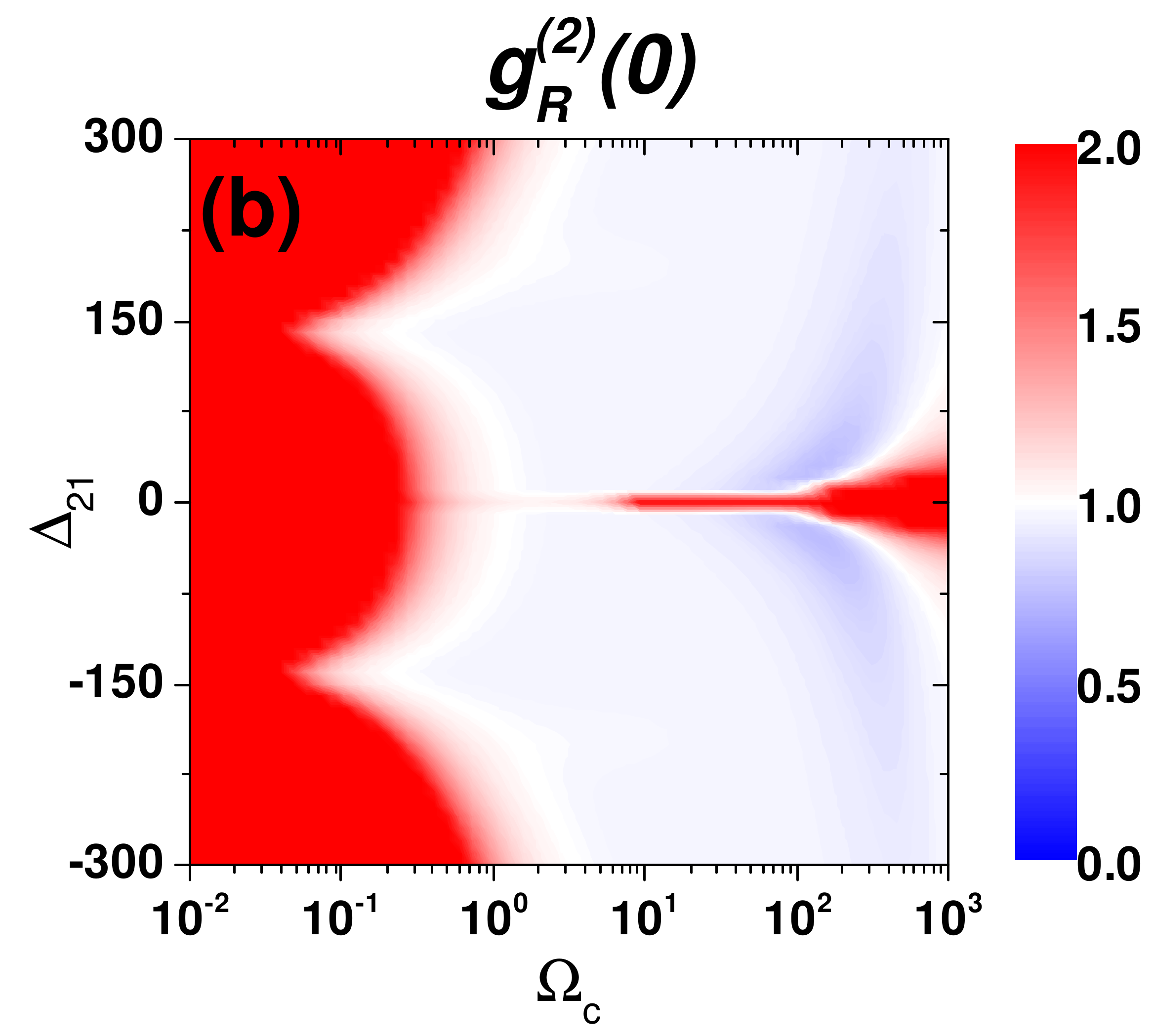} \\
\includegraphics[width=4cm]{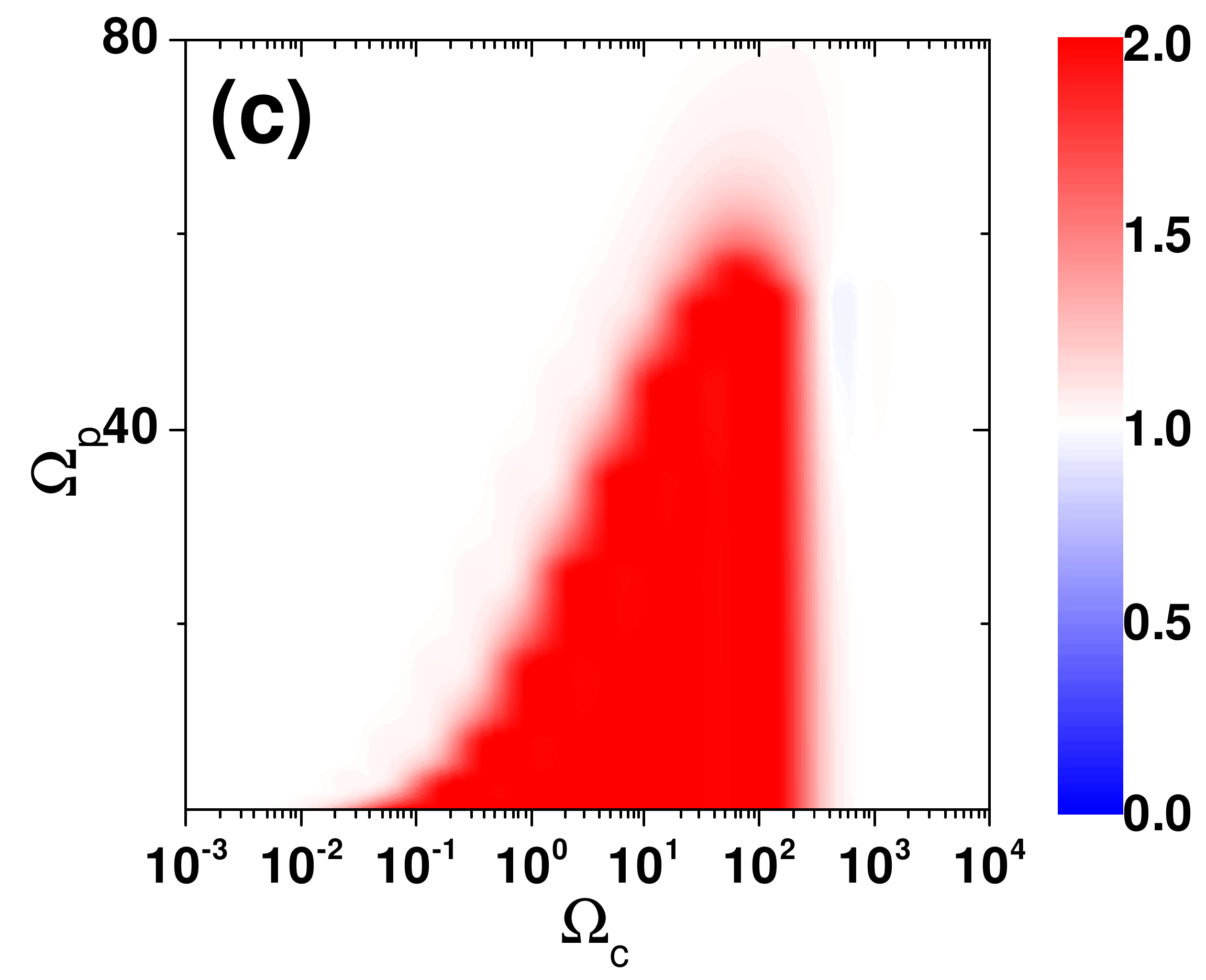} &
\includegraphics[width=4cm]{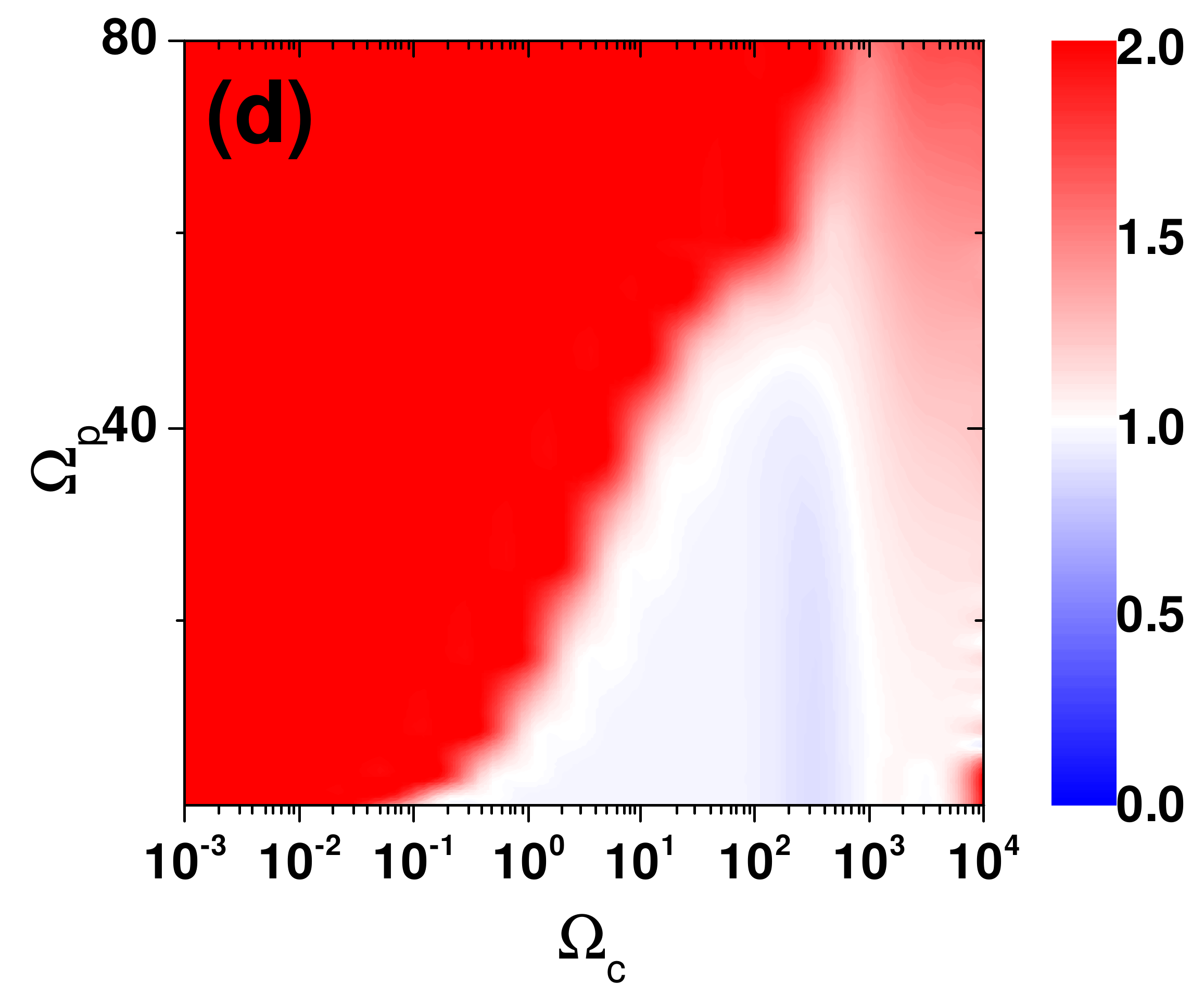} \\
\includegraphics[width=4cm]{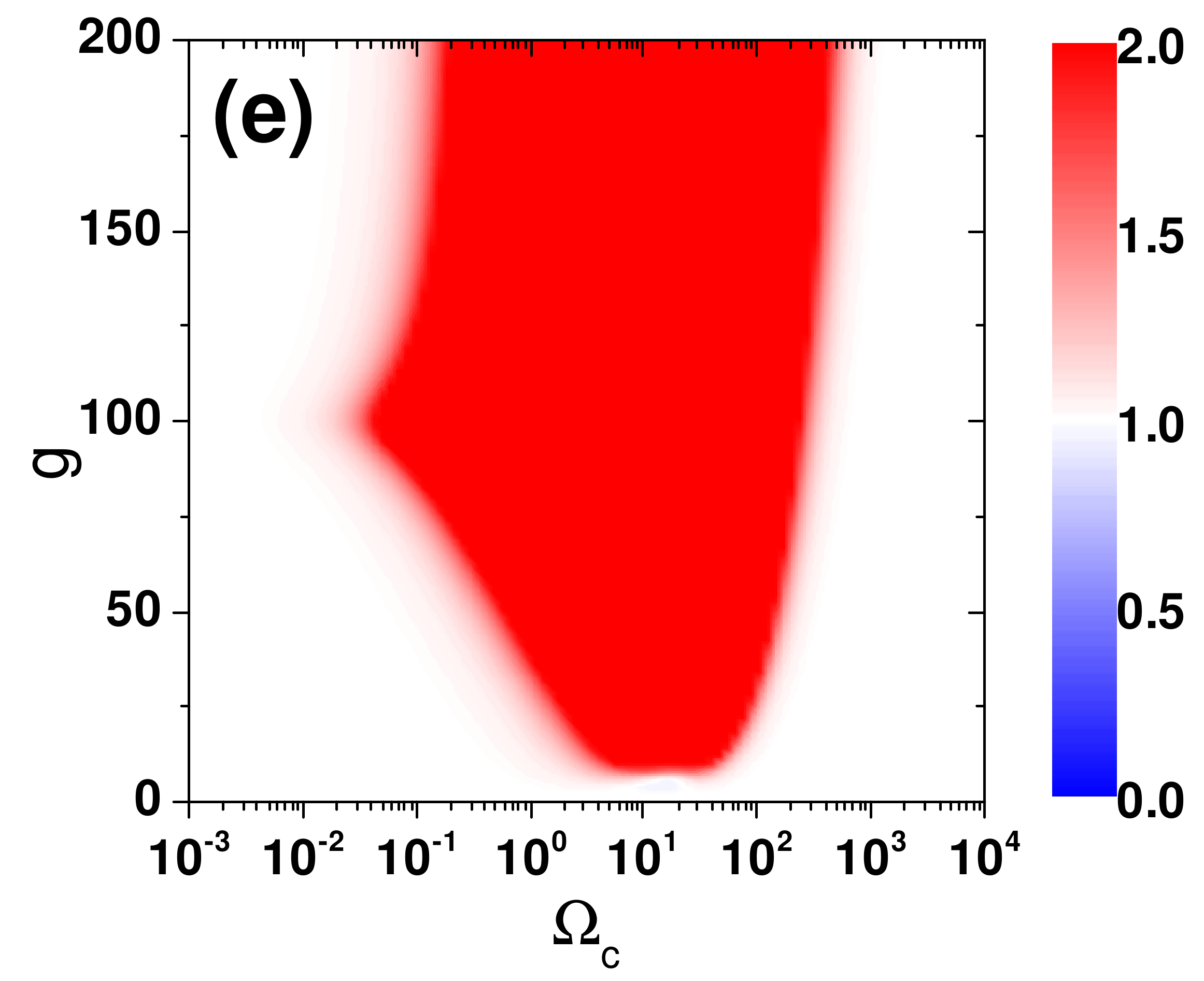} &
\includegraphics[width=4cm]{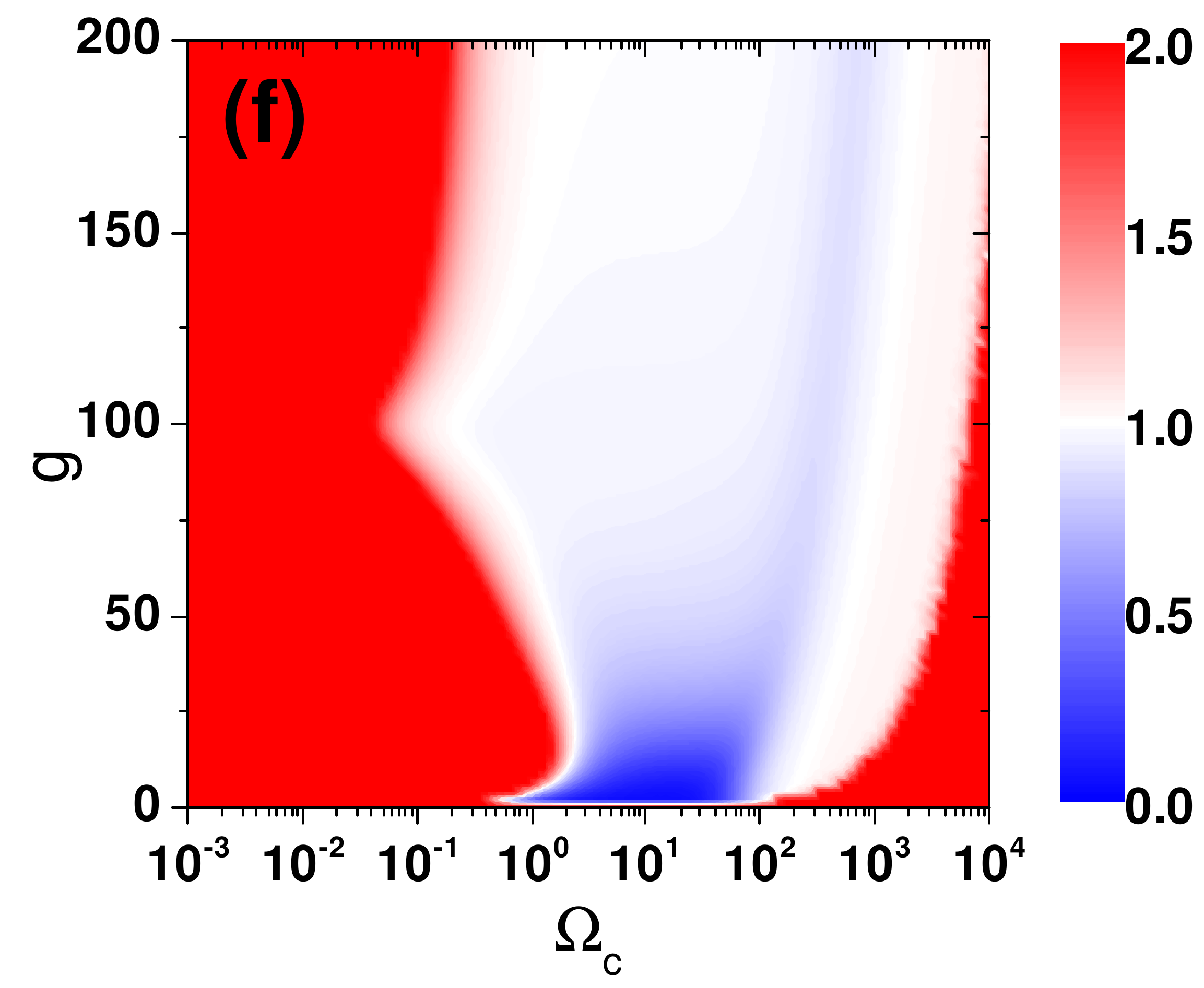} \\
\includegraphics[width=4cm]{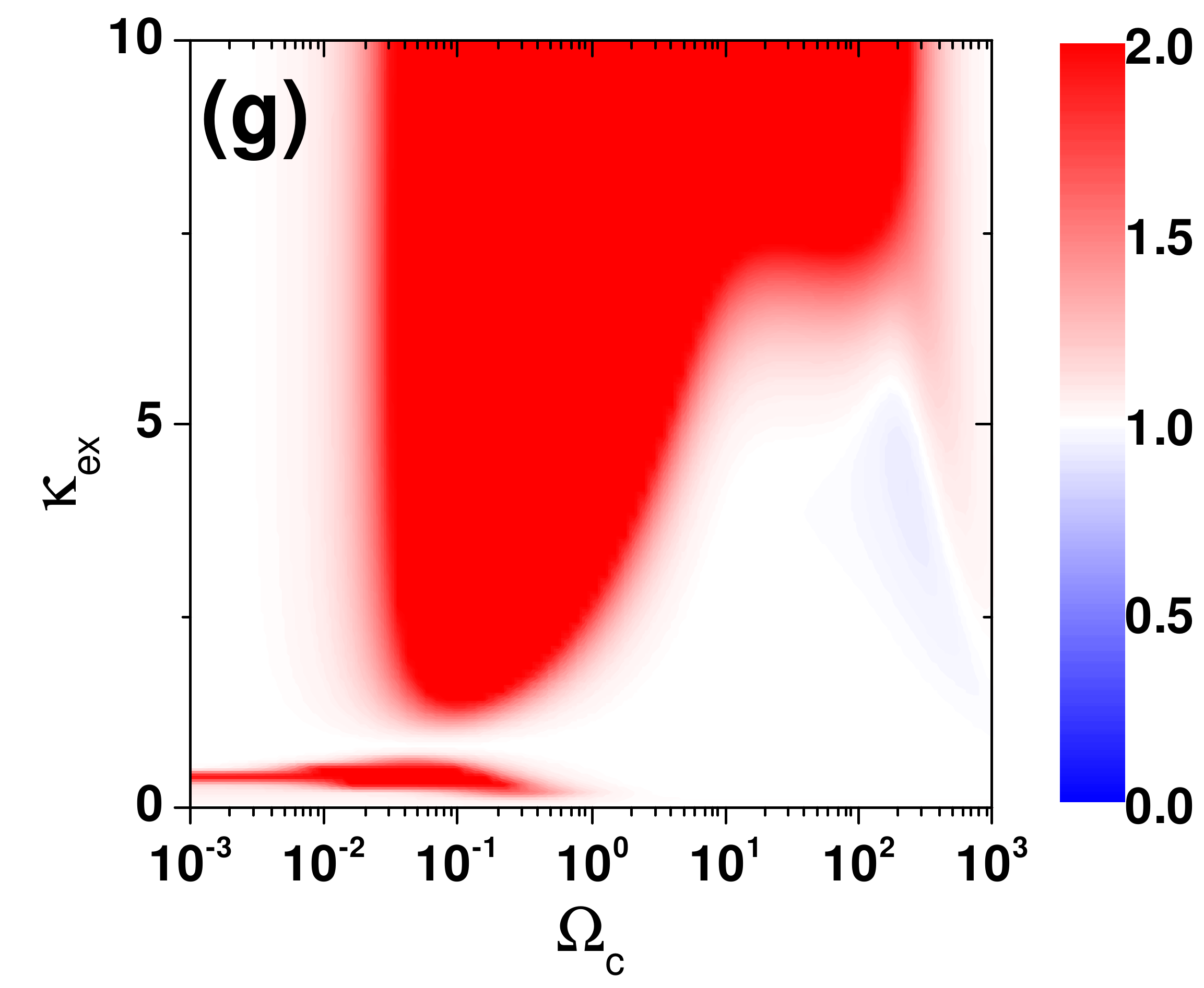} &
\includegraphics[width=4cm]{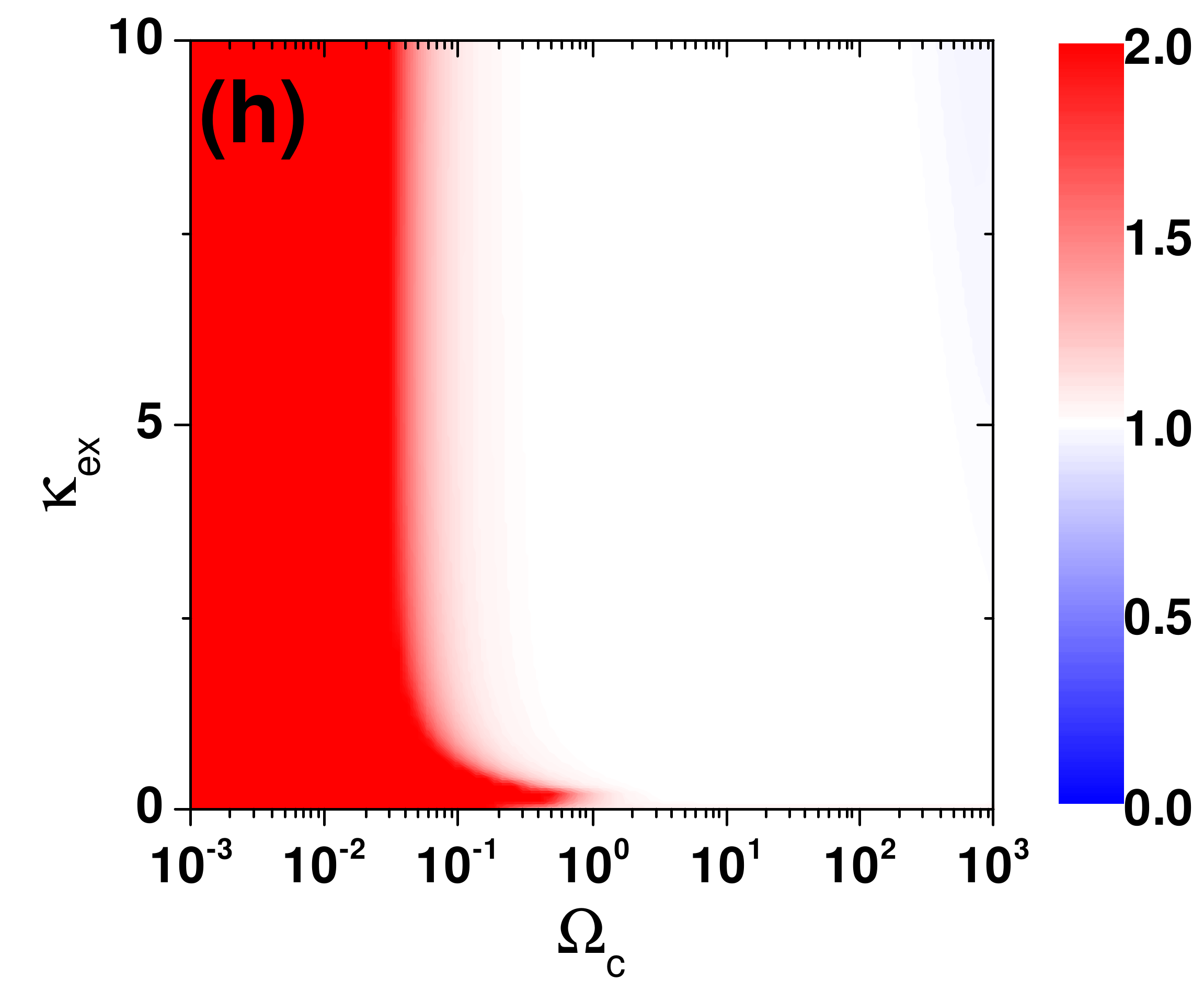} \\
\end{tabular}
\caption{(colour online). Contour plots for the photon statistics of transmitted and reflected photons in strong coupling regime. The parameters are the same as Fig. \ref{switch_demo}(c) except $\Omega_{p}=1$.}
\label{stat}
\end{figure}
In this section, we study the photon statistics  for the transmitted and reflected light fields, both in strong coupling and bad-cavity limits. Here we mainly focus on regions, where quantum switch is functioning, which means $ T \approx 0 $. In this region, most of the photon flux is reflected and our main focus here is to study the statistics of the reflected light, however since still small number of photons is passing in the forward direction  two-photon correlation function can be observed through photo-detection.
Two-photon correlation functions  for transmitted and reflected fields are defined through the output fields as follows:
\begin{equation}
\label{correlation}
\begin{split}
g_{\text{T}}^{(2)}(0)&=\frac{\langle{a_{\text{out},\text{ex}}^{\dag}a_{\text{out},\text{ex}}^{\dag}a_{\text{out},\text{ex}}a_{\text{out},\text{ex}}}\rangle_{\text{ss}}}{(\langle{a_{\text{out},\text{ex}}^{\dag}a_{\text{out},\text{ex}}}\rangle_{\text{ss}})^2},\\
g_{\text{R}}^{(2)}(0)&=\frac{\langle{b_{\text{out},\text{ex}}^{\dag}b_{\text{out},\text{ex}}^{\dag}b_{\text{out},\text{ex}}b_{\text{out},\text{ex}}}\rangle_{\text{ss}}}{(\langle{b_{\text{out},\text{ex}}^{\dag}b_{\text{out},\text{ex}}}\rangle_{\text{ss}})^2}.\\
\end{split}
\end{equation}
If $g^{(2)}(0)<1$ (e.g. for the field in the Fock state $|n \rangle$, it can be easily shown that $g^{(2)}(0)=1-1/n$), and the field has sub-Poissonian statistics. If $g^{(2)}(0)=1$ (e.g. .any coherent field  $| \alpha \rangle$), the field has a Poissonian statistics. Finally if $g^{(2)}(0)>1$, then the field has a super-Poissonian statistics (e.g. for the single-mode thermal field $g^{(2)}(0)=2$)\cite{SZ97}.

\begin{figure}
\begin{tabular}{ccccc}
\includegraphics[width=4cm]{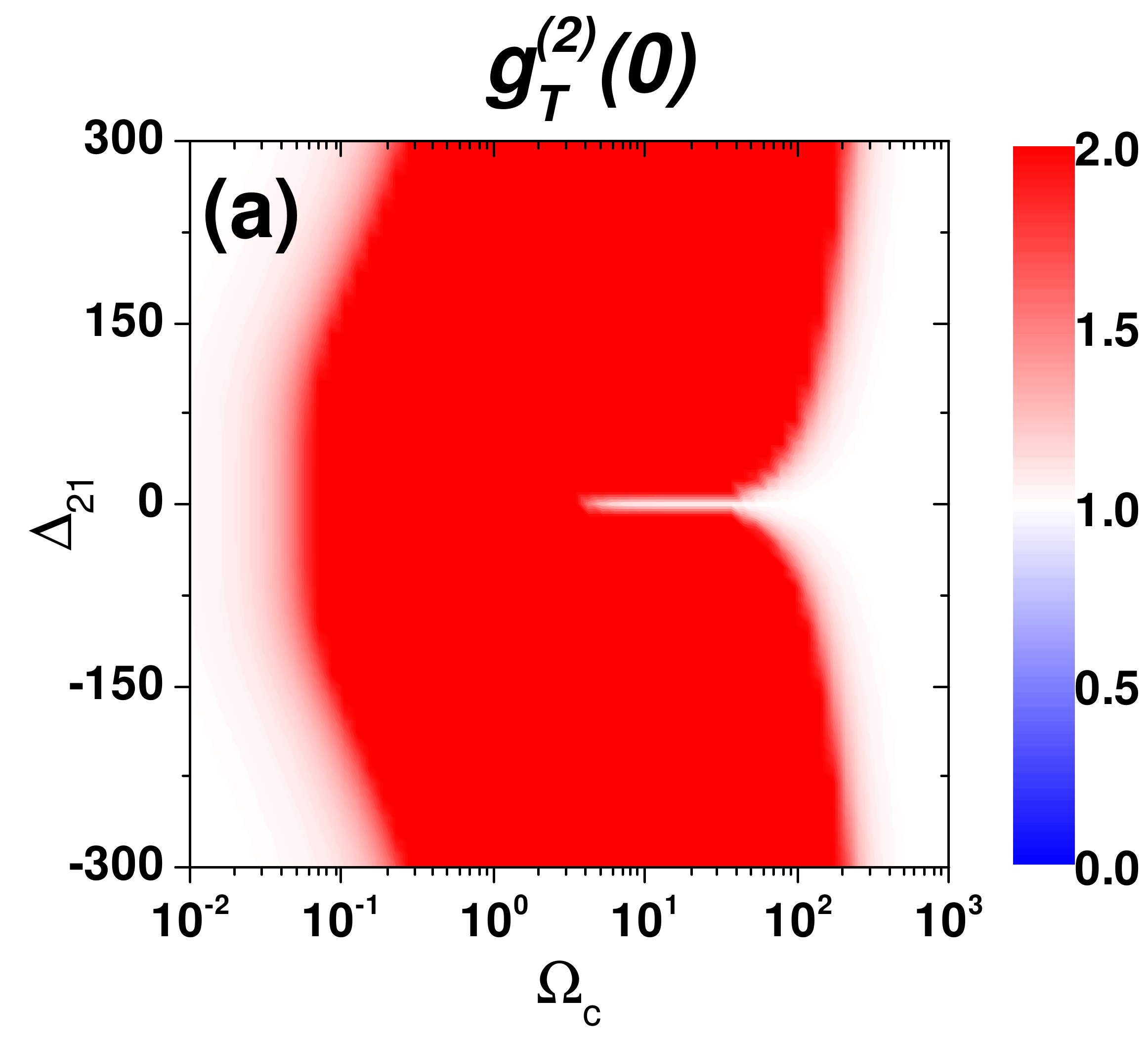} &
\includegraphics[width=4cm]{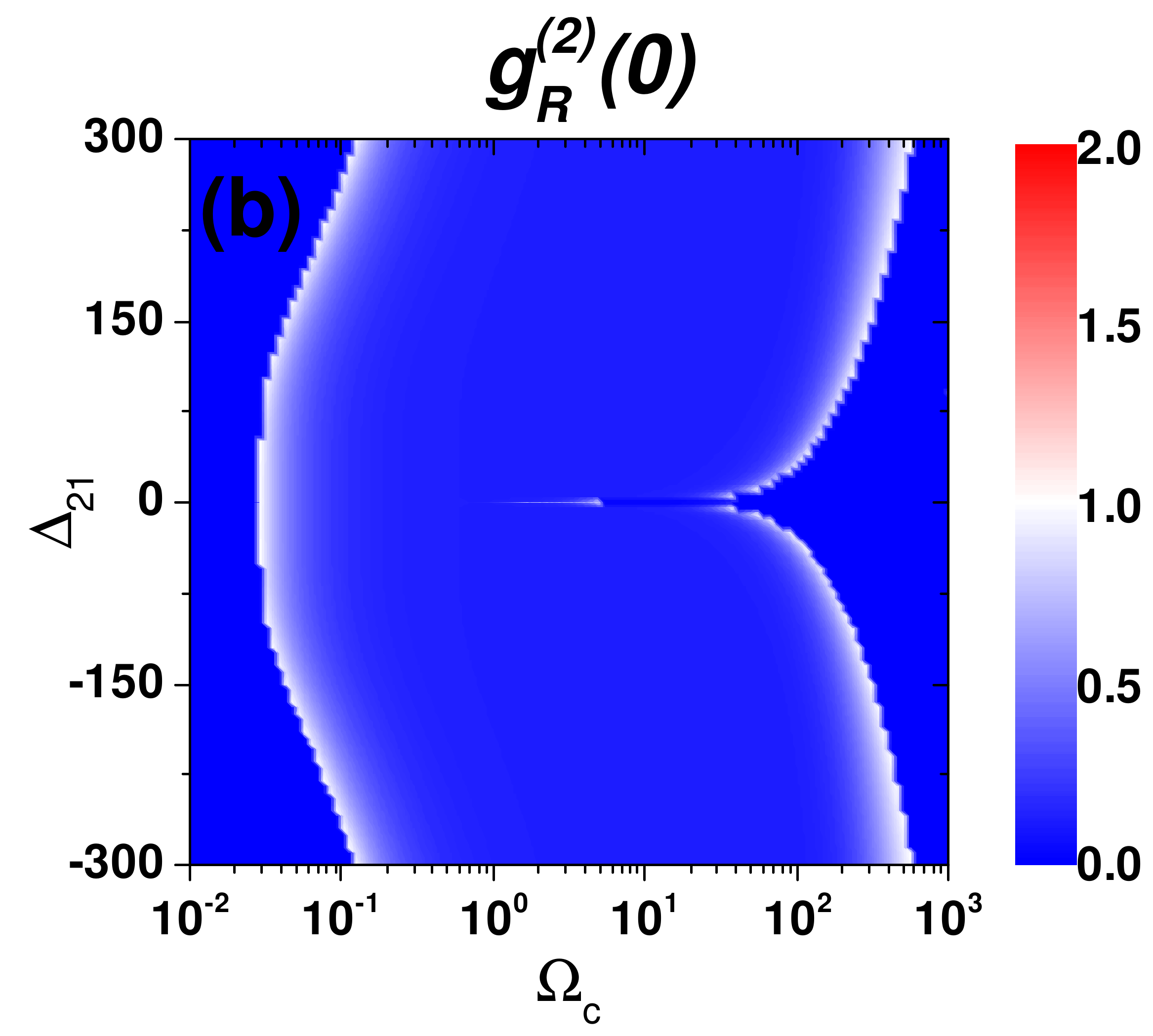} \\
\includegraphics[width=4cm]{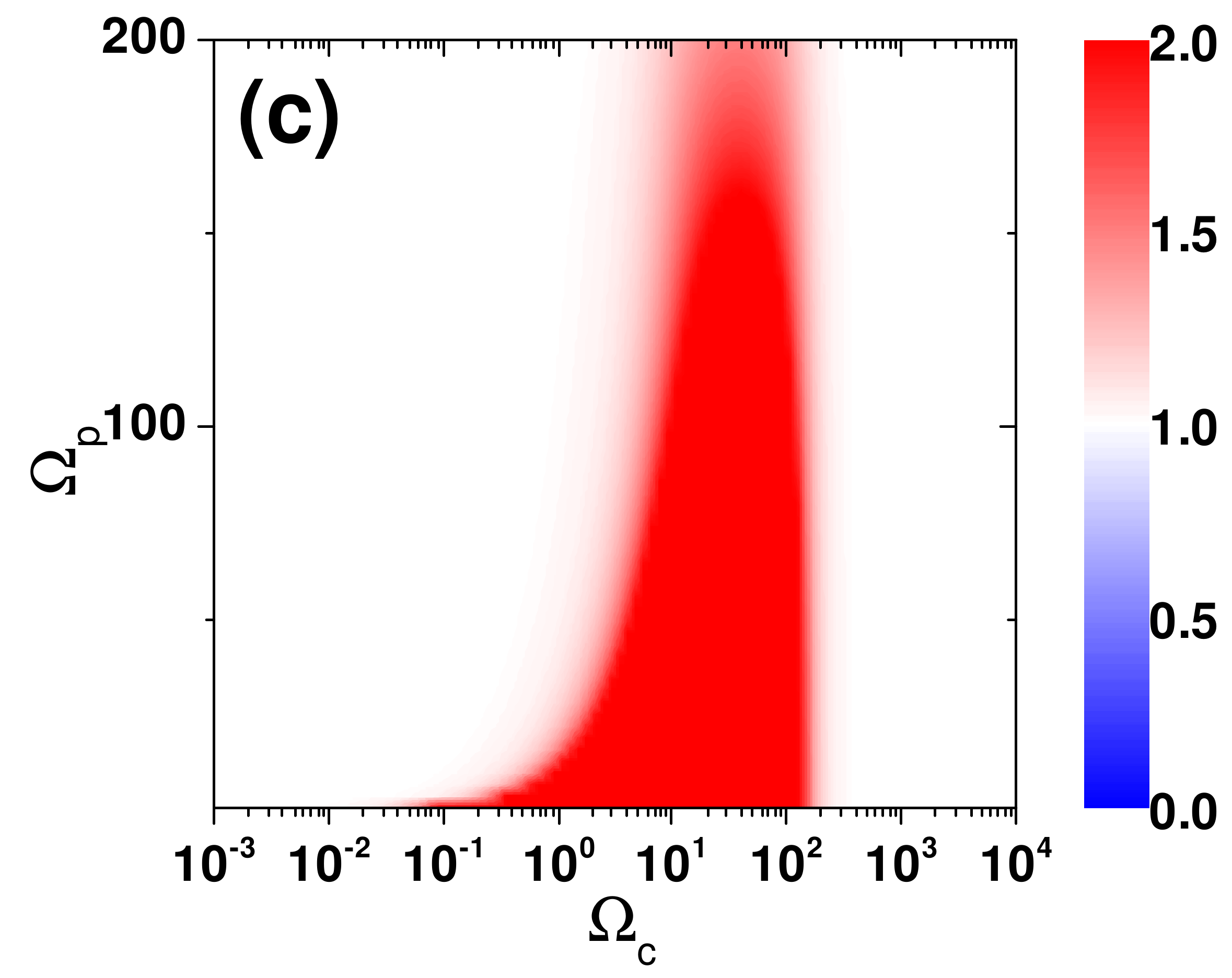} &
\includegraphics[width=4cm]{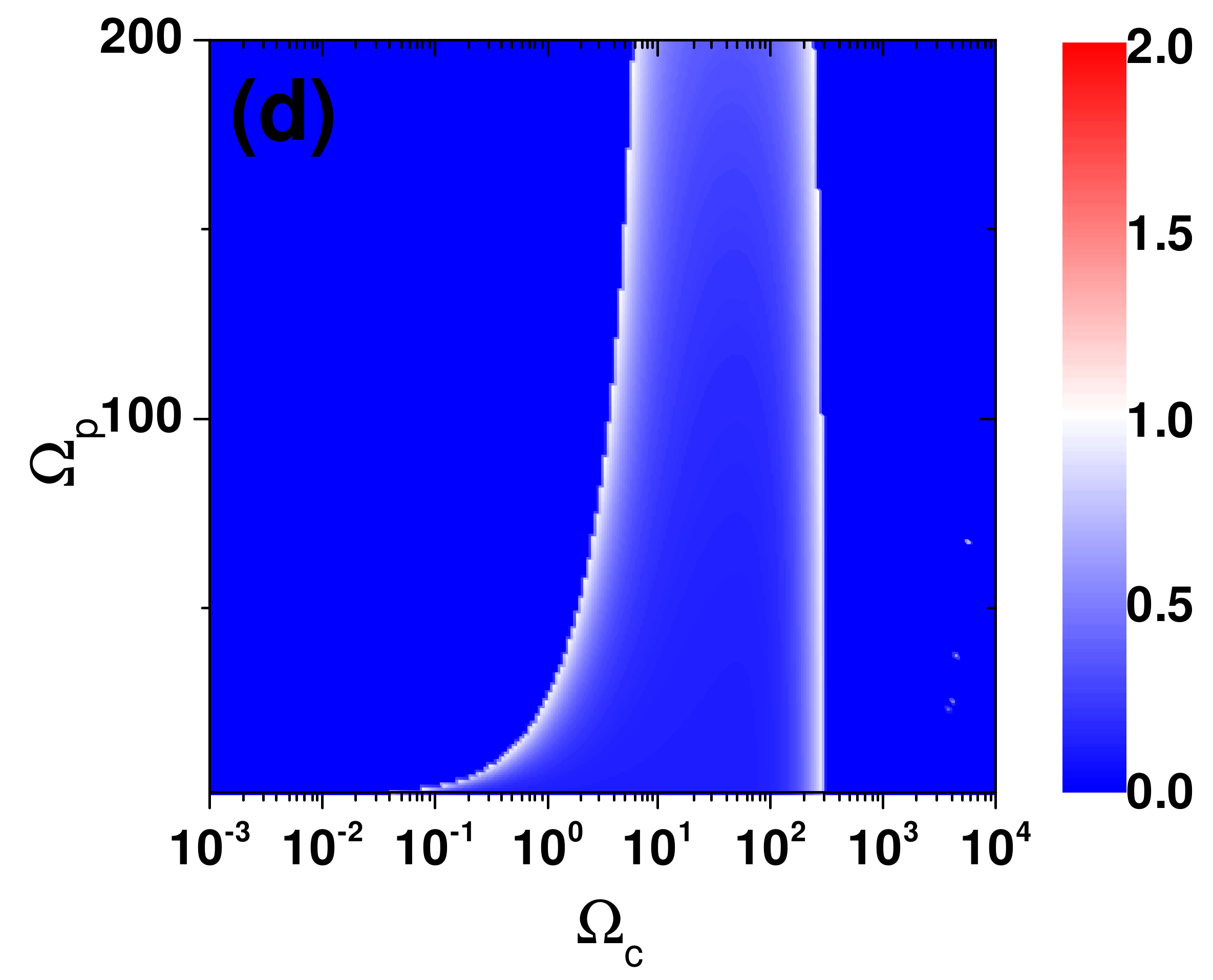} \\
\includegraphics[width=4cm]{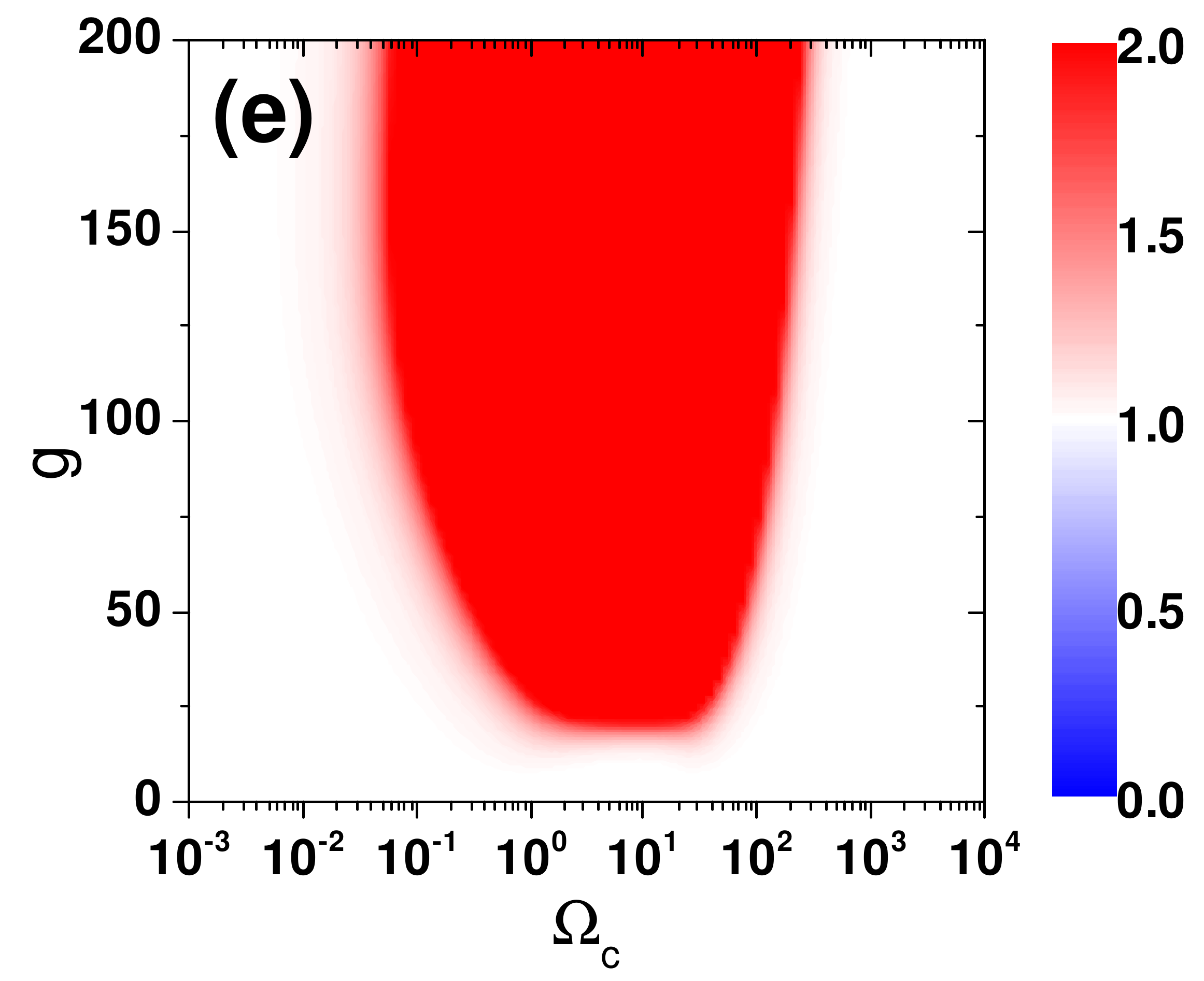} &
\includegraphics[width=4cm]{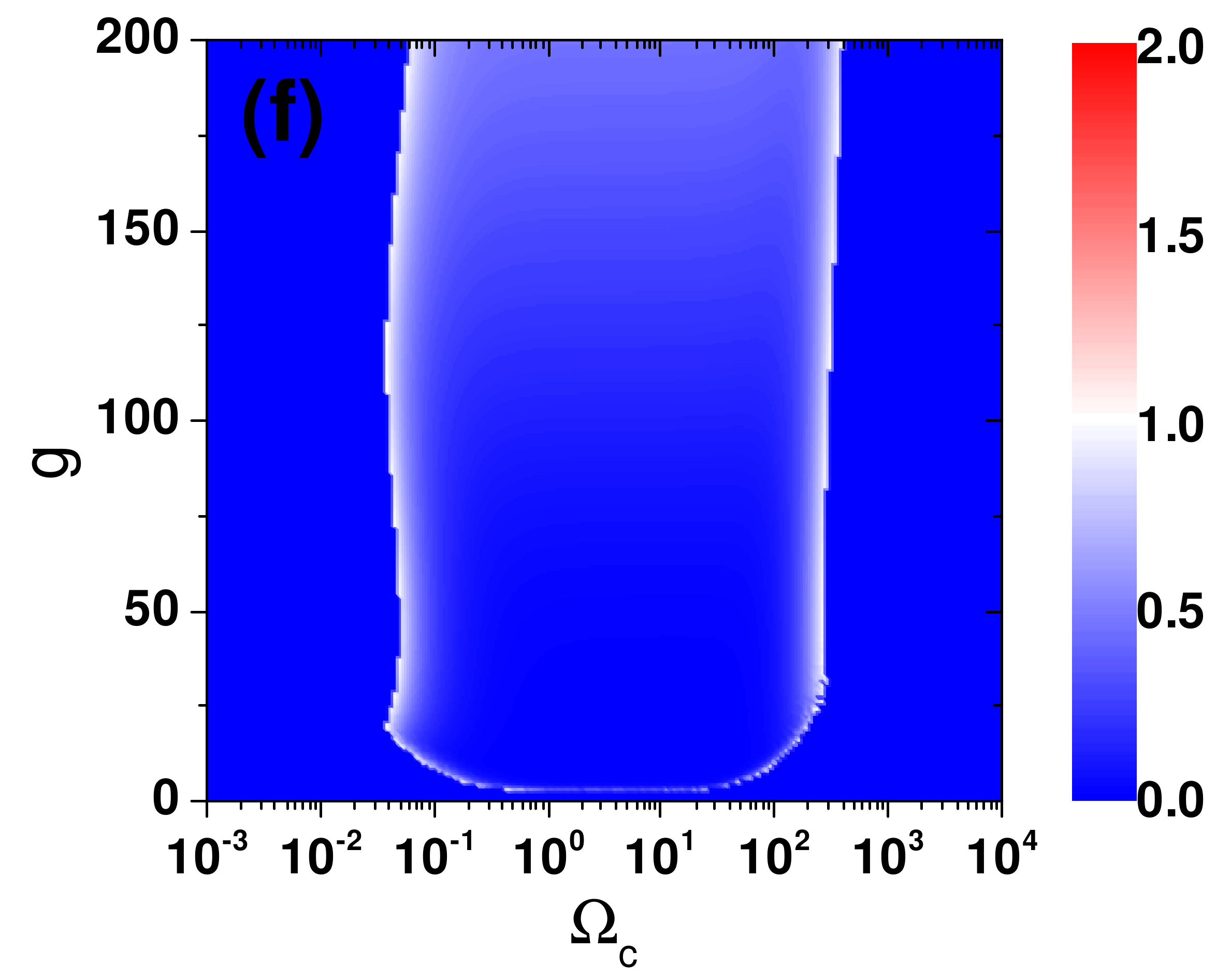} \\
\includegraphics[width=4cm]{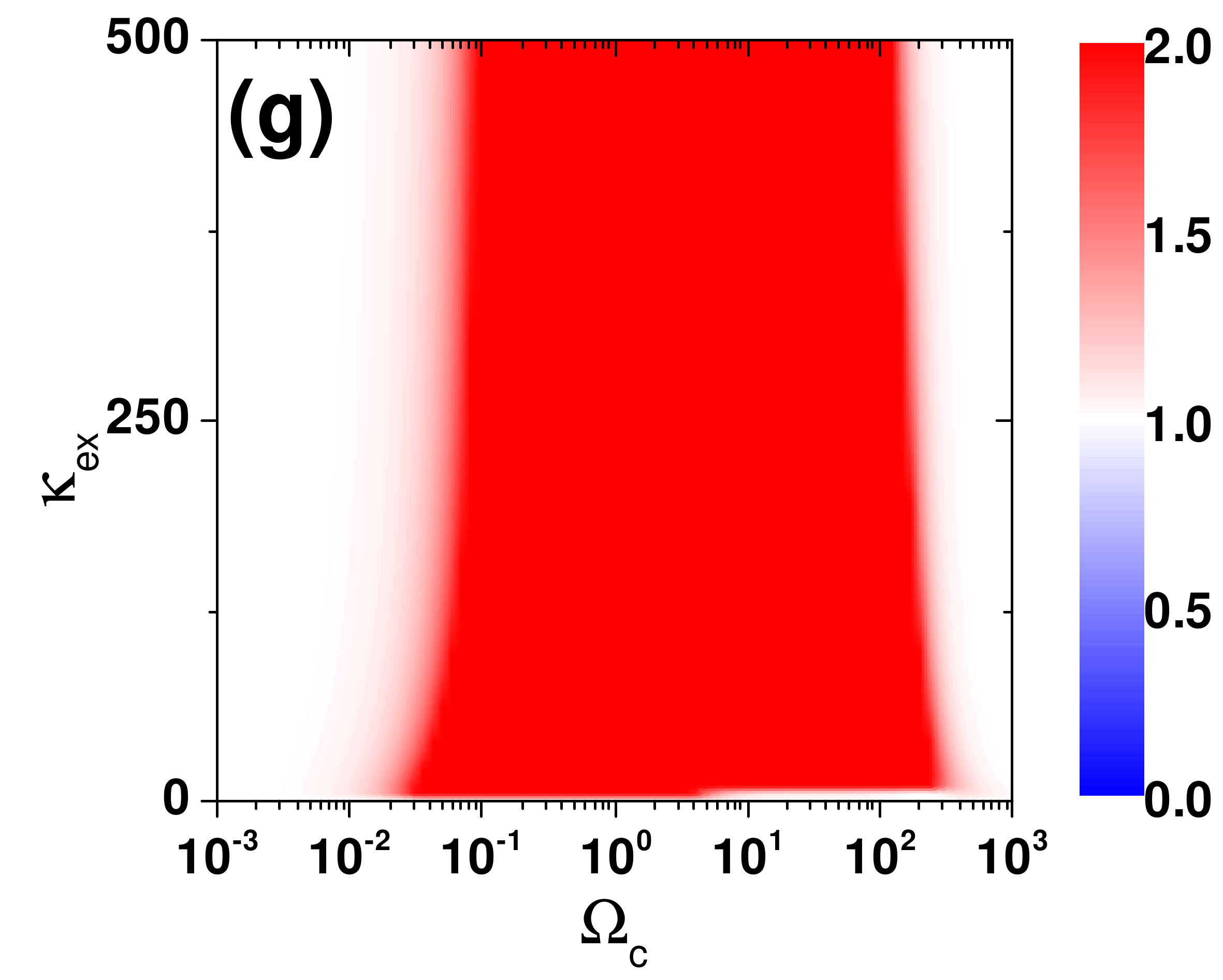} &
\includegraphics[width=4cm]{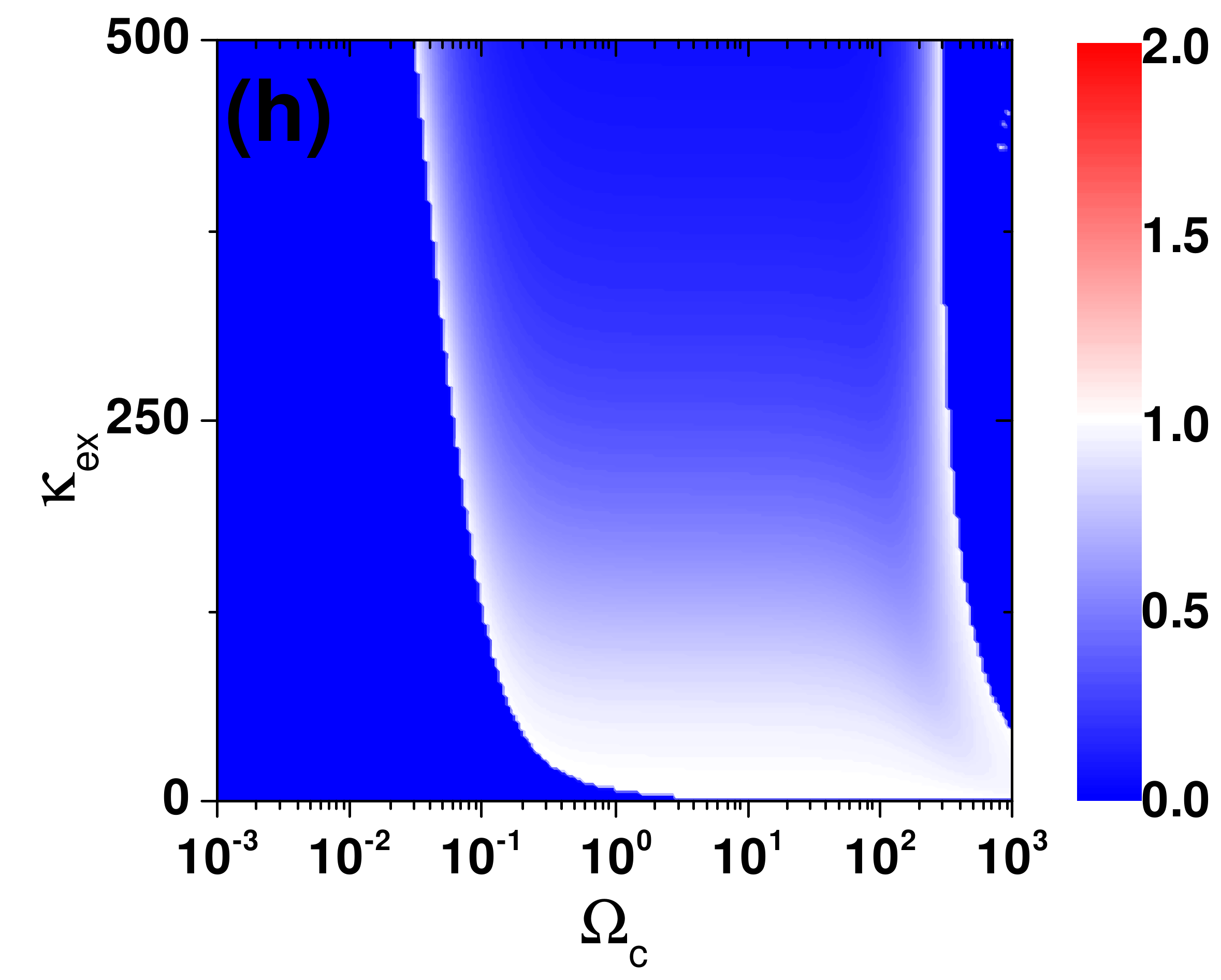} \\
\end{tabular}
\caption{(colour online). Contour plots for the photon statistics of transmitted and reflected photons in bad-cavity regime. The parameters are the same as Fig. \ref{switch_demo}(g) except $\Omega_{p}=1$.}
\label{statis}
\end{figure}
Correlation functions are plotted on Figs.\ref{stat} and \ref{statis}, respectively for the strong-coupling and bad cavity limits, varying on x-axis the control field and on y-axis the physical parameter of interest. Here we truncate $g^{(2)}(0)$ function for values higher than 2 for convenience of graphical representation, since the regions of quantum  light ($g^{(2)}(0)<1$) are easily noticeable in this case. We remark, that this kind of truncation still keeps all information about the statistics of the light only omitting the regions of extreme bunching which is not of interest in the current manuscript.

As we can see from the second column of Figs.\ref{stat}, in the regime of functional switch(here $g>k_{ex}$) reflected light remains in the coherent state. To understand why this is the case, we write an expression for the reflected output field and take into account the $\langle A \rangle \approx 0$, as was demonstrated in Fig.~\ref{intenisties}. Then we can estimate, that $b_{out,ex} \approx \sqrt{k}B$, here we took into account that $\langle b_{in,ex} \rangle = 0$.  The normal mode $\langle \sqrt{\kappa} B \rangle =-i\Omega_{p}\sqrt{2k}=\langle a_{in} \rangle$, which means the mode $B$ has the same statistics as the input field(this has been numerically demonstrated in the ref~\cite{SP07}) which is assumed to be in the coherent state.

In contrast, in bad cavity limit the reflected light becomes quantum as $g^{2}_{R}(0)<< 1$, which corresponds to the dark/blue regions of the Fig.\ref{statis}. This is a result of a non-conventional photon blockade \cite{Dayan2008}, and can be  understood by studying the output reflected field. After making the adiabatic elimination of cavity modes, which we outline in great detail in  Appendix~\ref{appendix}, for calculating average values the following mapping $b_{out,ex}\rightarrow \beta_{0}+\beta_{-}\sigma_{1e}$. Moreover, in the case when  $\Delta_{r}=0$ and $h=0$, the parameter $\beta_{0}=0$, which means reflected photons are solely generated by an atom. After an atom emits, the photon it is projected into its ground state, and it takes finite amount of time, given by $1/\Gamma$, where $\Gamma$ is the Purcell enhanced decay rate (See Appendix \ref{appendix} for the expression of $\Gamma$), for it to get re excited and emit a photon again. This also can be showed, from the analytical expression (\ref{badcavity}), from which it immediately follows that $g^{2}_{R}(0)=0$, which corresponds to the single photon statistics(as its been mentioned for the Fock state $|n \rangle$,$g^{(2)}(0)=1-1/n$, so when $n=1$,$g^{(2)}(0)=0$ ). It is important to notice, that $g^{2}_{R}(0)= 0$, does not hold in our numerical simulations, and $g^{2}_{R}(0)\approx 0.1$, since we are considering the case $k=2g$, and  not really in the  bad cavity limit, where $k>>g$.

For the transmitted field interesting features appear because of the interference between the straight-through transmission of the coherent driving field and the forward scattered fluorescence from the atom. The consequences of this interference on the photon statistics were first time noted in the Ref.\cite{Rice88}, for the single-atom interacting with a single mode-cavity in a bad-cavity limit. As we can see from the first columns of Figs.\ref{stat} and \ref{statis}, in the regions of switch functionality, transmitted light shows bunching behavior (dark/red regions) as a result of destructive interference between the field radiated by the atom and an intracavity field. This behavior in terms of normal modes $A$ and $B$, has been explained in great detail in Ref.~\cite{PA14}, and it turns out that bunching behaviour is a consequence of normal mode $A$ being strongly bunched(we have numerically verified that this holds for our system in both limits).

\begin{figure}
\begin{tabular}{ccccc}
\includegraphics[width=4cm]{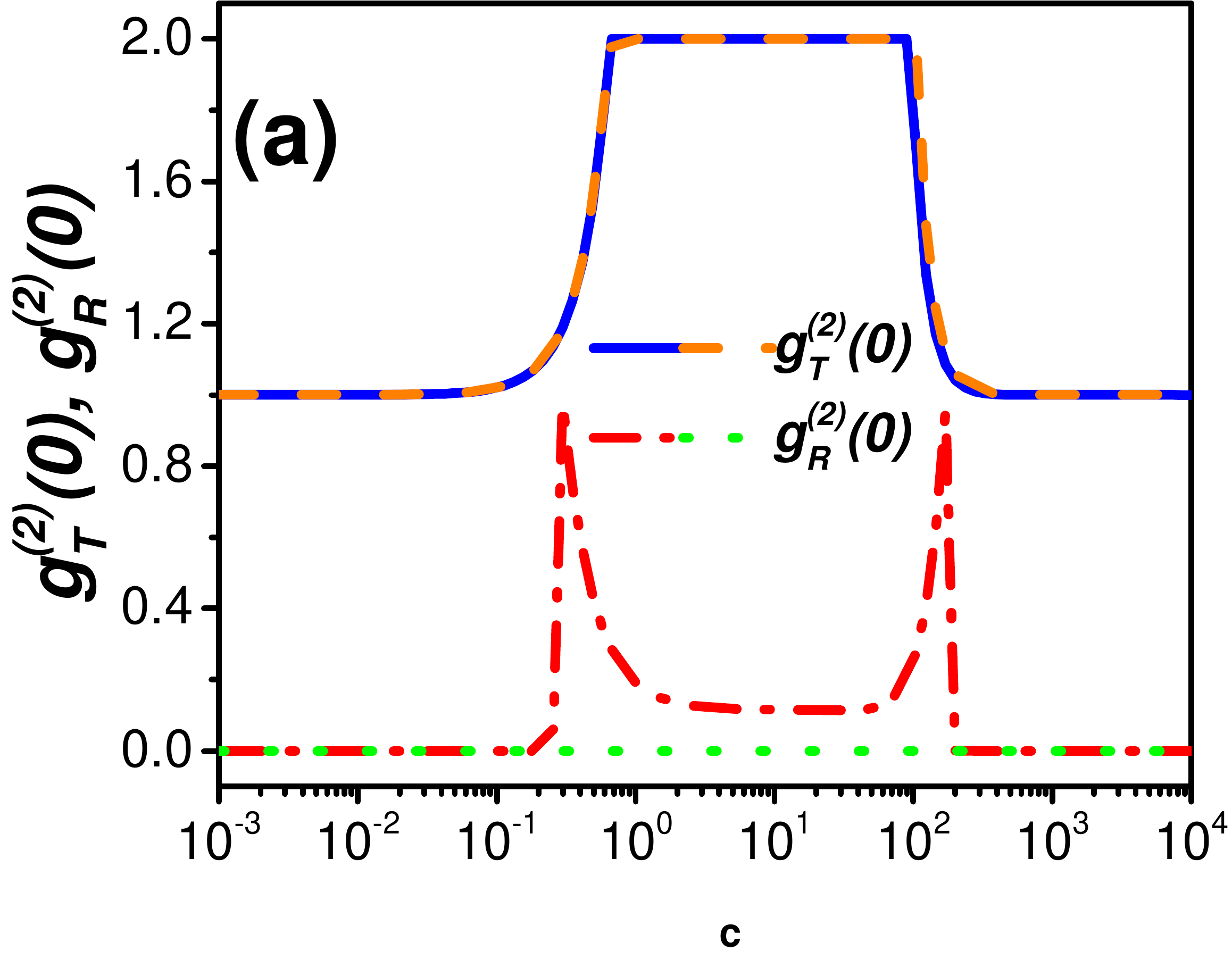} &
\includegraphics[width=4cm]{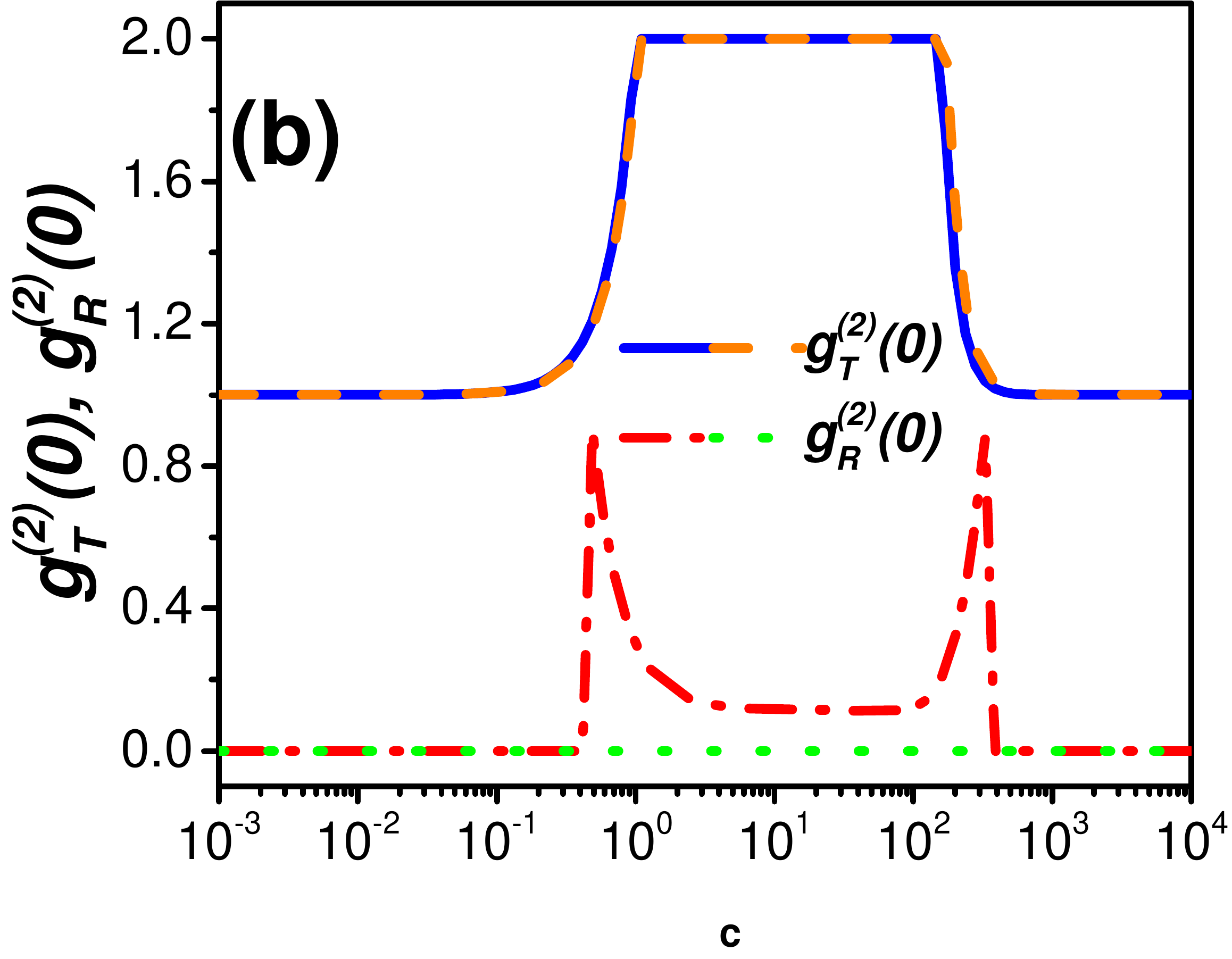} \\
\end{tabular}
\caption{(colour online). (a) and (b) are the comparison between numerical simulation and adiabatic elimination for Fig. \ref{switch_demo}(e) and \ref{switch_demo}(g).
}
\label{num_analyt}
\end{figure}

As it can be seen in Fig.~\ref{num_analyt}   analytical and numerical results for the two-photon correlation function agree well, in the bad-cavity limit, with  a drawback that numerical approach starts failing for obtaining $g^{(2)}_{R}(0)$ out of the switch functionality region, where we simply  set it to zero, when it obtains values bigger than one. This happens because of the divergence  of normalized two-photon correlation function, when photon flux is zero and no photons are detected(this gets even more apparent by considering correlation function for the Fock state $|n \rangle$,  $g2(0)=1-1/n$ which diverges when $n=0$). This remark is substantiated by the fact that for the finite value of $h$, analytical and numerical approaches start to  agree better with increasing value of $h$, because the photon flux for the reflected field never gets equal to zero. To summarize, $g^{(2)}_{R}(0)=0$ for the all values of $\Omega_{c}$, however, photodetection is going to reveal anti-bunched statistics in the region of switch functionality.




\section{Conclusions}
\label{conclusion}
In this paper, we suggested new scheme for realizing quantum transistor for the incoming coherent field. Our scheme is based on the coupling the fibre-coupled ring cavity with a single $\Lambda$-level atom. We have demonstrated that it is possible to tune the system by the external control field from a being fully transparent  to being a fully reflective through numerical simulations of the master equations. We emphasize, that our proposal has an advantage of being easily implemented experimentally, compared to other proposals which may require high control over the ``gate valve" parameter, which is not easily tunable in current experiments. Here, we have concluded, that the switch functions both in strong coupling and weak coupling limits, under the condition of strong fibre-over-coupling, (showing better performance in the strong-coupling limit) for the reasonably large amplitude of the incoming field(up to $\approx 50\text{MHz}$, when $g>k_{ex}$, up to $\approx 40\text{MHz}$, when $g<k_{ex}$). Moreover, we have demonstrated that the regime of functionality  can be extended by increasing two-photon detuning and found the optimal value for. Analytical results in bad-cavity limit are obtained through adiabatic elimination of cavity modes, and they are in good agreement with numerical simulations of respective master equations. Surprisingly, this approach works even in the strong coupling limit showing qualitative agreement for transmitted and reflected field intensities. It  is important to mention that our protocol works only for non-zero two-photon detuning which means that we are not using conventional EIT based approach.

By studying the statistics of transmitted and reflected fields, we have verified that quantum transistor does not modify the state of the incoming coherent field in the strong coupling limit, however in the bad-cavity limit our system can produce quantum states of light in the reflected field. So in bad-cavity limit our system can be used as a ``black box" which acts as a quantum device which takes as input coherent field and gives quantum light in the output.

Our proposal has a potential interest in realizing quantum information protocols with coherent light states. For future projects, it would be interesting to concatenate several ring cavities to fibre and study if the system can work as a photon router for a few-photon incoming state. It also would be of interest to implement quantum repeater schemes such as DLCZ\cite{Duan,Malak}.

In addition, there have been several interesting theoretical proposals on coupling NV centres with ring cavities for generating entangled states between the colour centres \cite{Shi2010,Yang1,Yang2,Liu2013}. Since colour centres are solid state systems there is no need to trap them as it is the case with cold atoms. Moreover, in the recent experimental realization a single photon source based on coupling ring cavity with SiV vacancies has been realized\cite{Jacques2018}. So it would be interesting to implement a quantum switch  by coupling ring resonators with colour centers, which have a multi-level structure and can be utilized as $\Lambda$-level systems.



\section{Acknowledgements}

D.A, L.C.K and L.K. acknowledges support from NRF grant 2014NRF-CRP002-042.
The IHPC A*STAR Team would like to acknowledge
the National Research Foundation Singapore (Grant
No. NRF2017NRF-NSFC002-015, NRF2016-NRF-ANR002, NRF-CRP
14-2014-04) and A*STAR SERC (Grant No. A1685b0005). D. A. would like to acknowledge  E. Munro and A. Chia for stimulating discussions, reading the manuscript and making useful comments and suggestions.

\appendix

\setcounter{figure}{0}
\renewcommand{\thefigure}{S\arabic{figure}}

\section{adiabatic elimination}
\label{appendix}
In the bad cavity limit ($\kappa\gg\gamma,g$), we can adiabatically eliminate the lossy cavity mode and obtain an effective model for the three-level atom \cite{Dayan2008,C91}. By expressing cavity modes through the normal modes as $a=\frac{A+B}{\sqrt{2}},b=\frac{A-B}{\sqrt{2}}$,  the Hamiltonian Eq. (\ref{atom-ring}) recasts into
\begin{equation}
\begin{split}
H&=(\Delta_{r}+h)A^{\dag}A+(\Delta_{r}-h)B^{\dag}B+\Delta_{e1}\sigma_{ee}+\Delta_{21}\sigma_{22}\\
&+g_{A}(A^{\dag}\sigma_{1e}+\sigma_{e1}A)-ig_{B}(B^{\dag}\sigma_{1e}-\sigma_{e1}B)\\
&+\frac{\Omega_{p}}{\sqrt{2}}(A+A^{\dag})+\frac{\Omega_{p}}{\sqrt{2}}(B+B^{\dag})+\Omega_{c}(\sigma_{2e}+\sigma_{e2}),\\
\end{split}
\end{equation}
where $g_{A}=\sqrt{2}\text{Re}[g]$ and $g_{B}=\sqrt{2}\text{Im}[g]$.
After writing the Heisenberg equations for the normal modes in the framework of the input-output theory\cite{WM94}, we obtain the following expressions:
\begin{equation}
\begin{split}
\dot{A}&=-i[(\Delta_{r}+h-i\kappa/2)A+g_{A}\sigma_{1e}+\frac{\Omega_{p}}{\sqrt{2}}]\\
&-\sqrt{\kappa_{\text{ex}}}A_{\text{in},\text{ex}}-\sqrt{\kappa_{\text{i}}}A_{\text{in},\text{i}},\\
\dot{B}&=-i[(\Delta_{r}-h-i\kappa/2)B-ig_{B}\sigma_{1e}+\frac{\Omega_{p}}{\sqrt{2}}]\\
&-\sqrt{\kappa_{\text{ex}}}B_{\text{in},\text{ex}}-\sqrt{\kappa_{\text{i}}}B_{\text{in},\text{i}}.\\
\end{split}
\end{equation}
It is easy to show that  the steady state amplitudes of empty cavity($g=0$) fields can be expressed as
\begin{equation}
\label{empty_cavity}
\begin{split}
\alpha_{A}&=\langle{A}\rangle=-\frac{\Omega_{p}}{\sqrt{2}}\frac{1}{\Delta_{r}+h-i\kappa/2},\\
\alpha_{B}&=\langle{B}\rangle=-\frac{\Omega_{p}}{\sqrt{2}}\frac{1}{\Delta_{r}-h-i\kappa/2}.\\
\end{split}
\end{equation}
To adiabatically eliminate the cavity modes, we formally integrate the operators of $A$ and $B$ and substitute them into the optical Bloch equation for the atom. After integration and taking into account that we are in the bad cavity limit, inside the integrals of for the normal cavity modes, the atomic correlation
functions may be evaluated at $t'=t$, thus we use the Markov approximation. After evaluating the integrals cavity normal modes takes the form:
\begin{equation}
\label{normalmodes}
\begin{split}
A(t)&=\alpha_{A}-\frac{ig_{A}\sigma_{1e}(t)+\sqrt{\kappa_{\text{ex}}}A_{\text{in},\text{ex}}(t)+\sqrt{\kappa_{\text{i}}}A_{\text{in},\text{i}}(t)}{i(\Delta_{r}+h-i\kappa/2)},\\
B(t)&=\alpha_{B}-\frac{g_{B}\sigma_{1e}(t)+\sqrt{\kappa_{\text{ex}}}B_{\text{in},\text{ex}}(t)+\sqrt{\kappa_{\text{i}}}B_{\text{in},\text{i}}(t)}{i(\Delta_{r}-h-i\kappa/2)}.\\
\end{split}
\end{equation}
The averages of $\sigma_{1e}$ and $\sigma_{12}$ are given by
\begin{equation}
\label{Blocheq}
\begin{split}
\langle{\dot{\sigma_{1e}}}\rangle&=-i(\Delta_{e1}-i\Gamma/2)\langle{\sigma_{1e}}\rangle+ig_{A}\langle{(\sigma_{ee}-\sigma_{11})A}\rangle\\
&-g_{B}\langle{(\sigma_{ee}-\sigma_{11})B}\rangle-i\Omega_{c}\langle{\sigma_{12}}\rangle,\\
\langle{\dot{\sigma_{12}}}\rangle&=-i\Delta_{21}\langle{\sigma_{12}}\rangle+ig_{A}\langle{\sigma_{e2}A}\rangle-g_{B}\langle{\sigma_{e2}B}\rangle-i\Omega_{c}\langle{\sigma_{1e}}\rangle.\\
\end{split}
\end{equation}
In the bad-cavity limit the cavity field correlation time is very short compared
to the atomic decay timescale. Thus we have, for example,
\begin{equation}
\begin{split}
&\langle\sigma_{ee}(t)A(0)\rangle=\langle\sigma_{ee}(t)A_{\text{in,ex}}(t')\rangle=\langle\sigma_{ee}(t)A_{\text{in,i}}(t')\rangle=0.\\
\end{split}
\end{equation}
After substituting Eqs.~(\ref{normalmodes}) into the Eqs.~(\ref{Blocheq}), we obtain:
\begin{equation}
\begin{split}
\langle\dot{\sigma_{1e}}\rangle&=-i(\Delta_{e1}^{'}-i\Gamma^{'}/2)\langle\sigma_{1e}\rangle+i\Omega_{p}^{'}\langle{\sigma_{ee}-\sigma_{11}}\rangle-i\Omega_{c}\langle{\sigma_{12}}\rangle,\\
\langle{\dot{\sigma_{12}}}\rangle&=-i\Delta_{21}\langle{\sigma_{12}}\rangle+i\Omega_{p}^{'}\langle{\sigma_{e2}}\rangle-i\Omega_{c}\langle{\sigma_{1e}}\rangle,\\
\end{split}
\end{equation}
where
\begin{equation}
\begin{split}
\Delta_{e1}^{'}&=\Delta_{e1}-\frac{g_{A}^{2}(\Delta_{r}+h)}{(\Delta_{r}+h)^2+(\kappa/2)^2}-\frac{g_{B}^{2}(\Delta_{r}-h)}{(\Delta_{r}-h)^2+(\kappa/2)^2},\\
\Gamma^{'}&=\Gamma_{e1}^{'}+\Gamma_{e2},\\
\Gamma_{e1}^{'}&=\Gamma_{e1}+\frac{g_{A}^{2}\kappa}{(\Delta_{r}+h)^2+(\kappa/2)^2}+\frac{g_{B}^{2}\kappa}{(\Delta_{r}-h)^2+(\kappa/2)^2},\\
\Omega_{p}^{'}&=g_{A}\alpha_{A}+ig_{B}\alpha_{B}.\\
\end{split}
\end{equation}
Therefore, the cavity modes are adiabatically eliminated. Notice that
\begin{equation}
\begin{split}
\langle\dot{\sigma_{22}}\rangle&=-i\Omega_{c}(\langle\sigma_{2e}\rangle-\langle\sigma_{e2}\rangle)+\Gamma_{e2}\langle{\sigma_{ee}}\rangle,\\
\end{split}
\end{equation}
the effective master equation for the three-level atom is
\begin{equation}
\label{eff_three_level}
\begin{split}
\dot{\rho}_{\text{a}}=-i[H_{\text{a}},\rho_{\text{a}}]+\frac{\Gamma_{e1}^{'}}{2}\mathcal{D}[\sigma_{1e}]\rho_{\text{a}}+\frac{\Gamma_{e2}}{2}\mathcal{D}[\sigma_{2e}]\rho_{\text{a}},\\
\end{split}
\end{equation}
where
\begin{equation}
\begin{split}
H_{\text{a}}&=\Delta_{e1}^{'}\sigma_{ee}+\Delta_{21}\sigma_{22}+\Omega_{p}^{'}\sigma_{e1}+\Omega_{p}^{'*}\sigma_{1e}+\Omega_{c}(\sigma_{2e}+\sigma_{e2}).\\
\end{split}
\end{equation}
To summarize, we mapped entire system of the atom coupled to the fiber-coupled microtoroidal cavity to the effective system which is represented by a three-level atom, which has Purcell enhanced decay rate and detuning on the $e-1$ leg of $\Lambda$ system , and is coupled to the classical fields $\Omega_{p}^{'}$ and $\Omega_{c}$, respectively  on the transitions $e-1$ and $2-e$.
We remark, that correlation functions for the output fields can be calculated by making following substitutions:
\begin{equation}
\begin{split}
&a_{\text{out},\text{ex}}\rightarrow\alpha_{0}+\alpha_{-}\sigma_{1e},\\
&b_{\text{out},\text{ex}}\rightarrow\beta_{0}+\beta_{-}\sigma_{1e},\\
&\alpha_{0}=\frac{i\Omega_{p}}{\sqrt{\kappa_{\text{ex}}}}+\sqrt{\kappa_{\text{ex}}/2}(\alpha_{A}+\alpha_{B}),\\
&\alpha_{-}=-\sqrt{\kappa_{\text{ex}}/2}\Big[\frac{ig_{A}}{i(\Delta_{r}+h-i\kappa/2)}+\frac{g_{B}}{i(\Delta_{r}-h-i\kappa/2)}\Big],\\
&\beta_{0}=\sqrt{\kappa_{\text{ex}}/2}(\alpha_{A}-\alpha_{B}),\\
&\beta_{-}=-\sqrt{\kappa_{\text{ex}}/2}\Big[\frac{ig_{A}}{i(\Delta_{r}+h-i\kappa/2)}-\frac{g_{B}}{i(\Delta_{r}-h-i\kappa/2)}\Big],\\
\end{split}
\end{equation}
after substituting these expressions into the  numerators of the Eq.~(\ref{transmission}) and Eq.~(\ref{correlation}) we obtain,
\begin{equation}
\label{badcavity}
\begin{split}
&\langle{a_{\text{out},\text{ex}}^{\dag}a_{\text{out},\text{ex}}}\rangle_{\text{ss}}\\
&=|\alpha_{0}|^2+\alpha_{0}^{*}\alpha_{-}\rho_{e1}^{\text{ss}}+\alpha_{0}\alpha_{-}^{*}\rho_{1e}^{\text{ss}}+|\alpha_{-}|^2\rho_{ee}^{\text{ss}},\\
&\langle{b_{\text{out},\text{ex}}^{\dag}b_{\text{out},\text{ex}}}\rangle_{\text{ss}}\\
&=|\beta_{0}|^2+\beta_{0}^{*}\beta_{-}\rho_{e1}^{\text{ss}}+\beta_{0}\beta_{-}^{*}\rho_{1e}^{\text{ss}}+|\beta_{-}|^2\rho_{ee}^{\text{ss}},\\
&\langle{a_{\text{out},\text{ex}}^{\dag}a_{\text{out},\text{ex}}^{\dag}a_{\text{out},\text{ex}}a_{\text{out},\text{ex}}}\rangle_{\text{ss}}\\
&=|\alpha_{0}|^2(|\alpha_{0}|^2+2\alpha_{0}^{*}\alpha_{-}\rho_{e1}^{\text{ss}}+2\alpha_{0}\alpha_{-}^{*}\rho_{1e}^{\text{ss}}+4|\alpha_{-}|^2\rho_{ee}^{\text{ss}}),\\
&\langle{b_{\text{out},\text{ex}}^{\dag}b_{\text{out},\text{ex}}^{\dag}b_{\text{out},\text{ex}}b_{\text{out},\text{ex}}}\rangle_{\text{ss}}\\
&=|\beta_{0}|^2(|\beta_{0}|^2+2\beta_{0}^{*}\beta_{-}\rho_{e1}^{\text{ss}}+2\beta_{0}\beta_{-}^{*}\rho_{1e}^{\text{ss}}+4|\beta_{-}|^2\rho_{ee}^{\text{ss}}),\\
\end{split}
\end{equation}
where $\rho_{e1}^{\text{ss}}=\langle{\sigma_{1e}}\rangle_{\text{ss}}$, $\rho_{1e}^{\text{ss}}=\langle{\sigma_{e1}}\rangle_{\text{ss}}$, $\rho_{ee}^{\text{ss}}=\langle{\sigma_{ee}}\rangle_{\text{ss}}$.
By solving  Eq.~(\ref{eff_three_level}) for the effective three-level system, the elements of the steady state density matrix are given by
\begin{equation}
\label{density}
\begin{split}
\rho_{ee}^{\text{ss}}&=\frac{a_{2}\Omega_{c}^2}{c_{0}+c_{2}\Omega_{c}^2+c_{4}\Omega_{c}^4+c_{6}\Omega_{c}^6},\\
\rho_{1e}^{\text{ss}}&=\frac{b_{2}\Omega_{c}^2+b_{4}\Omega_{c}^4}{c_{0}+c_{2}\Omega_{c}^2+c_{4}\Omega_{c}^4+c_{6}\Omega_{c}^6},\\
\end{split}
\end{equation}
where
\begin{equation}
\begin{split}
a_{2}&=|\Omega_{p}^{'}|^2\Delta_{21}^2\Gamma,\\
b_{2}&=\Omega_{p}^{'*}\Delta_{21}\left[\Gamma_{e2}|\Omega_{p}^{'}|^2-\Delta_{21}\Gamma_{e1}^{'}(\Delta_{e1}^{'}-i\Gamma/2)\right],b_{4}=\Omega_{p}^{'*}\Delta_{21}\Gamma_{e1}^{'},\\
c_{0}&=\Gamma_{e2}|\Omega_{p}^{'}|^2\left||\Omega_{p}^{'}|^2-\Delta_{21}^{2}+\Delta_{21}(\Delta_{e1}^{'}-i\Gamma/2)\right|^2,\\
c_{2}&=\Gamma_{e1}^{'}\Delta_{21}^2|\Delta_{e1}^{'}-i\Gamma/2|^2+2\Gamma\Delta_{21}^2|\Omega_{p}^{'}|^2+(\Gamma+\Gamma_{e2})|\Omega_{p}^{'}|^4,\\
c_{4}&=-2\Gamma_{e1}^{'}\Delta_{21}\Delta_{e1}^{'}+(\Gamma+\Gamma_{e2})|\Omega_{p}^{'}|^2,c_{6}=\Gamma_{e1}^{'},\\
\Gamma&=\Gamma_{e1}^{'}+\Gamma_{e2}.\\
\end{split}
\end{equation}
We remark, that when $\Delta_{21}=0$, $\rho_{1e}=0$ as it follows from Eq.~(\ref{density}) and is a consequence of coherent population trapping. This  clearly shows that for obtaining a quantum transistor we need a non-zero two photon detuning. To simplify the lengthy expressions for the intensities and correlation functions in certain limits, we resume to Taylor expanding the  density matrix elements in the limits of small and large driving fields, we find that when $\Omega_{c}\rightarrow0$, $\rho_{ee}^{\text{ss}}\approx\frac{a_{2}}{c_{0}}\Omega_{c}^2$ and $\rho_{1e}^{\text{ss}}\approx\frac{b_{2}}{c_{0}}\Omega_{c}^2$; on the other hand, when $\Omega_{c}\rightarrow\infty$, $\rho_{ee}^{\text{ss}}\approx\frac{a_{2}}{c_{6}}\Omega_{c}^{-4}-\frac{a_{2}c_{4}}{c_{6}^2}\Omega_{c}^{-6}$ and $\rho_{1e}^{\text{ss}}\approx\frac{b_{4}}{c_{6}}\Omega_{c}^{-2}+\frac{b_{2}c_{6}-b_{4}c_{4}}{c_{6}^2}\Omega_{c}^{-4}$. It is straightforward to demonstrate bys using these expressions that in both limits $\rho_{1e} \approx 0$, whic means that atoms is decoupling from the cavity in this limit(note that absorption is given by the imaginary part of the off-diagonal term of the density matrix).
\begin{figure}
\begin{tabular}{ccccc}
\includegraphics[width=4cm]{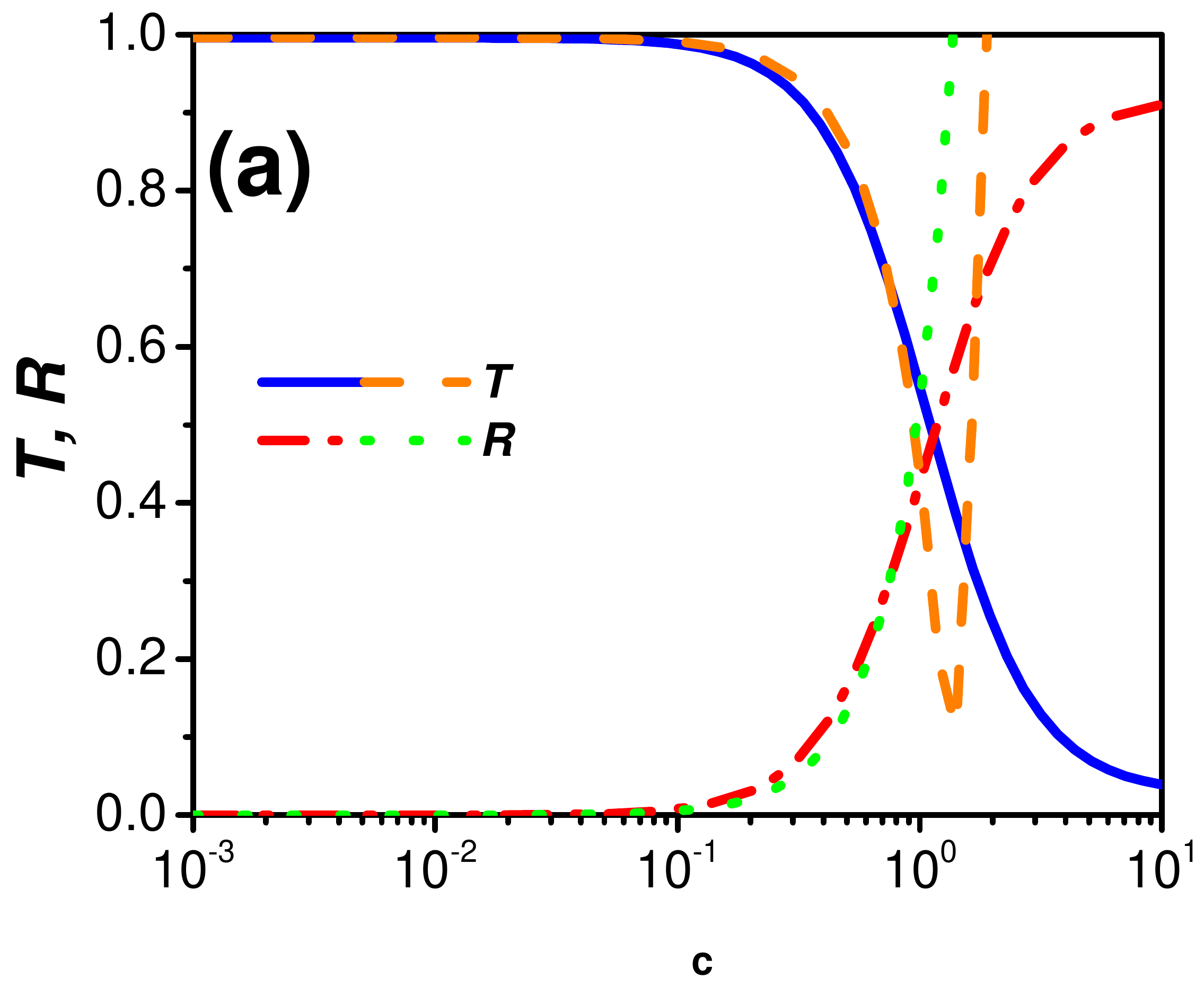} &
\includegraphics[width=4cm]{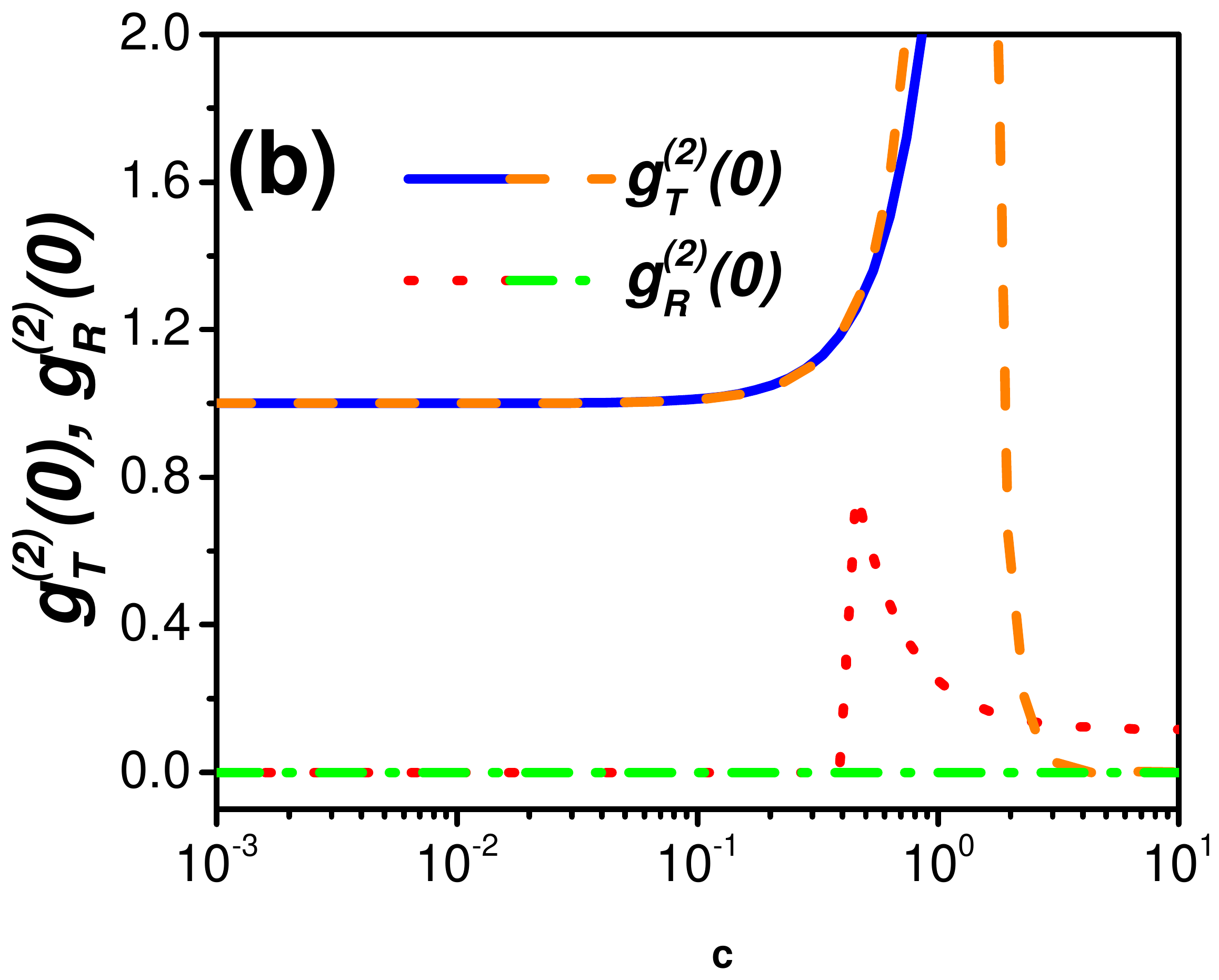} \\
\includegraphics[width=4cm]{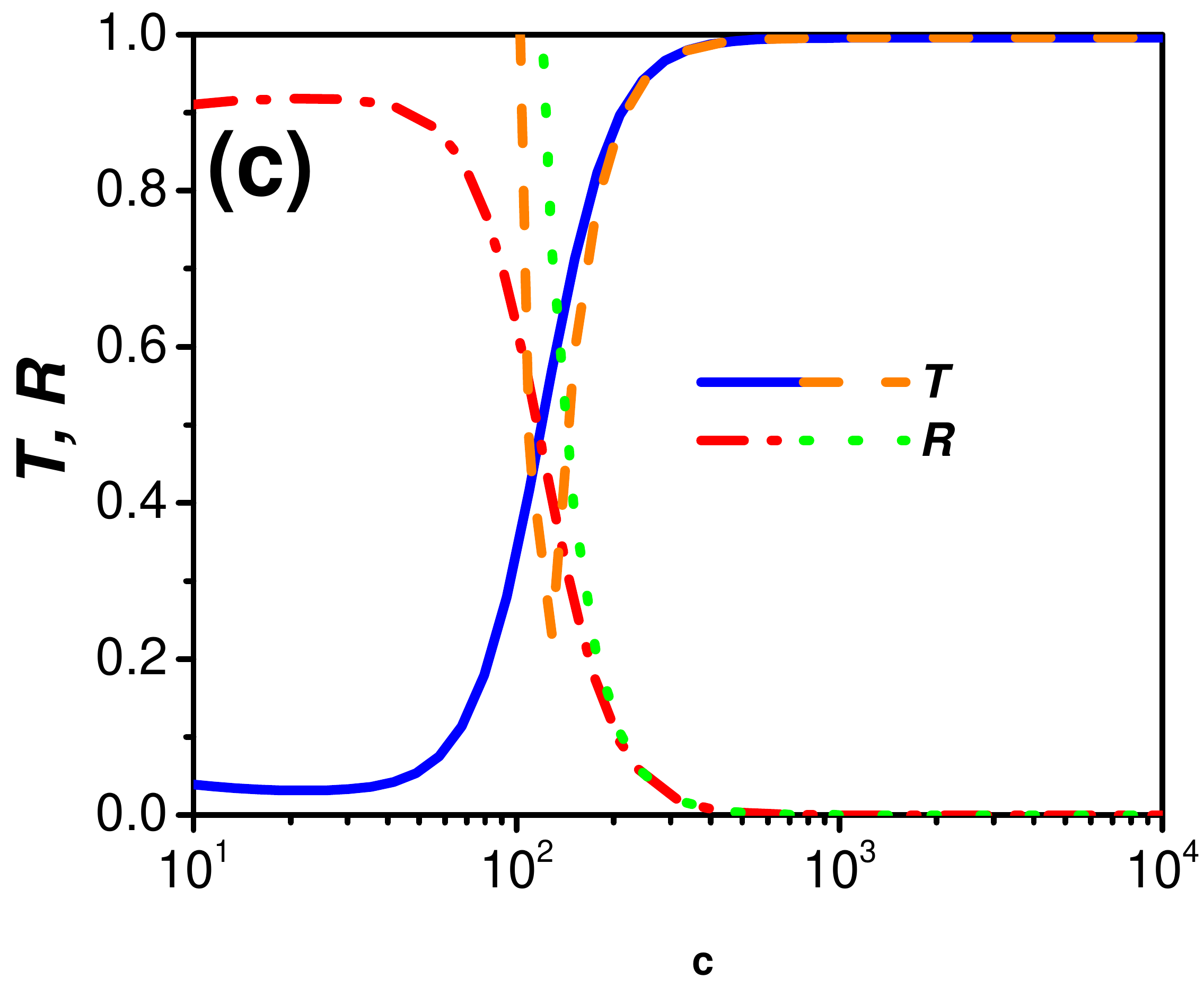} &
\includegraphics[width=4cm]{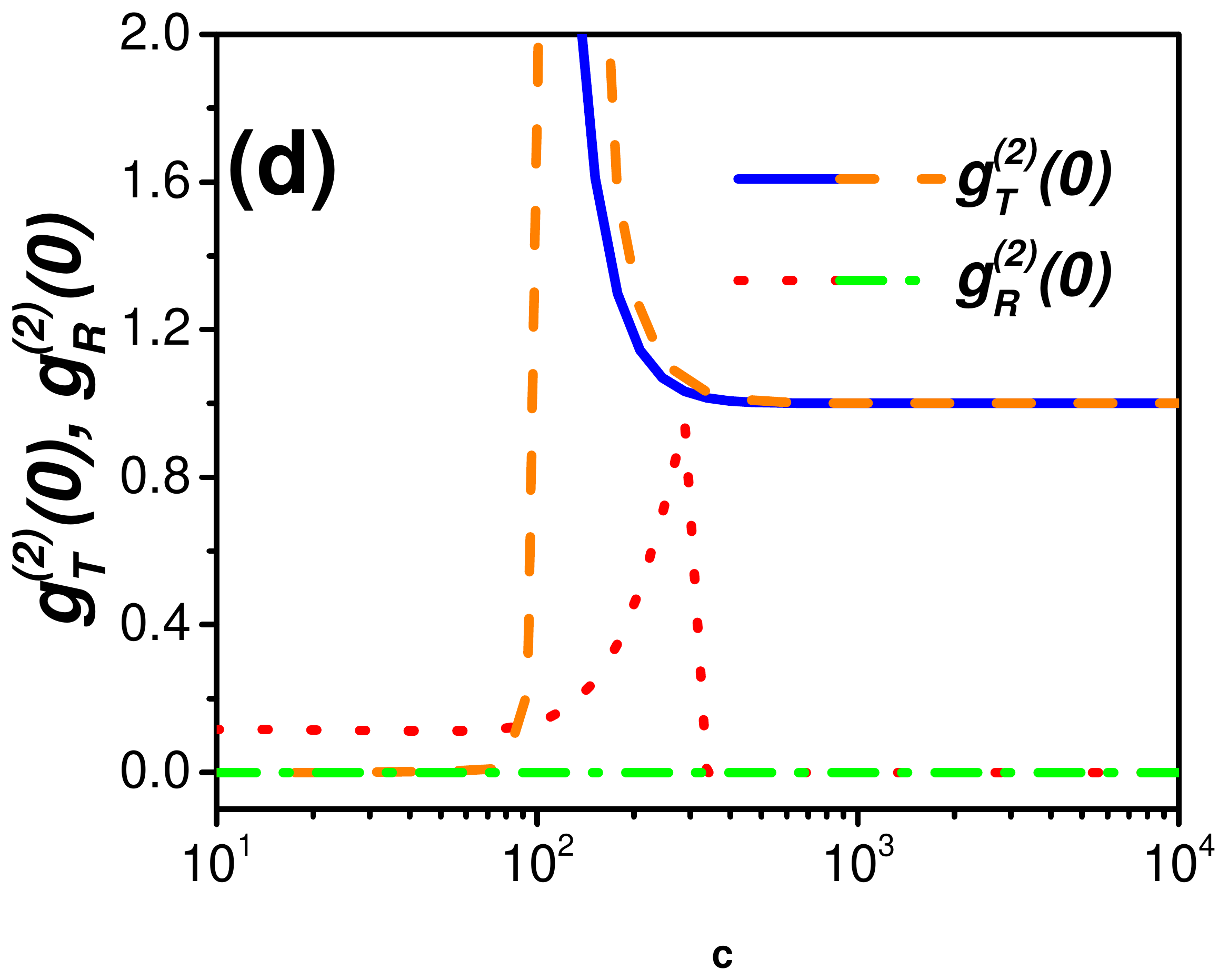} \\
\end{tabular}
\caption{(colour online). Taylor expansion for Fig. \ref{switch_demo}(g) in the main text. a and b are the Taylor expansion of transmission, reflection and correlations when $\Omega_{c}\rightarrow0$; c and d are the Taylor expansion of transmission, reflection and correlations when $\Omega_{c}\rightarrow\infty$.
}
\label{taylor}
\end{figure}

After substituting these expressions into Eqs.~{\ref{badcavity}}, we have in the limit $\Omega_{c}\rightarrow0$,
\begin{equation}
\begin{split}
&T=\frac{|\alpha_{0}|^2+(2\text{Re}[\alpha_{0}\alpha_{-}^{*}b_{2}/c_{0}]+|\alpha_{-}|^2a_{2}/c_{0})\Omega_{c}^2}{|\Omega_{p}|^{2}/2\kappa_{ex}},\\
&R=\frac{|\beta_{0}|^2+(2\text{Re}[\beta_{0}\beta_{-}^{*}b_{2}/c_{0}]+|\beta_{-}|^2a_{2}/c_{0})\Omega_{c}^2}{|\Omega_{p}|^{2}/2\kappa_{ex}},\\
&g_{\text{T}}^{(2)}(0)=\frac{|\alpha_{0}|^2(|\alpha_{0}|^2+4(\text{Re}[\alpha_{0}\alpha_{-}^{*}b_{2}/c_{0}]+|\alpha_{-}|^2a_{2}/c_{0})\Omega_{c}^2)}{(|\alpha_{0}|^2+(2\text{Re}[\alpha_{0}\alpha_{-}^{*}b_{2}/c_{0}]+|\alpha_{-}|^2a_{2}/c_{0})\Omega_{c}^2)^2},\\
&g_{\text{R}}^{(2)}(0)=\frac{|\beta_{0}|^2(|\beta_{0}|^2+4(\text{Re}[\beta_{0}\beta_{-}^{*}b_{2}/c_{0}]+|\beta_{-}|^2a_{2}/c_{0})\Omega_{c}^2)}{(|\beta_{0}|^2+(2\text{Re}[\beta_{0}\beta_{-}^{*}b_{2}/c_{0}]+|\beta_{-}|^2a_{2}/c_{0})\Omega_{c}^2)^2};\\
\end{split}
\end{equation}
and for the limit $\Omega_{c}\rightarrow\infty$ we obtain,
\begin{widetext}
\begin{equation}
\begin{split}
&T=\frac{|\alpha_{0}|^2+2\text{Re}[\alpha_{0}\alpha_{-}^{*}b_{4}/c_{6}]\Omega_{c}^{-2}+(2\text{Re}[\alpha_{0}\alpha_{-}^{*}(b_{2}c_{6}-b_{4}c_{4})/c_{6}^2]+|\alpha_{-}|^2a_{2}/c_{6})\Omega_{c}^{-4}-|\alpha_{-}|^2a_{2}c_{4}/c_{6}^{2}\Omega_{c}^{-6}}{|\Omega_{p}|^{2}/2\kappa_{ex}},\\
&R=\frac{|\beta_{0}|^2+2\text{Re}[\beta_{0}\beta_{-}^{*}b_{4}/c_{6}]\Omega_{c}^{-2}+(2\text{Re}[\beta_{0}\beta_{-}^{*}(b_{2}c_{6}-b_{4}c_{4})/c_{6}^2]+|\beta_{-}|^2a_{2}/c_{6})\Omega_{c}^{-4}-|\beta_{-}|^2a_{2}c_{4}/c_{6}^{2}\Omega_{c}^{-6}}{|\Omega_{p}|^{2}/2\kappa_{ex}},\\
&g_{\text{T}}^{(2)}(0)=\frac{|\alpha_{0}|^2(|\alpha_{0}|^2+4\text{Re}[\alpha_{0}\alpha_{-}^{*}b_{4}/c_{6}]\Omega_{c}^{-2}+4(\text{Re}[\alpha_{0}\alpha_{-}^{*}(b_{2}c_{6}-b_{4}c_{4})/c_{6}^2]+|\alpha_{-}|^2a_{2}/c_{6})\Omega_{c}^{-4}-4|\alpha_{-}|^2a_{2}c_{4}/c_{6}^{2}\Omega_{c}^{-6})}{(|\alpha_{0}|^2+2\text{Re}[\alpha_{0}\alpha_{-}^{*}b_{4}/c_{6}]\Omega_{c}^{-2}+(2\text{Re}[\alpha_{0}\alpha_{-}^{*}(b_{2}c_{6}-b_{4}c_{4})/c_{6}^2]+|\alpha_{-}|^2a_{2}/c_{6})\Omega_{c}^{-4}-|\alpha_{-}|^2a_{2}c_{4}/c_{6}^{2}\Omega_{c}^{-6})^2},\\
&g_{\text{R}}^{(2)}(0)=\frac{|\beta_{0}|^2(|\beta_{0}|^2+4\text{Re}[\beta_{0}\beta_{-}^{*}b_{4}/c_{6}]\Omega_{c}^{-2}+4(\text{Re}[\beta_{0}\beta_{-}^{*}(b_{2}c_{6}-b_{4}c_{4})/c_{6}^2]+|\beta_{-}|^2a_{2}/c_{6})\Omega_{c}^{-4}-4|\beta_{-}|^2a_{2}c_{4}/c_{6}^{2}\Omega_{c}^{-6})}{(|\beta_{0}|^2+2\text{Re}[\beta_{0}\beta_{-}^{*}b_{4}/c_{6}]\Omega_{c}^{-2}+(2\text{Re}[\beta_{0}\beta_{-}^{*}(b_{2}c_{6}-b_{4}c_{4})/c_{6}^2]+|\beta_{-}|^2a_{2}/c_{6})\Omega_{c}^{-4}-|\beta_{-}|^2a_{2}c_{4}/c_{6}^{2}\Omega_{c}^{-6})^2}.\\
\end{split}
\end{equation}
\end{widetext}
To demonstrate these results we compare Taylor expansion results with analytical results for the adiabatic elimination on  Fig. \ref{taylor} .


\end{document}